\documentclass{article}

\usepackage{subcaption}
\usepackage{PRIMEarxiv}

\usepackage{lscape} 
\usepackage{rotating}
\usepackage[utf8]{inputenc} 
\usepackage[T1]{fontenc}    
\usepackage[hidelinks]{hyperref}       
\usepackage{url}            
\usepackage{booktabs}       
\usepackage{amsfonts}       
\usepackage{nicefrac}       
\usepackage{microtype}      
\usepackage{lipsum}
\usepackage{fancyhdr}       
\usepackage{graphicx}       
\graphicspath{{media/}}     
\usepackage[square,sort,comma,numbers]{natbib}

\usepackage{float}
\usepackage{amsmath,amssymb}
\usepackage[ruled,vlined]{algorithm2e}

\usepackage{xcolor}

\newcommand{\revision}{\textcolor{black}}

\usepackage{grffile}

\usepackage{bbm}

\title{Detectability of varied hybridization scenarios using genome-scale hybrid detection methods
}

\author{
  Marianne Bj{\o}rner \\
  Department of Computer Sciences \\
  Wisconsin Institute for Discovery \\
  University of Wisconsin-Madison\\
  Madison, WI\\
   \\
   \And
   Erin K. Molloy\\
   Department of Computer Sciences\\
   University of Maryland\\
   College Park, MD\\
   \And
   Colin N. Dewey\\
   Department of Biostatistics \\ and Medical Informatics\\
     University of Wisconsin-Madison\\
  Madison, WI\\
  \And
  Claudia Sol\'{i}s-Lemus\thanks{Corresponding author:  \texttt{solislemus@wisc.edu}} \\
  Department of Plant Pathology \\
  Wisconsin Institute for Discovery \\
  University of Wisconsin-Madison\\
  Madison, WI\\
}

\date{}

\begin{document}

\maketitle
\begin{abstract}
Hybridization events complicate the accurate reconstruction of phylogenies, as they lead to patterns of genetic heritability that are unexpected under traditional, bifurcating models of species trees. This phenomenon has led to the development of methods to infer these varied hybridization events, both methods that reconstruct networks directly, as well as summary methods that predict individual hybridization events from a subset of taxa. 
However, a lack of empirical comparisons between methods -- especially those pertaining to large networks with varied hybridization scenarios -- hinders their practical use. 
Here, we provide a comprehensive review of popular summary methods: TICR, MSCquartets, HyDe, Patterson's D-Statistic (ABBA-BABA), D3, and D$_p$. 
TICR and MSCquartets are based on quartet concordance factors gathered from gene tree topologies and \revision{HyDe, Patterson's D-Statistic, D3, and D$_p$ use site pattern frequencies to identify hybridization events between sets of three taxa}. 
We then use simulated data to address questions of method accuracy and ideal use scenarios by testing methods against
complex networks which depict gene flow events that differ in depth (timing), quantity (single vs. multiple, overlapping hybridizations), and rate of gene flow ($\gamma$). 
We find that deeper or multiple hybridization events may introduce noise and weaken the signal of hybridization, leading to higher relative false negative rates across all methods.
Despite some forms of hybridization eluding quartet-based detection methods, MSCquartets displays high precision in most scenarios. While HyDe results in high false negative rates when tested on hybridizations involving \revision{extinct or unsampled ghost lineages, HyDe is the only method able to identify the direction of hybridization, distinguishing the source parental lineages from recipient hybrid lineages.}
Lastly, we test the methods on a dataset of ultraconserved elements from the bee subfamily Nomiinae, finding
\revision{possible} hybridization events between clades which correspond to regions of poor support in the species tree estimated in \revision{a previous} study.
\end{abstract}

\section{Introduction}
Phylogenetics studies the evolutionary history between organisms. 
In many popular phylogenetic inference models, these relationships are assumed to be best represented as a binary tree, where each child node arises from only one direct parent 
\cite{raxml, beast2, iq-tree}. However, a binary tree model ignores the possibility of a reticulation 
or gene flow event. 
Gene flow occurs when members of one population reproduce or otherwise exchange genetic information with another population, \revision{which leads to the formation of admixed populations or new hybrid species lineages}
\cite{barton1985-hyb-zone}. 
These reticulation events transform bifurcating
phylogenetic trees into network structures, wherein \revision{the} taxa affected have more than one parental lineage \cite{phylo-networks-rice}. 
Non-tree-like evolution is common across the tree of life, found in groups such as insects \cite{introgression-drosophila}, plants \cite{introgression-wild-tomatoes} and mammals \cite{introgression-humans}. 
    
The study of gene flow events in the tree of life have been aided by recent advances in 
sequencing technology, granting evolutionary researchers access to genome-scale information. This abundance of information 
can be used to infer reticulation events such as introgression, hybrid speciation, and horizontal gene transfer.
Each mechanism for gene flow leaves behind various traces in a population's genetic information, and may be identified through hybridization detection methods that leverage gene trees or sequence information \cite{phylog-approaches-review}. \revision{Though the exact biological processes originating gene flow might differ, in this paper, we broadly refer to descendants of reticulation events as hybrids, and the methods that detect them as hybridization methods.}
	
Many \revision{existing} methods to infer phylogenies are based on a binary tree, where they do not account for reticulations between taxa. While this assumption limits the search space to only trees, 
it might \revision{be an unreasonable} assumption, especially for populations where gene flow is common or expected.
Methods to infer phylogenetic networks\revision{, such as those that use maximum likelihood, Bayesian inference, and combinatorial techniques}   \revision{\cite{solisane, zhang2018bayesian, wen2018inferring, Zhang2017, WenNakleh2018, Allman2019NANUQ}} are becoming increasingly popular for their ability to overcome the strictly bifurcating assumption.
\revision{While valuable for studies of few taxa, these methods become very computationally expensive with increasingly large datasets.} 
Furthermore, most network methods require specification 
of the number of expected hybrid events, as any increase in number of reticulations artificially increases the likelihood of the network \cite{RF-Net2}. 
    
Alternatively, to find evidence for individual hybridization events, 
summary methods analyze subsets of triples or quartets of taxa -- which is an intrinsically more scalable endeavor than the search in network space -- without any predetermination of the total of number of hybrids in the phylogeny \cite{phylog-approaches-review}. 
Despite these advantages, summary methods still require comparisons to each other in order to address questions of method accuracy.
The hybrid detection methods compared in this simulation study (Table \ref{tab:summary} \revision{in Supplementary Material}) are \revision{MSCquartets \cite{mscquart}, TICR \cite{TICR-original},} HyDe \cite{hyde-paper}, Patterson's D-Statistic \cite{pattersonD}, also known as the ABBA-BABA test, as well as methods derived from the D-Statistic such as D$_p$ \cite{Dp} and D3 \cite{D3}. \revision{These methods (excluding TICR)} identify specific hybrid relationships within either \revision{subsets of four or three taxa. TICR, in contrast,} tests for how well a binary population tree fits the data \revision{(with failure to reject the population tree suggesting no hybridization).}
    
Here, we address questions of method accuracy and ideal use scenarios by testing these five summary methods against complex 
networks which depict gene flow events that differ in depth (timing), quantity (number of hybridization events: single vs. multiple),
and proportion of genes transferred through the hybridization event (inheritance probability $\gamma$). We note that we are not treating hybridization events as continuous flow of genes over a period of time\revision{; i}nstead, we \revision{treat each} hybridization event \revision{as being instantaneous, representing it} by a single arrow (see Figure \ref{fig:abba-baba}).
The hybridization \revision{scenarios also} differ in terms of time consistency, 
a characteristic of reconstructed networks that, when violated, arises in the presence of incomplete sampling or 
ghost lineages \cite{phylo-networks-rice, Tricou2022, Pang2022}.
Finally, we use hybridization detection methods to analyze published empirical data from the bee subfamily Nomiinae, a dataset complete with sequences, and \revision{estimated} gene and species trees.

\section{Materials and Methods}

\revision{We begin this section by reviewing the five methods evaluated in our study, breaking them into two classes: methods that take gene trees as input and methods that take molecular sequences as input. We then describe the generation of synthetic data and the metrics for evaluating methods. Lastly, we outline our re-analysis of a dataset an ultraconserved elements (UCEs) for the subfamily Nomiinae.}

\subsection{Methods based on gene trees}
\begin{figure*}
    \centering
    \includegraphics[width=0.9\linewidth]{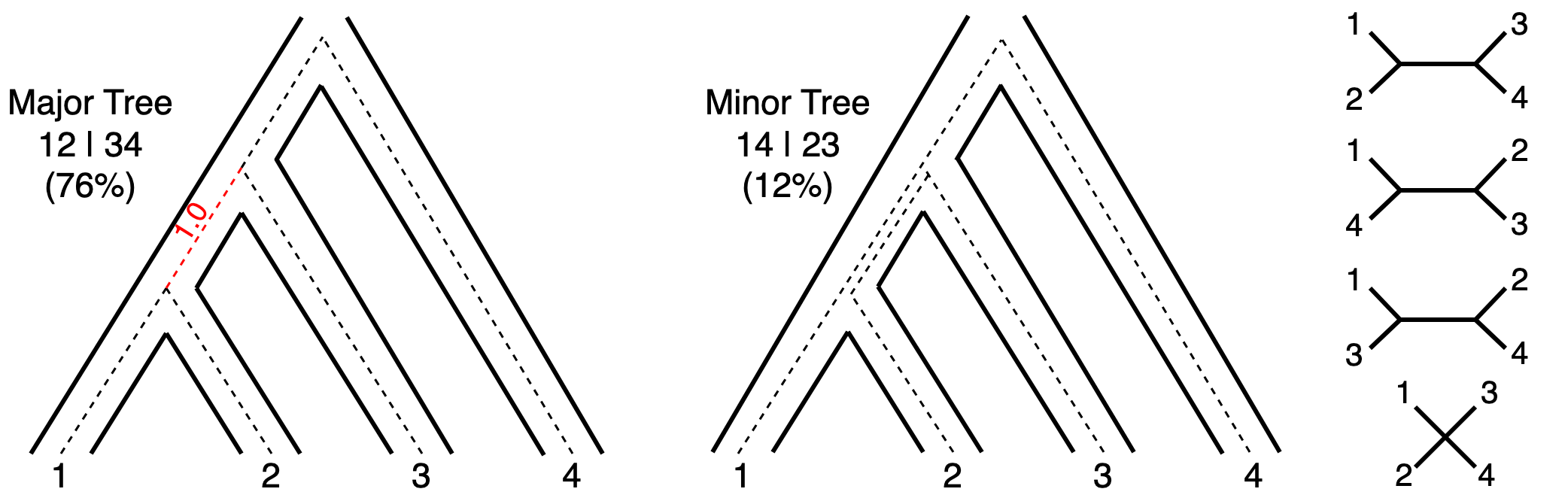} 
    \caption{\revision{Concordant (left) and discordant (middle) gene and species histories. Solid lines denote species history, and dotted lines denote an individual gene history. 
    The middle tree shows an example of incomplete lineage sorting (ILS) when the gene history fails to coalesce with the ancestral speciation event. The tree on the left (after unrooting) corresponds to the bipartition $12|34$ and agrees with the unrooted species tree; thus it is referred to as the "major tree". Under the MSC model, this tree is expected to have a frequency of 76\% in a given sample of gene trees because the internal branch of $1.0$ coalescent unit. The tree in the center (after unrooting) corresponds to the bipartition $14|23$ and disagrees to the unrooted species tree; thus, it is referred to as a "minor tree".
    Under the MSC model, the minor trees  have an expected frequency of 12\%. The far right depicts four possible quartet relationships, used by MSCquartets \cite{Allman-simplex-plot-MSC} and TICR \cite{TICR-original} depicted from top to bottom: the major tree ($12|34$), the minor trees ($14|23$ and $13|24$), and the star tree ($1234$), which neither agrees nor disagrees with the species tree.}}
    \label{fig:ils_msc_splits}
\end{figure*}

MSCquartets \revision{\cite{Allman-simplex-plot-MSC, mscquart}} and TICR \cite{TICR-original, TICR-GOF-Julia} both rely on frequencies of quartet gene tree topologies to conduct tests into how well a tree\revision{-like evolution} fits the data and whether there is evidence for hybridizations. 
These tests are based on the multispecies coalescent model (MSC) which defines expected distributions of gene tree topologies under incomplete
lineage sorting (ILS) \revision{\cite{MSC-Allman, rannalayang2003}}. ILS, also called deep coalescence, occurs when individual gene histories fail to coalesce at the same time as their given species history \cite{deep-coalescence}. An example is shown in Figure \revision{\ref{fig:ils_msc_splits}} \revision{(note that this figure displays gene trees that are rooted and ultrametric; however, only the unrooted gene tree topologies are used by MSCquartets and TICR (i.e., the input to these methods is not required to be ultrametric).}

    \revision{A resolved tree on four taxa (called a quartet) can take on one of three unrooted} topologies\revision{: one} concordant and \revision{two} discordant
    with the species tree (Figure \ref{fig:ils_msc_splits}).
    \revision{Under the MSC model, the probability of the concordant gene tree is strictly greater than that of the two discordant gene trees, which have equal probability \cite{Allman-Phylogenetic-Invariants} (note that for five or more taxa, the most probable unrooted gene tree may not be concordant with the unrooted species tree \cite{degnan2013anomalous}).
    Given a model species tree and a number of gene trees, we can compute the expected frequencies of each quartet, referred to as} quartet count concordance factors (qcCFs), where the \revision{qc}CFs of concordant topologies are called major \revision{qc}CFs, as
    they align with the major tree, or species tree, and the \revision{qc}CFs of discordant topologies are called minor \revision{qc}CFs \cite{Allman-simplex-plot-MSC}. 
    \revision{The observed qcCFs} are gathered \revision{by} calculating the frequency of each of three possible resolved quartet trees (Figure \ref{fig:ils_msc_splits}) across \revision{the input} set of gene trees \cite{Allman-simplex-plot-MSC}, \revision{typically normalizing by the number of gene trees displaying any one of the three possible quartets}.
    \revision{Major qcCFs are expected to be greater than the two minor qcCFs, and the two m}inor CFs are expected \revision{to be} equal under MSC.
    Deviation from this expectation violates an ILS-only model of genetic inheritance\revision{, and MSCquartets leverages this invariant, looking at singularities in the space of possible topologies for each subset of four species}.     
    TICR, \revision{on the other hand,} uses qcCFs to conduct a $\chi^2$ goodness-of-fit test against a specific \revision{model species tree}.
    
    
    \subsubsection{MSCquartets}
        MSCquartets \cite{Allman-simplex-plot-MSC} is a\revision{n R} package \cite{MSCquartets-RPackage} 
        \revision{that takes as input previously inferred} gene trees\revision{, from which it computes the observed qcCFs. If the input gene trees are not fully resolved, it is possible to have a star tree when looking at four taxa} (Figure \revision{\ref{fig:ils_msc_splits}}).
        Unresolved (star) trees \revision{can optionally be removed or redistributed among the resolved topologies}.
        \revision{The resulting} qcCFs are \revision{then} compared to \revision{the} expected 
        \revision{invariant}
        defined in \cite{mscquart} derived from the \revision{MSC} model to test the hypothesis of whether a specific four-taxon subset follows a tree-like pattern (in agreement with the MSC model) or not. 
        Each hypothesis test  produces a p-value. 
        \revision{We note that MSCquartets does not require a specific tree topology to test the qcCF expectations against, instead utilizing the information on all three qcCFs per four-taxon subset to compute the test statistic}.
When \revision{expectations are} violated, it leads to low support for a tree-like species history
between the four taxa. 
In other words, significant results fail to support an ILS-only model of evolution, 
and hybridization becomes a possible explanation for the imbalanced relationship between quartets.
        
In terms of \revision{computational} efficiency, 
\revision{MSCquartets is used in a pipeline that requires costly preprocessing steps to produce the input, including aligning sequences and estimating gene trees. With these precalculated, the two primary factors that influence the speed of this approach are the number $n$ of taxa and the number $g$ of gene trees. Just consider that computing the observed qcCFs can be done by identifying the quartet displayed by a gene tree for each of the $\frac{n!}{4!(n-4)!}$ possible subsets of four species, repeating across all gene trees. This procedure alone would give the time complexity of MSCquartets a lower bound of $o(n^{4}g)$. Thus, MSCquartets may be time consuming for large numbers of taxa.}

  \subsubsection{TICR}
In its original implementation, TICR -- Tree Incongruence Checking in R -- \cite{TICR-original} was used as part of a pipeline that begins with a \revision{set of} alignment\revision{s (one per gene), estimates} gene trees, calculates \revision{qcCFs} from \revision{the (estimated)} gene trees, and finally calculates a population tree  \revision{based on the qcCFs} using the software \texttt{Quartet-Max-Cut} \cite{TICR-original, quartet-max-cut}. However, the only inputs required by TICR are \revision{the observed} qcCFs \revision{computed} from gene trees and \revision{the} expected \revision{qc}CFs calculated from a hypothesized population tree. Recently, this method was extended to test goodness-of-fit on a given population network, rather than population tree \cite{TICR-GOF-Julia}.
We note that in our experiments we use the TICR version implemented in the Julia package called \texttt{QuartetNetworkGoodnessFit.jl} \cite{TICR-GOF-Julia} though we restrict our tests to the case of population trees, not networks.

Given a \revision{fully resolved} population tree \revision{with branch lengths} in coalescent time, the expected probabilities of observing quartet relationships can be directly computed. For example, in Figure \revision{\ref{fig:ils_msc_splits} assuming the} internal branch \revision{in red has length} $t=1.0$ coalescent units, the probability of the major gene tree is given by $1-\frac{2}{3}e^{-t}=0.76$.
TICR computes a $\chi^2$ goodness-of-fit test statistics that evaluates the fit of the \revision{observed qcCFs} to the expected \revision{qc}CFs under the ILS-only model.
TICR can also be used to test for the likelihood of panmixia, or a star tree which occurs when all taxa arise from the same common ancestor and diverge at the same time, \revision{though any occurrences of star trees in the input gene trees are ignored when calculating qcCFs}. 
TICR uses the p-values of the individual tests to form an overall test that inspects whether the distribution of observed qcCFs falls within the expected qcCFs of the input tree or network. This overall test indicates whether the proposed population tree fits the observed \revision{qcCF}s. \revision{Although} TICR does not directly test for the presence or absence of specific hybridizations,
by failing to reject a specific tree model, it provides lack of evidence for hybridization.

In terms of \revision{computational} efficiency, \revision{the remarks made above for MSCquartets apply to TICR, although it is worth noting that the TICR pipeline additionally needs to estimate a species tree and compute the expected qcCFs based on it.}

\subsection{Methods based on \revision{(aligned)} sequences}
HyDe \cite{hyde-paper}, Patterson's D-Statistic \cite{pattersonD}, D3 \cite{D3}, and D$_p$ \cite{Dp} are all methods that use site pattern frequencies or pairwise differences to test the null hypothesis of tree-like evolutionary patterns (ILS-only). 
This eliminates the need for the estimation of gene trees, such that sequences can be used directly as input. \revision{However, it is important to use the input of many gene sequences, as opposed to few long sequences, as these tests are based on a model of coalescent independent sites, or designed for use at the allele-level \cite{hyde-paper, pattersonD}. These sequences must also come from equidistant gene trees, where each taxon is equidistant from the root}.
While HyDe, Patterson's D-Statistic, and D$_p$ 
use rooted triples \revision{plus} an outgroup, D3 uses a rooted triple without an outgroup. 
This is advantageous as a poorly chosen, or distant
outgroup can result in inclusion of ghost hybridizations and lead to false interpretations \cite{Tricou2022, D3}. Ghost hybridizations are defined as hybridization events when one (or both) of the parent lineages that provide genetic material to the hybrid node are either extinct or unsampled. 
Note that the Patterson's D-statistic, D3, and D$_p$ are intended as tests for introgression between two species, but do not indicate directionality. 
Additionally, they require that the species relationship between the triple and its outgroup (if any) is known.

\subsubsection{HyDe}
Distinct from other methods which evaluate for the overall presence of hybridization, HyDe \cite{hyde-python-package} identifies a singular parent-hybrid relationship between a triple, given its outgroup. It can also be used to estimate a mixing parameter $\gamma$, depicting the proportion of genetic material contributed by each parental lineage. 
        
HyDe is based on phylogenetic invariants, or a function of site pattern probabilities, which evaluate to zero when consistent with given associated tree models \cite{MSC-Allman, hyde-paper}. 
The linear invariants ($f_1$ and $f_2$) are themselves representative of mixing parameters $\gamma$ and $1 - \gamma$, respectively, 
\begin{align*} 
    f_1 &= p_{iijj|(S\gamma ,\tau )} - p_{ijij|(S\gamma ,\tau )} \\
    f_2 &=  p_{ijji|(S\gamma ,\tau )} - p_{ijij|(S\gamma ,\tau )}
\end{align*}
where $p_{iijj|(S\gamma ,\tau )}$ represents the probability for the site pattern $iijj$ under the species triple tree $S\gamma$ with $\gamma$ as inheritance probability and $\tau$ as divergence times.
As a result, HyDe can also be used to estimate the mixing parameter $\gamma$ between two taxa that are putative parent lineages of a proposed hybrid as the ratio of $\frac{f_1}{f_2}$ represents $\frac{\gamma}{1-\gamma}$. When there is no \revision{hybridization}, $\gamma$ is 0, \revision{so} the ratio is expected to be zero.

Site pattern probabilities observed in the sample are used to form estimates of $f_1$ and $f_2$, along with means and variance. These are then rearranged with the Geary-Hinkley transformation to form the Hils statistic:
\begin{equation*} 
    H:= \frac{\hat{f}_2(\frac{\hat{f}_1}{\hat{f}_2}-\frac{\mu_{f_1}}{\mu_{f_2}})} 
    {\sqrt{{\hat{\sigma}^2}_{f_2} (\frac{\hat{f}_1}{\hat{f}_2})^2 - 2\hat{\sigma}_{f1, f2}\frac{\hat{f}_1}{\hat{f}_2} + \hat{\sigma}_{f1}^2 }}
\end{equation*}
which follows the normal distribution with $N(0,1)$ for large $n$ under the null hypothesis of ILS-only, or no hybrid speciation \cite{hyde-paper}. This allows direct interpretation of HyDe test results without the need for resampling by bootstrapping. However, significance levels should be adjusted with a Bonferroni correction due to multiple hypothesis testing, as HyDe considers all possible combinations of sets of three taxa, where one hybrid taxon is tested for every two distinct parent taxa.
        
As a C-backed python package, HyDe is designed for use with multiple individuals per taxon, from a phylogeny where the outgroup is specified, and aligned sequence information is provided in PHYLIP format. Power increases with increasing sequence length, with a recommendation that sequence length is at minimum 50 kbp \cite{hyde-python-package, KongKubatko}.
In addition, HyDe outputs counts of site pattern observations, namely AABB, ABBA, AABC,... that can be used for calculations of other statistics based on site pattern frequencies such as the Patterson's D-Statistics (ABBA-BABA) \cite{pattersonD}, D$_p$ \cite{Dp}, and can be rearranged for use in pairwise distance metrics, such as D3 \cite{D3}.
        
\revision{In terms of computational efficiency, HyDe calculates site pattern frequencies \revision{from subsets of sequences in an} alignment\revision{; thus,} its speed is impacted by the number of taxa and alignment length.} \revision{HyDe is capable of processing alignments of 20 taxa \revision{and} 100,000 {sites}  in under a minute \cite{hyde-paper}.} For each set of three taxa $(t_1, t_2, t_3)$, \revision{HyDe labels two as parental populations, and one as a hybrid population. HyDe evaluates each of the} three possible hybrid relationships \revision{between the taxa}.
        
        
Sites from each taxa $t_1$, $t_2$, and $t_3$ are then compared to sites from an outgroup to calculate the Hils statistic from $f1$ and $f2$. For a set of $n + 1$ taxa, (where $n$ is \revision{number of} taxa \revision{ex}cluding the outgroup), each \revision{with an aligned} sequence \revision{of} length $L$, HyDe performs $3\frac{n!}{3!(n-3)!}$ tests, where each test involves comparison of $L$ sites, across each of the three taxa for hybrid testing, plus the outgroup resulting in a \revision{time} complexity of
$O(n^{3} L)$.
        

\subsubsection{Patterson's D-Statistic (ABBA-BABA)}
\begin{figure*}
    \centering
    \includegraphics[width=.95\linewidth]{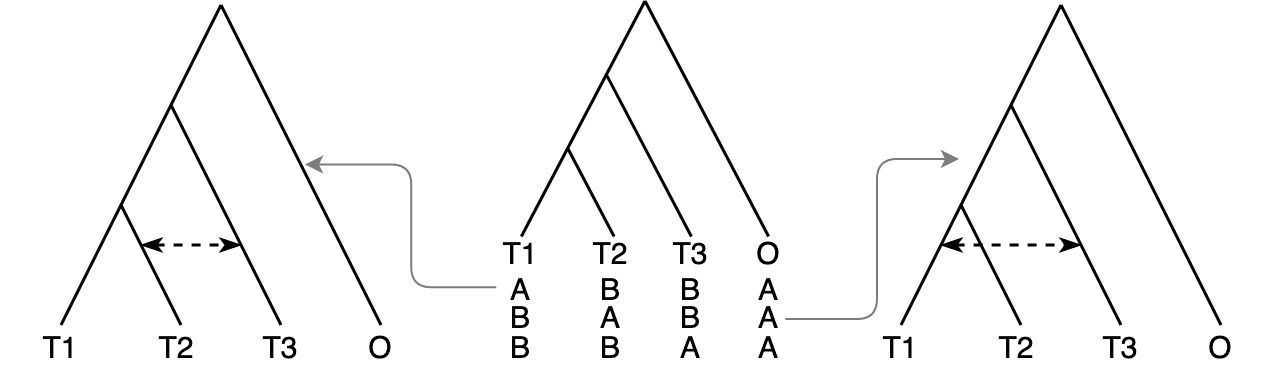}
    \caption{\textbf{Center:} Population tree of the form (((T1,T2),T3),O). Under each taxa, ABBA, BABA, and BBAA are three possible site patterns, with positions in the site pattern sequence corresponding to particular taxa. Under ILS, the proportion of sites where T3 and T1 match should be equal to the proportion of sites where T2 and T3 match, given that T3 is equidistant on the species tree from T1 and T2. \textbf{Left:} An excess of ABBA, corresponds to a gene flow event between T2 and T3. \textbf{Right:} An excess of BABA, corresponds to a gene flow event between T1 and T3.}
    \label{fig:abba-baba}
\end{figure*}

Patterson's D-Statistic \cite{pattersonD}, much like HyDe, involves the calculation of site pattern frequencies from SNPs and scales in a similar manner with respect to sequence length and number of taxa to compare. Patterson's D-statistic is defined as
$D = \frac{ABBA - BABA}{ABBA + BABA}$
where $A$ and $B$ are \revision{distinct alleles}. Each of the four positions in these nucleotide sequences belong to one of four taxa: three of these taxa are compared to each other for hybrid identification, and the fourth is used as an outgroup. Under the null hypothesis of no hybridization, the D-statistic is expected to be zero, as the frequency of ``ABBA'' and ``BABA'' \revision{patterns} should be equal. Significant deviation from 0 \revision{are attributed to} gene flow events (Figure \ref{fig:abba-baba}).
\revision{We n}ote that the Patterson's D-statistic does not indicate \revision{the} directionality of \revision{gene flow}. 
    
In terms of \revision{computational} efficiency,
testing on all permutations of three taxa, with a specified outgroup, computing time of Patterson's D-Statistic on an alignment of length $L$, containing $n$ taxa scales with respect to
$4L \times \frac{n!}{(n-3)!}$, giving \revision{a time complexity is}
        $O(n^{3}L)$.
However, \revision{it is worth noting that} Patterson's D-statistic is symmetric, such that the resulting D-statistic of the topology ((a,b),c) is equal to the negative D-statistic given the topology ((b,a),c).
Further reduction of the number of tests occurs when provided the correct topology of the underlying species tree, \revision{as is necessary, as the D-statistic relies upon the structure of the major tree}.
As a result, only one test is necessary per \revision{subset of three taxa}.

\subsubsection{D3}
        
Motivated by the original Patterson's D-Statistic, D3 \revision{was created as an alternative method that does not require an outgroup, relying only on three taxa} \cite{D3}.
It uses pairwise distances instead of site patterns frequencies, where the pairwise distances between taxa $t_3, t_2$ and between taxa $t_3, t_1$ are 
expected to be equal, given a known species tree relationship, where $t_1$ and $t_2$ are sisters (such as in $((t_1,t_2),t_3)$).
        
D3 can be calculated as a ratio of pairwise differences between three sequences
$D3 = \frac{d_{BC} - d_{AC}}{d_{BC} + d_{AC}}$ where $d_{ij}$ corresponds to the distance between taxa $i$ and $j$.
Here, significant deviation of D3 from 0 may imply gene flow between taxa C and B, in the case of a \revision{negative result, and between A and C, in the case of a positive result.} \revision{The distance used in this calculation can either be the uncorrected distance, i.e. Hamming, or a measure of distance corrected for multiple hits.} 
This operates much like the original D-statistic to test for the presence, but not \revision{the} directionality, of gene flow events.
\revision{However, unlike Patterson's D, D3} can only distinguish hybridization between non-sister lineages \cite{D3,phylog-approaches-review}. 
        
In terms of \revision{computational} efficiency,
unlike the Patterson's D-Statistic and D$_p$, D3 does not include comparison to an outgroup. As a result, \revision{the method is slightly faster} as there is 25\% less sequence information to analyze and compare to, as there are now three sequences instead of four. 
As with the Patterson's D-Statistic, prior knowledge of the species topology must be known, and due to symmetry, only one test per combination of three taxa is necessary\revision{; these methods have the same time complexity:}
            $O(L n^{3})$.

\subsubsection{D$_p$}
D$_p$ adds the site pattern frequency BBAA to the denominator of the original
Patterson's D-Statistic in order to estimate the net proportion of the genome resulting from introgression \cite{Dp}. This feature provides comparability with HyDe's computation of $\gamma$. The denominator in D$_p$ accounts for the total number of variable sites: 
$D_p = \frac{|ABBA - BABA|}{BBAA + ABBA + BABA}$.
    
In terms of \revision{computational} efficiency,
as with other forms of the D-test, D$_p$ tests each combination of $n$ taxa given an alignment of length $L$ and a specified topology. Its \revision{time} complexity is equivalent to that of Patterson's D-Statistic: 
$O(n^{3} L)$.

\subsection{Simulations}
All methods were tested on the same proposed networks and compared in their ability to test for the presence of hybridization events \revision{in relation to} either the \revision{three or four}  taxa used as input.
Networks used for simulation are illustrated in Figures \ref{fig:fig-kongkubatkonets}, \ref{fig:fig-n10n25nets}, and \ref{fig:fig-n15nets}. These vary in number of reticulations, number of taxa, depth of reticulations and their mixing parameter $\gamma$\revision{,} which denotes how much ancestral DNA is passed from the minor hybrid edge to the hybrid node. Size ranges from 4--25 taxa,
where the number of reticulation events for networks with 10, 15, and 25 taxa are 20\% of the number of taxa. Each reticulation event can have singular or multiple affected taxa
downstream of the hybridization event. We name the networks based on the number of taxa ($n$) and number of hybridizations ($h$). For example, the first network in Figure \ref{fig:fig-kongkubatkonets} is denoted n4h1 as it has four taxa ($n=4$) and one hybridization ($h=1$). Six out of the twelve networks are replicated from earlier studies\revision{; specifically,} four networks (n4h1, n4h1$_{\textrm{introgression}}$, n8h3 and n5h2) in Figure \ref{fig:fig-kongkubatkonets} were used in \cite{KongKubatko}\revision{,} and  \revision{two} networks (n10h2 and n15h3) in Figures \ref{fig:fig-n10n25nets} and \ref{fig:fig-n15nets}\revision{, respectively,} were used in \cite{solisane}.

We separate networks labeled n10h2 (with two hybridization events) and n15h3 (with three hybridization events) into singular reticulation events (n10h1$_{\textrm{deep}}$, n10h1$_{\textrm{shallow}}$, n15h1$_{\textrm{deep}}$, n15h1$_{\textrm{intermediate}}$, n15h1$_{\textrm{shallow}}$ in Figures \ref{fig:fig-n10n25nets} and \ref{fig:fig-n15nets}). These single-hybridization networks allow us to measure the ability of methods to detect singular hybridization events without the possible influence of overlapping hybridizations, and to compare the performance of the methods on shallow vs deep hybridizations as it has been reported that deep hybridizations are more difficult to detect \cite{phylog-approaches-review}. Note that here we use the term "overlapping hybridizations" not to refer to hybridizations that share edges (e.g. level-2 networks), but to hybridizations that affect the same set of taxa downstream.
Finally, to represent how well methods perform at a larger scale, both in terms of \revision{computational} efficiency and accuracy, we evaluate their performance on a larger network labeled n25h5, shown in Figure \ref{fig:fig-n10n25nets}.

We simulated gene trees under each network with the software \texttt{ms} \cite{ms}, with a single individual per taxon. Note that this approach reduces the power of HyDe, when compared to simulations with multiple individuals per population \cite{KongKubatko, hyde-paper}. The software \texttt{ms} allows hybridization events to be modeled with {\fontfamily{qcr}\selectfont -es t i p} and {\fontfamily{qcr}\selectfont -ej t i j} events which correspond to population admixture and population splitting, respectively \cite{ms}. 
With this approach, we circumvent the alternative of decomposing each network into $2^r$ trees as in \cite{KongKubatko}, where $r$ is the number of reticulations, and sampling a proportion of each tree to represent the mixing parameter $\gamma$. \revision{We note that the gene tree distribution under a network is not equivalent as the gene tree distribution under $2^r$ displayed trees, unless there is only one taxon sampled beneath the hybrid node, and thus, } directly modeling reticulation events from a network \revision{ensures} that simulated gene trees follow their underlying network structure, which has a different probability density than the combination of individual trees, especially under complex reticulation events \cite{gene-tree-topology-probability-networks}. The decomposition to individual trees may produce reticulate gene tree patterns that are artificially clearer, and may not be as accurately representative of the timing of natural gene flow events.

\begin{figure*}
\centering
\includegraphics[scale=0.55]{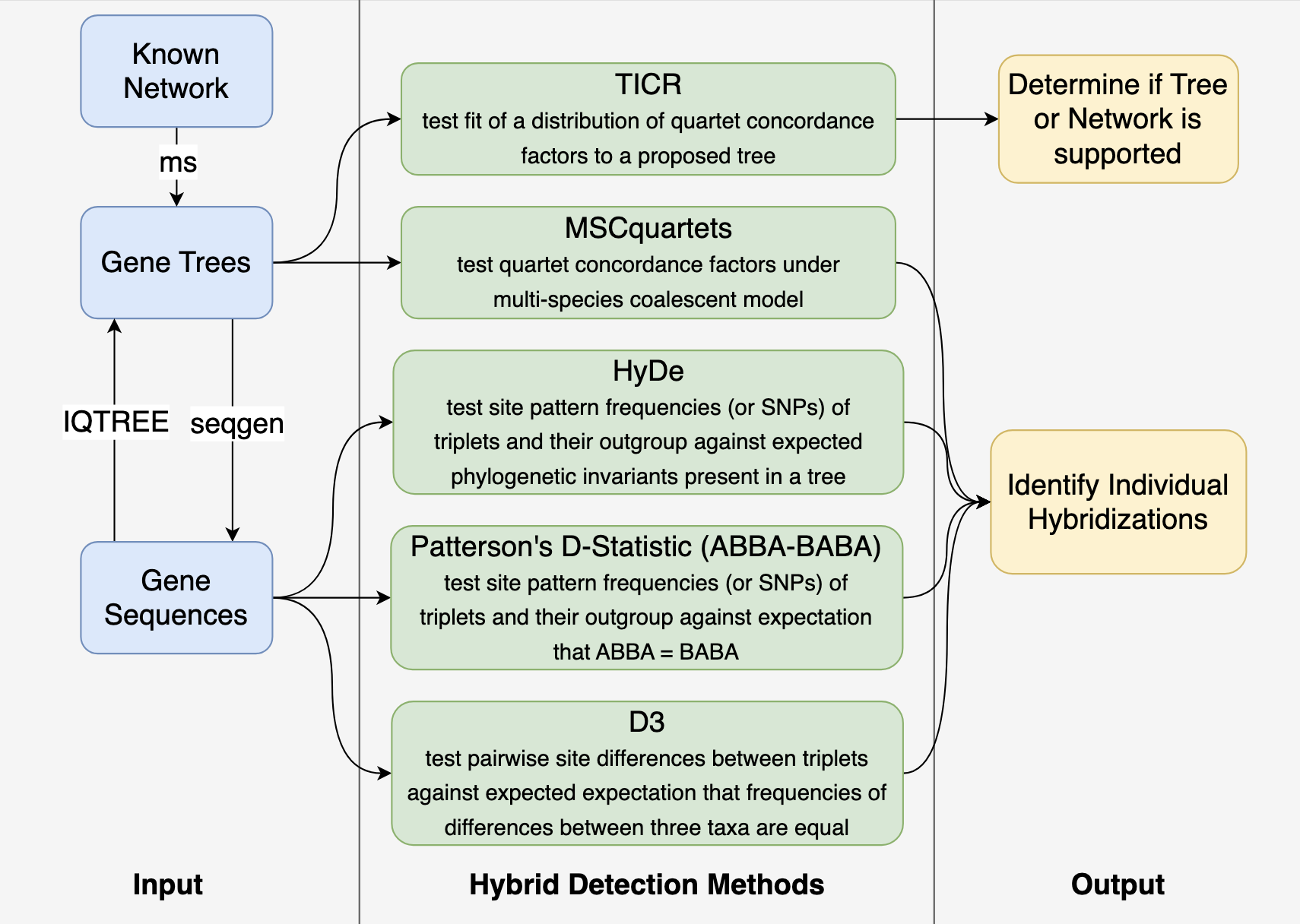}
    \caption{Simulation Pipeline. The software \texttt{ms} is used to simulate $g$ gene trees from a known network where $g \in \{30, 100, 300, 1000, 3000\}$. These gene trees are used as input for the hybrid detection methods TICR and MSCquartets. Additionally, sequences of length $L$ are generated using \texttt{seqgen} from $g$ gene trees where $g \in \{30, 100, 300, 1000, 3000, 10000\}$. Each gene tree is used to generate sequences with 100 base pairs generated per gene tree as in \cite{KongKubatko}. These sequences are used as input for HyDe, D3, Patterson’s D-Statistic and D$_p$. \revision{Additionally, these sequences were used as input for IQTREE, in order to transform them into estimated gene trees. The estimated gene trees were again used as input for TICR and MSCquartets.} This process was repeated for each network structure in Figures \ref{fig:fig-kongkubatkonets}, \ref{fig:fig-n10n25nets}, and \ref{fig:fig-n15nets} with 30 replicates each.}
    \label{fig:simulation_flow}
\end{figure*}

For TICR and MSCquartets, we simulate gene trees for $g \in \{30, 100, 300, 1000, 3000\}$ to be used directly as input.
For methods which require sequences, \texttt{ms} generates $g$ unlinked gene trees where $g \in \{30, 100, 1000, 3000, 10000\}$ and from each gene tree, \texttt{seqgen} \cite{rambaut-seqgen} is used to generate sequences with 100 base pairs generated per gene tree, similar to the approach used in \cite{KongKubatko} and then concatenated to form sequences of total length $L$. 
The \texttt{seqgen} parameters used to generate these base pairs are {\fontfamily{qcr}\selectfont -mHKY -s0.036 -f0.300414,0.191363,0.196748,0.311475 -n1 -l100}. \revision{Additionally, \texttt{IQ-TREE} was used to estimate gene trees from sequences with 100bp lengths using parameters {\fontfamily{qcr}\selectfont -m HKY85 -s} . 
The estimated gene trees are also used as input for TICR and MSCquartets.}
Full simulation details including \texttt{ms} commands and newick structures of these networks can be found on the GitHub repository \url{https://github.com/mbjorner/hybrid-detection-comparison}. 
\revision{Thirty} trials were simulated for each combination of network and gene tree number or sequence length. A pipeline of this simulation is \revision{shown} in Figure \ref{fig:simulation_flow}.

Note that the D-derived tests rely on pre-specification of topology and thus, we can expect increased false positives when testing on inputs of $(((t_1,t_3),t_2),O)$ when $(((t_1,t_2),t_3),O)$ is the true topology. \revision{For D3, we use the uncorrected genetic distance, as all simulations have stationary mutation rates.}
In addition, HyDe relies on the existence of concurrent parental lineages to test for hybrid speciation. Where only one parental lineage is sampled \revision{(see} network n10h2 in Figure \ref{fig:fig-n10n25nets} \revision{for an example)}, we investigate the influence of introgression events from these "ghost" lineages on HyDe's output. \revision{To run MSCquartets, we chose to remove unresolved star trees, and use the T3 model, which represents an unspecified tree topology.} Last, TICR requires an input \revision{estimated} population tree to be used for the expected qcCFs. \revision{The estimated population tree that we use is the major tree from the input network. This is equivalent to the network with any minor hybrid edges removed.} Deviations from the expected qcCFs could indicate deviations from the ILS-only model, but also, it could indicate that the wrong population tree was used for comparison. In our simulation studies, we use the known major tree as the input population tree for TICR so that any significant TICR results are interpreted with the possibility of hybridization.

Each method is also evaluated in terms of computing time, as measured in CPU time in seconds, given their gene tree or sequence data inputs, for the purpose of predicting how well each summary method accommodates the addition of sampled taxa. As genetic sequence information has become more widely available, so too have the datasets that biologists use to construct phylogenies and infer these reticulation events. In practice, often tens or hundreds of taxa are compared \cite{bees, introgression-drosophila}.
Since the Patterson's D-Statistic, D3, and D$_p$ were computed from the output of HyDe, which describes all possible site pattern frequencies, timing was omitted for D-statistic related tests.

\subsubsection{Evaluation of Accuracy on Simulated Datasets}
\begin{figure}[!ht]
    \centering
    \includegraphics[scale=0.7]{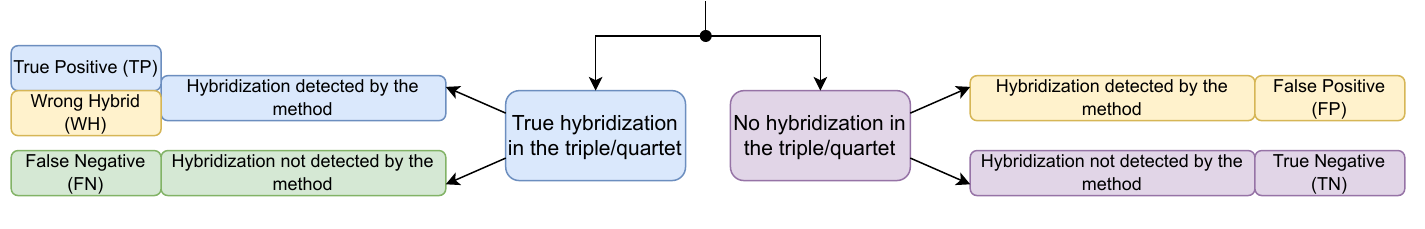}
    \caption{Visual description of false positives, false negatives, wrong hybrids (for HyDe only), true positives and true negatives.}
    \label{fig:flowTP}
\end{figure}

We now describe the computation of the false positive/negative rates and precision (see also Figure \ref{fig:flowTP}). Every triple (or quartet for MSCquartets) could have a hybrid or not. For example, in the n4h1 network in Figure \ref{fig:fig-kongkubatkonets}, the triple $\{1,2,3\}$ contains a hybrid (taxon $2$), but the triple $\{1,3,4\}$ does not contain any hybrid. If the triple (quartet) contains a hybrid, and the method detects it (pvalue$<\alpha$ for significance level $\alpha$), we consider this a true positive (TP). If the triple (quartet) contains a hybrid, but the method does not detect it (pvalue$>\alpha$), we consider this a false negative (FN). If the triple (quartet) does not contain a hybrid, and the method finds no hybrid (pvalue$>\alpha$), we consider this a true negative (TN). If the triple (quartet) does not contain a hybrid, but the method detects a hybrid (pvalue$<\alpha$), we consider this a false positive (FP). The False Positive Rate \revision{(FPR)} is computed as $FP/(FP+TN)$. 
\revision{The recall is computed as $TP / (TP + FN)$.}
The precision is computed as $TP/(TP+FP)$.
For HyDe, an additional metric, Wrong Hybrid Rate (WHR) describes the rate at which hybridization is detected but is falsely attributed to the incorrect hybrid taxon. That is, if the triple contains a hybrid, and HyDe detects it, but identifies the wrong taxon as the hybrid taxon, we consider this a wrong hybrid (WH). For example, in the n4h1 network in Figure \ref{fig:fig-kongkubatkonets}, the triple $\{1,2,3\}$ contains a hybrid (taxon $2$). HyDe could test whether $1$ and $3$ are the parents of hybrid taxon $2$ (correct hybrid), or whether $2$ and $3$ are parents of hybrid taxon $1$ (wrong hybrid). If the latter test is significant, then HyDe correctly identified that there is a hybrid relationship among these taxa, but wrongly identified the hybrid taxon. We define the Wrong Hybrid Rate (WHR) as $WHR = WH/(WH+TP+FP)$.
We use \texttt{PhyloNetworks} \cite{PhyloNetworks}, a Julia package that allows for efficient manipulation of phylogenetic networks to easily identify triples or quartets with hybrid relationships in all networks under study.

\subsection{Hybridizations in the bee subfamily Nomiinae}
To demonstrate the use of these hybrid detection methods on real data, we compare method performance on a dataset of ultraconserved elements (UCEs) from the bee subfamily Nomiinae. This data originates from a paper investigating the impacts of gene tree estimation error on species tree reconstruction \cite{bees}, 
and was used to demonstrate improved tree reconstruction with weighted \revision{ASTRAL} \cite{ASTRAL}, a new version of ASTRAL that weights quartets based on their uncertainty (branch support) and terminal branch lengths in input gene trees.
The dataset is available for download on \url{https://datadryad.org/stash/dataset/doi:10.5061/dryad.z08kprrb6}.

This dataset contains sequences and gene trees of up to 852 UCEs, for a total concatenated sequence length of 576,041 base pairs for each of 32 taxa.
In the original study \cite{bees}, gene trees were estimated using six different methods, (1) IQ-Tree2 with the GTR-G substitution model, (2) IQ-Tree2 with the substitution model chosen
by ModelFinder, (3) MrBayes with the GTR-G substitution model, (4) MrBayes with reversible jump MCMC, (5) PhyloBayes, and (6) RAxML. 
The original investigation found a consensus tree using PhyloBayes on concatenated UCEs. 

Here, we use each of the proposed sets of gene trees created using the six different methods, as input for MSCquartets, and apply a Bonferroni correction to evaluate significant quartets which may contain hybridization. 
We use the proposed species tree and gene trees in combination for TICR, for which we interpret a poor fit of \revision{the observed qcCFs} to either possibility of incorrect species tree, presence of hybridization, or a combination of the two.
Next, we use the original UCE sequences and concatenate them to run HyDe, using \textit{Lasioglossum albipes} as the outgroup, as indicated by the consensus tree constructed with PhyloBayes \cite{bees}, and a Bonferroni correction for significance. As the original study included two outgroups, \textit{Lasioglossum albipes} and \textit{Dufourea novaeangilae}, we removed \textit{Dufourea novaeangilae} from all gene trees and sequences prior to running hybrid detection methods because these methods require only one outgroup.

\section{Results}

\subsection{Simulations}
\begin{figure*}
    \centering
    \includegraphics[scale=0.75]{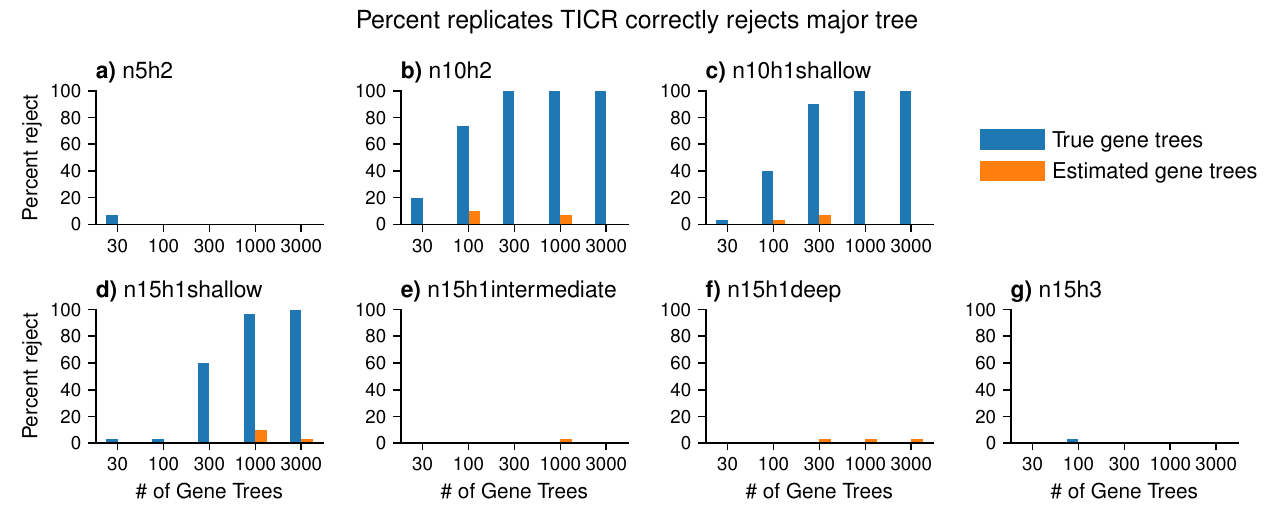} 
    \caption{\revision{Y axis corresponds to the proportion replicates in which TICR correctly rejects the true major tree at a level of $0.05$ and X axis corresponds to the number of gene trees. With true gene trees as input (blue bars), TICR rejects the null hypothesis of the major tree (equivalent to the input network sans hybridization events) for networks with single, shallow hybridizations. TICR fails to detect deeper hybridizations such as n10h1$_{\textrm{deep}}$, n15h1$_{\textrm{intermediate}}$, and n15h1$_{\textrm{deep}}$. For networks with multiple hybridization events (n5h2, n8h3, n10h2, n15h3, and n25h5), TICR is also unable to reliably detect non-treelike patterns. With estimated gene trees as input (orange bars), TICR is unable to reject the major tree across tested networks.}}
\label{fig:ticr}
\end{figure*}

Figure \ref{fig:ticr} shows the proportion of times that TICR correctly rejects the major tree \revision{from true and estimated gene trees}, and thus, detects the presence of hybridizations under the different networks under study. We highlight that TICR accurately detects hybridizations for the case of single shallow hybridizations, n10h1$_{\textrm{shallow}}$ and n15h1$_{\textrm{shallow}}$. However, TICR \revision{does not} detect deeper hybridizations as in n10h1$_{\textrm{deep}}$, n15h1$_{\textrm{intermediate}}$, and n15h1$_{\textrm{deep}}$ or multiple hybridizations in the same network as in n5h2, n8h3, n10h2, n15h3, and n25h5. 
TICR \revision{also does not} detect hybridizations on networks with four taxa (as n4h1 or n4h1$_{\textrm{introgression}}$) and those results are not included in the figure. \revision{We highlight the decreased accuracy in performance when using estimated gene trees across all tested networks.} 

\begin{figure*}[!ht] 
    \centering
    \includegraphics[width=.95\linewidth]{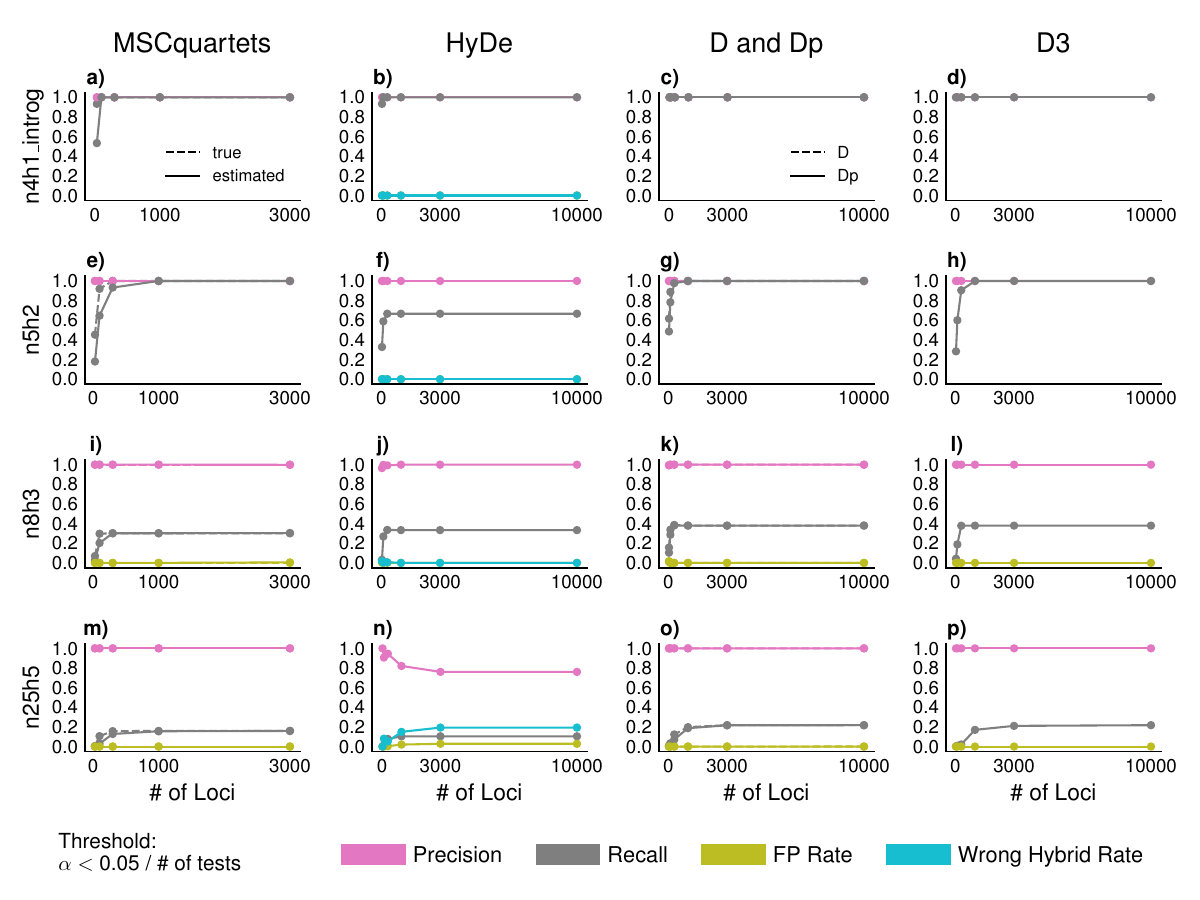}
     \caption{\revision{Precision, recall, and false positive rate for MSCquartets (from true and estimated gene trees), HyDe, Patterson's D-Statistic, Dp, and D3 on the networks: n4h1$_{\textrm{introgression}}$ (network with single shallow hybridization event), n5h2 (network with two overlapping hybridization events), n8h3 (network with three overlapping hybridization events), and n25h5 (network with 5 overlapping hybridization events). All tests are Bonferroni-corrected at a level of significance $\alpha = 0.05 / \textrm{number of tests}$ (see the Supplementary materials for results on $\alpha=0.05$). For HyDe, an additional metric, Wrong Hybrid Rate describes the rate at which hybridization is detected but is falsely attributed to the incorrect hybrid taxon.
      For the introgression event (n4h1$_{\textrm{introgression}}$), MSCquartets, HyDe, Patterson's D-Statistic, and D3 all behave similarly to the hybrid speciation scenario in n4h1, with near-perfect recovery and identification of hybrid scenarios. High recall is noted for n5h2 across MSCquartets and D-related methods. For n5h2, HyDe correctly recovers parent-hybrid relationships, but has a slightly decreased recall. Network n8h3 displays low recall across all methods. However, precision is high, and false positive rate across all methods are comparable. For n25h5, high precision is noted for all tests that evaluate for hybridization within a triple or quartet. All methods also show a low recall, and low false positive rate. Given that HyDe tests specific parent-hybrid relationships, it has lower precision scores than other methods.
      X axis corresponds to the number of loci. For MSCquartets, the number of gene trees is the same as the number of loci, ranging from 30 to 3000 gene trees.  For site-based methods, the number of sites is 100 times the number of loci, so ranging from 3,000 to 1,000,000 sites.
      }}
      \label{fig:n4n5n8}
   \end{figure*}

Figure \ref{fig:n4n5n8} shows the false positive rate (yellow), precision (pink) and \revision{recall} (gray) for MSCquartets \revision{(from true and estimated gene trees)}, HyDe, Patterson's D-Statistic, Dp, and D3 on the networks: n4h1$_{\textrm{introgression}}$ (network with single shallow introgression event), n5h2 (network with two overlapping hybridization events), n8h3 (network with three overlapping hybridization events), \revision{and n25h5 (network with five overlapping hybridization events)}. As in \cite{KongKubatko}, an overlapping hybridization event is defined as a hybridization where one taxa is the parent of multiple hybridization events. For HyDe, an additional metric, wrong hybrid rate (blue) describes the rate at which hybridization is detected but is falsely attributed to the incorrect hybrid taxon. The network n4h1$_{\textrm{introgression}}$ displays an introgression event which is easily detected by all methods (high precision and \revision{high recall}). All methods also display no false positive rates on this network, as all triples or quartets tested contain a hybrid relationship. For the case of two hybridizations (n5h2), all methods display a high precision and \revision{high recall}, except for HyDe which has a \revision{lower recall} than others. False positive rate is low and comparable for all methods in this network. For three hybridizations (n8h3), all methods have high precision and \revision{lower recall}. As more taxa become part of the network, certain combinations contain hybrids that arise from ghost lineages, which may not have strong signal to detect the hybridization events. 
In this figure, all test are Bonferroni-corrected at a level of significance $\alpha = 0.05 / \textrm{number of tests}$, but we also show the uncorrected version ($\alpha=0.05$) in Figure \ref{fig:N458-05} in the \revision{Supplementary materials}, \revision{as well as Figure \ref{fig:n4n5n8-supp} with a different presentation of the results}.

Figure \ref{fig:n4n5n8} also shows the results on the largest network under study (n25h5). Again, all methods show a \revision{low recall} and low false positive rate both of which could be explain by a weakening of the hybridization signal when multiple hybridizations are affecting the same taxa. All methods have a high precision which means that when a hybrid is detected, it is very likely a true hybrid. HyDe has slightly lower precision compared to other methods, but this is due to the fact that HyDe (unlike other methods) test for a very specific parent-hybrid relationship. When HyDe is tested in the setup of clear parent-hybrid relationships (Figures \ref{fig:n4h1mix}, \ref{fig:n4n5n8-supp}, and \ref{fig:hyde_hybridlambda}), HyDe indeed displays high precision.
It is notable that HyDe's precision is better for n25h5 compared to n15h3 or n10h2. This is due to the fact that the hybridizations in n10h2 and n15h3 involve ghost lineages which is not accounted for in HyDe. In this figure, all test are Bonferroni-corrected at a level of significance $\alpha = 0.05 / \textrm{number of tests}$, but we also show the uncorrected version ($\alpha=0.05$) in Figure \ref{fig:N25-05} in the \revision{Supplementary materials}.

Figure \ref{fig:n4h1mix} \revision{in the Supplementary Material} shows the false positive rate (yellow), precision (pink) and \revision{recall} (gray) for MSCquartets \revision{(from true and estimated gene trees)}, HyDe, Patterson's D-Statistic, Dp, and D3 on the n4h1 network with mixing parameters $\gamma \in 0, 0.1, 0.2, 0.3, 0.4, 0.5$. For HyDe, an additional metric, wrong hybrid rate (blue) describes the rate at which hybridization is detected but is falsely attributed to the incorrect hybrid taxon. This network is among the simplest cases with a single shallow hybridization, and thus, all methods have a high precision, low false positive rate, and high recall for as little as 100,000 sites or 300 gene trees. In this figure, all test are Bonferroni-corrected at a level of significance $\alpha = 0.05 / \textrm{number of tests}$, but we also show the uncorrected version ($\alpha=0.05$) in Figure \ref{fig:N4H1-05} in the Supplementary Material.
    
Figure \ref{fig:n10} shows the false positive rate (yellow), precision (pink) and \revision{recall} (gray) for MSCquartets \revision{(from true and estimated gene trees)}, HyDe, Patterson's D-Statistic, and D3 on the networks: n10h2 (network with two hybridization events), n10h1$_{\textrm{shallow}}$ (network with single shallow hybridization event) and n10h1$_{\textrm{deep}}$ (network with a single deep hybridization event). For HyDe, an additional metric, wrong hybrid rate (blue) describes the rate at which hybridization is detected but is falsely attributed to the incorrect hybrid taxon. All methods report a \revision{lower recall} compared to the simpler networks (n4h1 and n5h2), although precision continues to be high for all methods, except for HyDe. HyDe's lower precision is due to the fact that some hybridizations involve ghost lineages (hybridizations when one or both parental lineages contributing to the hybrid node are extinct or unsampled) and HyDe cannot account for this scenario. False positive rate is controlled in all methods. This combined with the \revision{lower recall} allows us to conclude that multiple overlapping hybridizations result in loss of signal for hybridization, rather than contradicting signal pointing at wrong hybrids. In addition, \revision{the recall is low} across all methods for the single deep hybridization case (n10h1$_{\textrm{deep}}$) which means that it is not only multiple hybridizations that result in loss of signal, hybridizations occurring in deeper parts of the tree have also lost signal to be detected. We also note that unlike previous cases (Figures \ref{fig:n4h1mix} and \ref{fig:n4n5n8}) where HyDe's wrong hybrid rate (blue) and false positive rate (yellow) were overlapped, for these cases, the wrong hybrid rate is much higher than the false positive rate. This implies that HyDe is better able to identify hybrid relationships for these networks, but not the correct hybrid taxon. In this figure, all test are Bonferroni-corrected at a level of significance $\alpha = 0.05 / \textrm{number of tests}$, but we also show the uncorrected version ($\alpha=0.05$) in Figure \ref{fig:N10-05} in the \revision{Supplementary materials}.
    
\begin{figure*}[!ht]
    \centering
    \includegraphics[width=.95\linewidth]{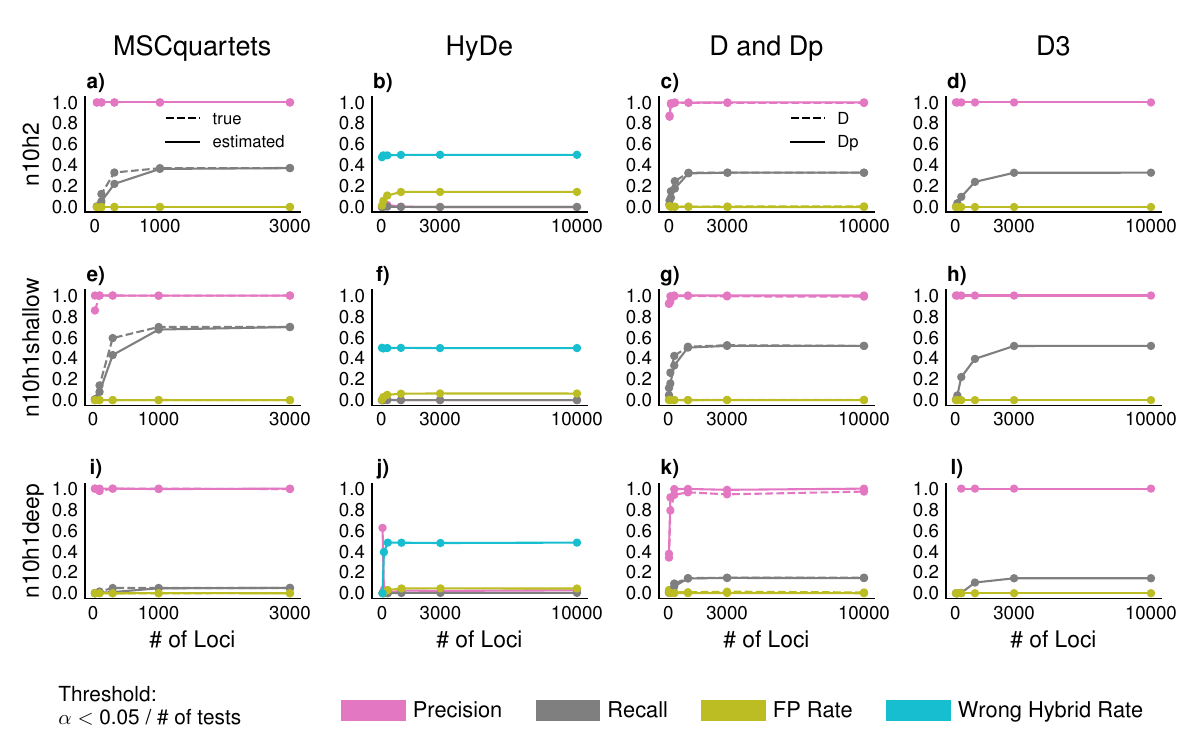}
    \caption{
    False positive rate (yellow), precision (pink) and \revision{recall} (gray) for MSCquartets \revision{(from true and estimated gene trees)}, HyDe, Patterson's D-Statistic, and D3 on the networks:
      n10h2, n10h1$_{\textrm{shallow}}$ (n10h2 with only the shallow hybridization event), n10h1$_{\textrm{deep}}$ (n10h2 with only the deeper hybridization event). All tests are Bonferroni-corrected at a level of significance $\alpha = 0.05 / \textrm{number of tests}$ (see Figure \ref{fig:N10-05} in the \revision{Supplementary materials} for results on $\alpha=0.05$). For HyDe, an additional metric, Wrong Hybrid Rate (blue) describes the rate at which hybridization is detected but is falsely attributed to the incorrect hybrid taxon. D3, Patterson's D-Statistic, and MSCquartets behave comparably across network structures, with high precision, low false positive rate, but \revision{lower recall}, which is seen in the deepest hybridization (n10h1$_{\textrm{deep}}$). A shallower hybridization that creates a single hybrid taxon (n10h1$_{\textrm{shallow}}$) is readily detected by these tests, but still with \revision{low recall}. HyDe has the \revision{lowest recall} with the n10h1$_{\textrm{deep}}$ hybridization event, and the lowest with n10h1$_{\textrm{shallow}}$. While HyDe correctly detects hybridization in triples, it often misattributes the hybridization to the wrong parent or child. When both hybridization events are present, as in n10h2, all methods result in \revision{low recall} with HyDe also having a higher number of false positives in this case.}
    \label{fig:n10}
\end{figure*}
  
\begin{figure*}[!ht] 
    \centering
    \includegraphics[width=.95\linewidth]{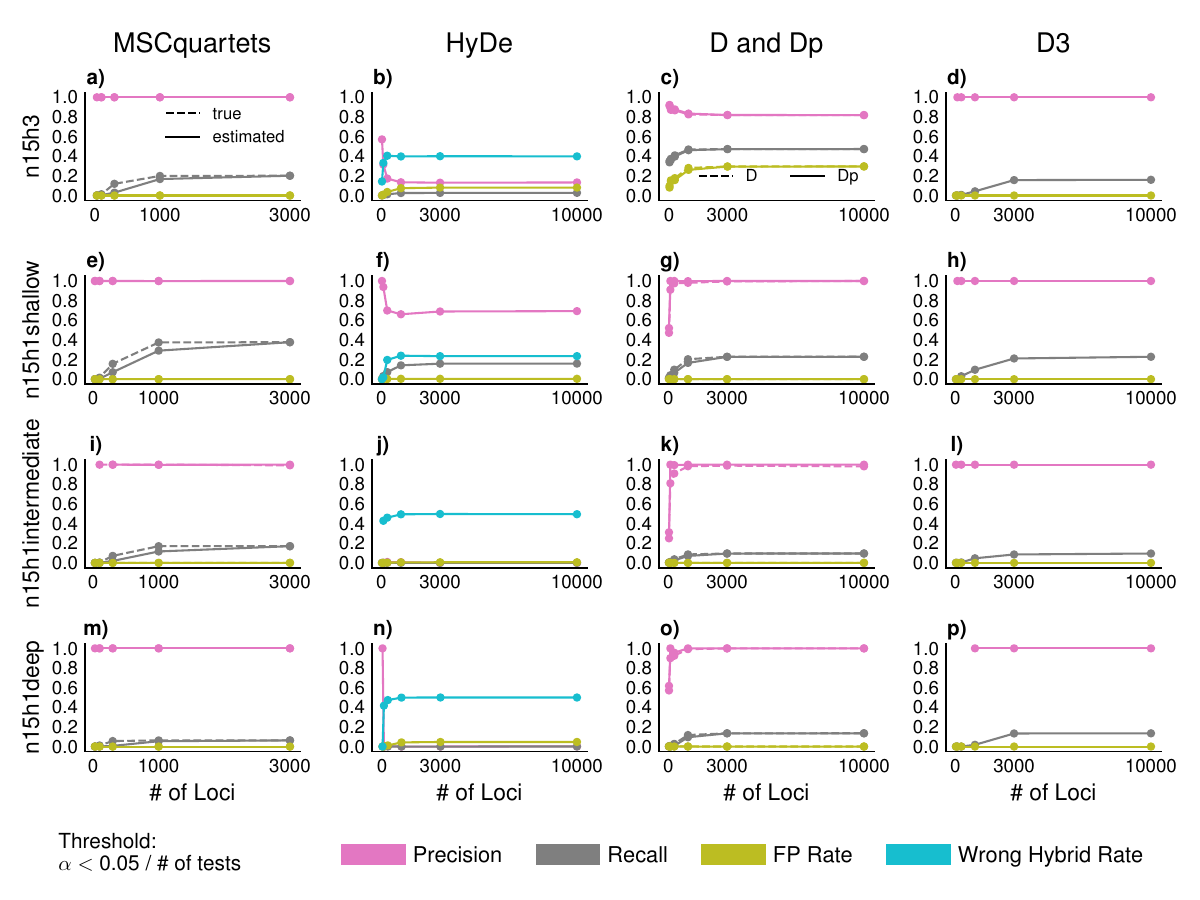}
    \caption{
    False positive rate (yellow), precision (pink) and \revision{recall} (gray) for MSCquartets \revision{(from true and estimated gene trees)}, HyDe, Patterson's D-Statistic, and D3 on the networks:
    n15H3, n15h1$_{\textrm{deep}}$ (n15h3 with only the deepest hybridization event), n15h1$_{\textrm{shallow}}$ (n15h3 with only the shallowest hybridization event), and n15h1$_{\textrm{intermediate}}$ (n15h3 with a hybridization event of intermediate depth). All tests are Bonferroni-corrected at a level of significance $\alpha = 0.05 / \textrm{number of tests}$ (see Figure \ref{fig:N15-05} in the \revision{Supplementary materials} for results on $\alpha=0.05$). For HyDe, an additional metric, Wrong Hybrid Rate (blue) describes the rate at which hybridization is detected but is falsely attributed to the incorrect hybrid taxon. MSCquartets results in \revision{lower recall} given deeper hybridization events, with a stable, high precision rate given more than 30 gene trees as input. All methods have a controlled false positive rate, except Patterson's D-Statistic with n15h3. HyDe has a similar or \revision{lower recall} when compared to other tests, and again misidentifies the exact hybrid-parent relationship in which there is a hybridization present. However, in the cases of single hybridizations, its false positive rate remains low. 
    For all single hybridization events, Patterson's D-Statistic and D3 perform comparably, with the same pattern in \revision{recall} given hybridization depth. However, in the case of n15h3, D3 outperforms Patterson's D-Statistic in terms of its higher precision and lower false positive rate. In this case, Patterson's D-Statistic has a very high false positive rate, of approximately 25\%.}
    \label{fig:n15}
\end{figure*}

Figure \ref{fig:n15} shows the false positive rate (yellow), precision (pink) and \revision{recall} (gray) for MSCquartets \revision{(from true and estimated gene trees)}, HyDe, Patterson's D-Statistic, and D3 on the networks:
n15h3 (network with three hybridization events), n15h1$_{\textrm{shallow}}$ (network with a single shallow hybridization event), n15h1$_{\textrm{intermediate}}$ (network with a single intermediate hybridization event), and n15h1$_{\textrm{deep}}$ (network with a single deep hybridization event). For HyDe, an additional metric, wrong hybrid rate (purple) describes the rate at which hybridization is detected but is falsely attributed to the incorrect hybrid taxon. As already shown in the case of n10h2 (Figure \ref{fig:n10}), all methods have a higher false negative rate, but controlled false positive rate except for Patterson's D-statistic with a high false positive rate for the case of three hybridizations (n15h3). HyDe shows lower precision compared to other methods which is due to the fact that the hybridizations in n15h3 involve ghost lineages which HyDe cannot account for. The fact that there is \revision{low recall} even for the single shallow hybridization (n15h1$_{\textrm{shallow}}$) across of methods could provide some evidence that ghost lineages create challenges, not just for HyDe. In this figure, all test are Bonferroni-corrected at a level of significance $\alpha = 0.05 / \textrm{number of tests}$, but we also show the uncorrected version ($\alpha=0.05$) in Figure \ref{fig:N15-05} in the \revision{Supplementary materials}.
 
\subsection{\revision{Empirical running time}}

\begin{figure*}[!ht]
    \centering
    \includegraphics[width=.99\linewidth]{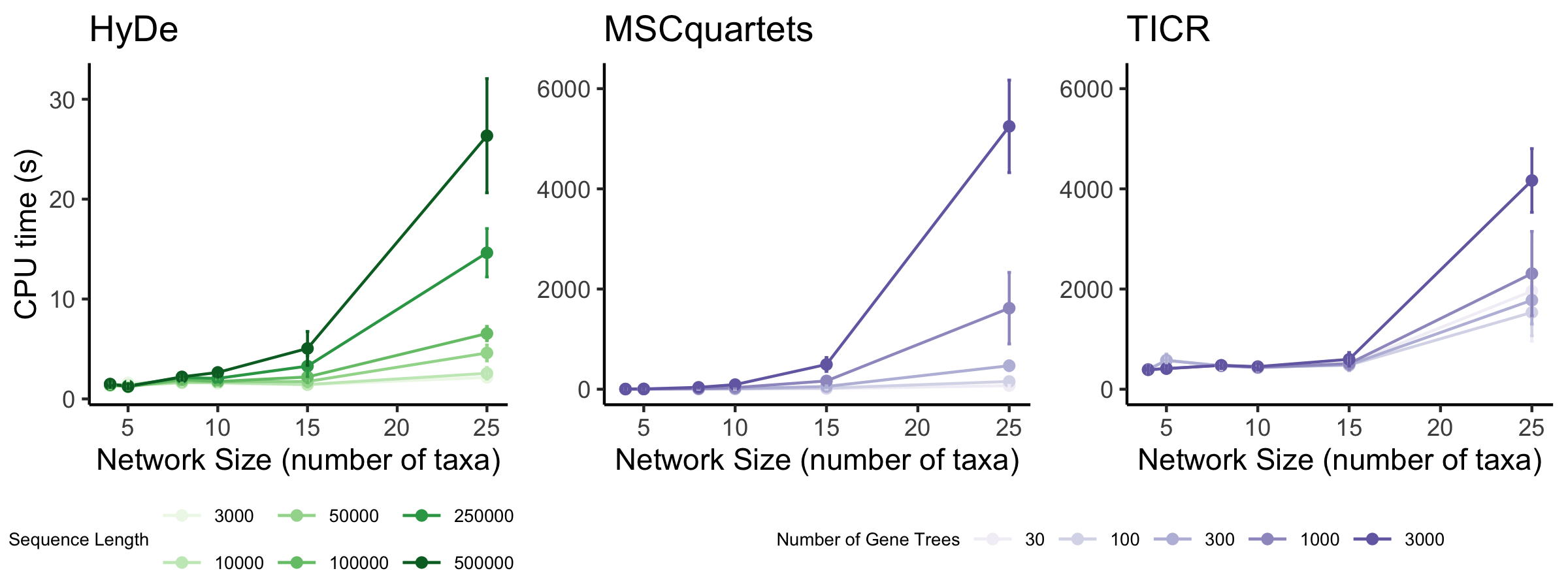}
    \caption{CPU time of HyDe (left), MSCquartets (center), and TICR (right) as network size and sequence length or number of gene trees change. Note the different limits on the Y axis for HyDe, which is the fastest of all three methods. For HyDe, there is a cubic increase in time is observed with respect to network size. For MSCquartets, there is a quartic increase in time is observed with respect to network size.}
    \label{fig:timing}
\end{figure*}

Each method is predicted to increase linearly with respect to the number of gene trees or sequence length used as input. Due to the nature of summary \revision{methods'} triple- or quartet-wise analysis, an increase in network size corresponds to a cubic or quartic increase in time\revision{, respectively, and indeed, the running time of methods (HyDe, MSCquartets, and TICR) dramatically increase with the number of taxa and the number of gene trees (Figure \ref{fig:timing}). It is worth noting that time complexity does not account for many practical issues, like memory locality and cache performance, that greatly impact runtime in practice.} 

\subsection{Nomiinae bee subfamily}

\begin{figure*}[!ht]
    \centering
    \includegraphics[width=.9\linewidth]{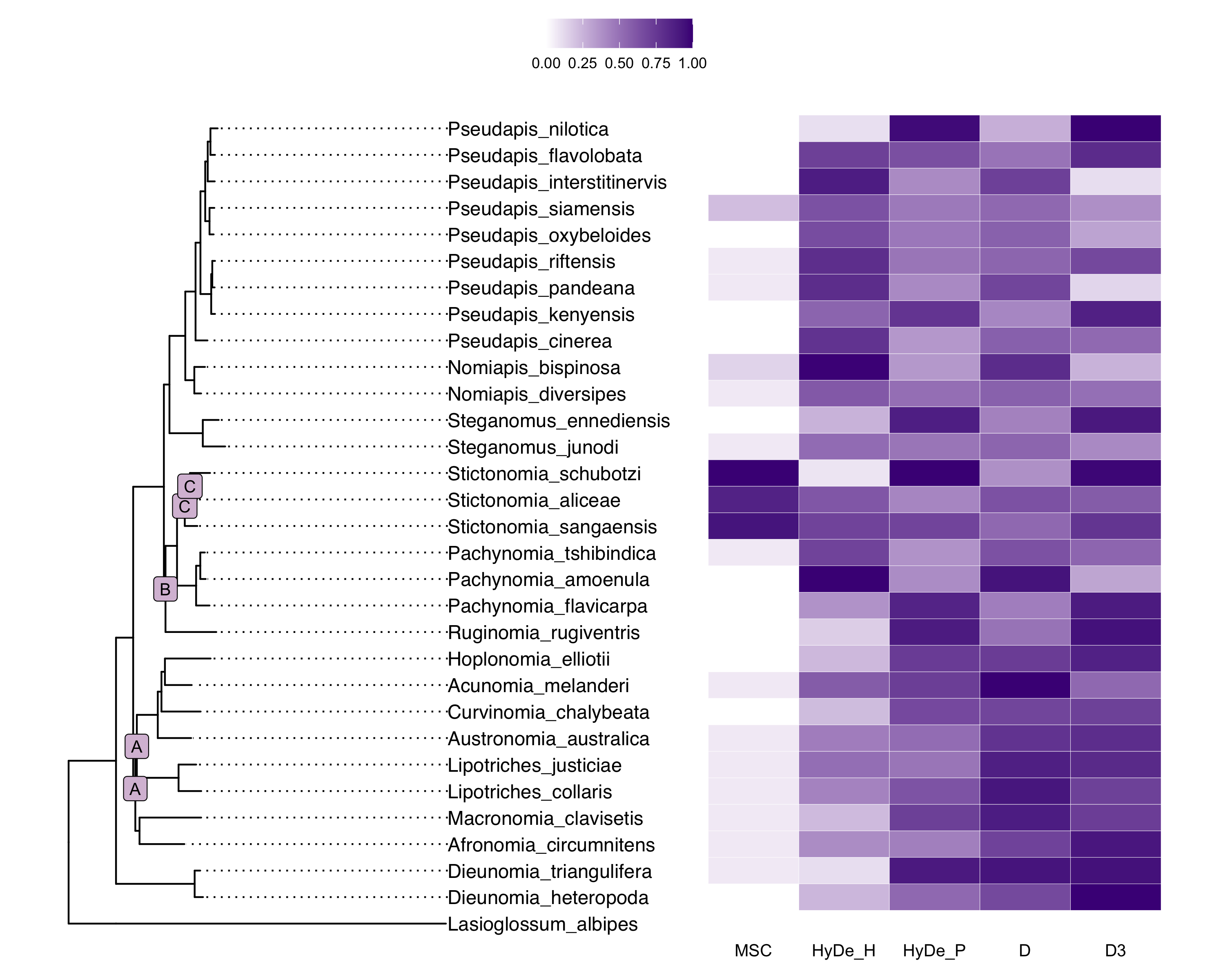}
    \caption{Left: Phylogenetic tree of selected species from the bee subfamily Nomiinae, as provided by \cite{bees}. Letters A-C identify nodes that were identified as conflicting relationships by the analyses conducted in \cite{bees}. Right: Heatmap of frequency of taxa identified as part of hybridization events (scaled between 0 and 1 for comparison across methods). MSCquartets was conducted with gene trees estimated using the IQTree2-GTRG model. Tests are Bonferroni-corrected (see Figure \ref{fig:bee-fig-alpha} in the \revision{Supplementary materials} for significance level $\alpha=0.05$). The clade with the highest proportion of proposed hybridizations contains \textit{Stictonomia spp.}}
    \label{fig:bee-fig-bonferroni}
\end{figure*}

Figure \ref{fig:bee-fig-bonferroni} displays the heatmap of the proportion of times that each taxon is involved in a significant hybridization event as identified by MSCquartets, HyDe, Patterson's D-statistics and D3. This figure was created using \texttt{ggtree} \cite{ggtree}.

We selected the estimated gene trees from the IQtree2-GTRG model to use as input in MSCquartets and display the proportion of times that each taxa is involved in a significant hybridization (Figure \ref{fig:bee-fig-bonferroni} for the Bonferroni-corrected significance level of $0.05/14950$, or $3.3 \times 10^6$ and Figure \ref{fig:bee-fig-alpha} in the \revision{Supplementary materials} for significance level of $\alpha=0.05$) 
\textit{Stictonomia spp.} is implicated in all significant quartets, with \textit{Stictonomia schubotzi} appearing with the highest frequency.
In addition, HyDe detects \textit{Stictonomia spp.} implicated 923 times over 2851 significant hybrid speciation events. We also show the proportion that each taxon is identified as a parent (ancestral lineage contributing genetic material to the hybrid) in the HyDe tests which provides a broader picture of hybridization than any of the other methods.
The results of Patterson's D-statistics or D3 align with those of HyDe and MSCquartets in the identification of \textit{Stictonomia spp.} (especially \textit{Stictonomia schubotzi}) as involved in hybridization events. The D-related tests, however, cannot separate the hybrid taxon from the parents as HyDe.
Though MSCquartets primarily implicates \textit{Stictonomia spp.} in hybridization, other methods find widespread hybridization across the tree. These differences may be due to methods' sensitivity to the depth of hybridization events.

These results suggest that these closely related species may not be reproductively isolated, which can lead to gene tree estimation error, and difficulty in reconstructing the phylogenetic tree. In the original study \cite{bees}, gene tree estimation error was identified as a source of the discordance and conflict. However, here we identify hybridization events as a plausible explanation for the gene tree discordance.

    
\section{Discussion}

Here, we present a deep investigation of the performance of genome-wide hybrid detection methods. We found that all five methods compared (TICR \cite{TICR-original}, MSCquartets \cite{ mscquart}, HyDe \cite{hyde-paper}, Patterson's D-Statistic \cite{pattersonD}, D$_p$ \cite{Dp} and D3 \cite{D3}) have similar good performance (high precision and low false positive/negative rates) on single shallow hybridizations involving few taxa (n4h1 or n5h2). \revision{Our investigation confirms previous findings \cite{KongKubatko}, and extends the conclusions to previously untested scenarios.}

\revision{By design, both MSCQuartets and TICR should also be able to detect complex hybridization of more than one instance of gene flow among four taxa by relying on the rejection of a tree hypothesis. However, }
as more hybridizations are added involving similar groups of taxa (n8h3, n10h2 and n15h3), all methods have a higher false negative rate which suggests that combinations of gene flow events weaken the signal to detect such hybridizations as opposed to creating discordant signal to identify wrong hybridizations (which would have been evidenced by an increased false positive rate). This is also confirmed by the results of TICR which is \textit{unable} to reject the major tree in most cases, except for those involving single shallow hybridizations \revision{even if they involve ghost lineages. This finding for true gene trees did not hold for estimated gene trees 
, as TICR was rarely able to reject the major tree when given estimated gene trees, even for this easier model condition.} 

HyDe had a lower precision that other methods when ghost lineages were involved (n10h2 and n15h3) which aligns with previous studies on the subject \cite{Tricou2022, Pang2022}. It also showed a higher rate of wrong hybrid identified within the hybrid triple. However, HyDe is the only method able to detect the hybrid taxon and the parent taxa involved in the hybridization as shown in the bee dataset. 
    The results of methods using site pattern frequencies or pairwise differences is highly influenced by the topology of the underlying species tree from which taxa arise. Longer coalescent times introduce noise to sequence data such that comparison to a distant outgroup or comparison between distant species is no longer advantageous, as the infinite-sites mutation model on which Patterson's D-Statistic is based expects a single mutation per site \cite{phylog-approaches-review}. With increased branch lengths, or increased distance between taxa, convergent substitutions can cause ABBA and BABA (and other) site patterns to accumulate \cite{phylog-approaches-review}. 
    
    
    
Finally, we re-analyzed the dataset of the bee subfamily Nomiinae. While the original study \cite{bees} concludes that gene \revision{tree} estimation error could be the source of discordance in the clade, here we show that hybridization is another plausible explanation for the discordant patterns with all methods identifying \textit{Stictonomia spp.} (especially \textit{Stictonomia schubotzi}) as involved in hybridization events.
    As tools for discovering reticulations across the tree of life are being developed and improved, so are tools to analyze and manipulate complex networks. Method developers should seek to assess both scalability and accuracy given growing, complex datasets. 

\paragraph{Practical advice for evolutionary biologists.} From our investigation, we can conclude that MSCquartets \cite{Allman-Phylogenetic-Invariants} is an accurate \revision{method} to detect hybridization events under a variety of different scenarios. HyDe is the only method that can \revision{identify which taxon is the hybrid taxon among the taxa involved in the hybridization event}. However, HyDe cannot perform well when the parents of hybridization are unsampled or extinct. Furthermore, all methods are unable to detect hybridizations when multiple events are affecting the same set of taxa or when hybridizations are deep\revision{. That is, when the hybridization event is close to the root. In this situation, we recommend taking samples of taxa (suspected to be involved a single hybridization event) to test at a time.} 
When a given parent-hybrid relationship is to be tested, HyDe outperforms D-Statistics-like tests by allowing the identification of hybrid taxa vs parent taxa. Finally, we conclude that TICR is a powerful method to detect single shallow hybridization events, even if they involve ghost lineages \revision{provided that gene trees can be accurately estimated}.


\paragraph{Limitations and future work.}
\revision{All the networks in the simulation study are ultrametric on coalescent units, which implicitly assumes equal population sizes across lineages. While this assumption is unrealistic, it is convenient to disentangle the causes that create differences across methods. A more thorough investigation of the interaction between population structure and hybridization patterns is needed.}
Along these lines, multiple sequence alignment errors can occur in phylogenomics data sets \cite{taper} \revision{and} could impact the performance of all methods tested. 
We simulated data under a substitution-only model and thus all methods were given true alignments as input.
Lastly, the substitution model is well behaved (homogeneous, stationary, and reversible), which is assumed by HyDe; 
however, these assumptions can be violated in practice  \cite{naser2019prevalence}. 
\revision{All sequences generated from gene trees were the same length, which does not reflect real-life variation in gene length.}
\revision{Differences in performance based on substitution model were also not explored here. We point at a recent manuscript that explores the effect of rate variation on the performance of introgression tests \cite{Frankel2023}.}
\revision{Lastly, our evaluation of TICR allowed it to use the major tree derived from the true network, rather than an estimated species tree.}
It is not clear to what extent these practicalities will impact methods, and future work should explore them in simulations and in real data sets.

\paragraph{Acknowledgements.} This work was supported by the Department of Energy [DE-SC0021016 to CSL] and by the National Science Foundation [DEB-2144367 to CSL]. We thank Laura Kubatko and Sungsik Kong for meaningful discussions about HyDe.

\bibliographystyle{plain}
\bibliography{biblio}

\begin{thebibliography}{10}

\bibitem{Allman2019NANUQ}
Elizabeth~S. Allman, Hector Ba{\~n}os, and John~A. Rhodes.
\newblock {NANUQ}: a method for inferring species networks from gene trees
  under the coalescent model.
\newblock {\em Algorithms for Molecular Biology}, 14(1):24, 2019.

\bibitem{MSC-Allman}
Elizabeth~S. Allman, James~H. Degnan, and John~A. Rhodes.
\newblock Identifying the rooted species tree from the distribution of unrooted
  gene trees under the coalescent.
\newblock {\em Journal of Mathematical Biology}, 62(6):833--862, 2011.

\bibitem{Allman-simplex-plot-MSC}
Elizabeth~S. Allman, Jonathan~D. Mitchell, and John~A. Rhodes.
\newblock Gene tree discord, simplex plots, and statistical tests under the
  coalescent.
\newblock {\em Systematic Biology}, 71(4):929--942, 02 2021.

\bibitem{Allman-Phylogenetic-Invariants}
Elizabeth~S. Allman and John~A. Rhodes.
\newblock Phlogenetic invariants.
\newblock In {\em Reconstructing Evolution: New Mathematical and Computational
  Advances}, pages 108--146. Oxford University Press, 2007.

\bibitem{barton1985-hyb-zone}
N.~H. Barton and G.~M. Hewitt.
\newblock Analysis of hybrid zones.
\newblock {\em Annual Review of Ecology and Systematics}, 16:113--148, 1985.

\bibitem{hyde-python-package}
Paul~D. Blischak, Julia Chifman, Andrea~D. Wolfe, and Laura~S. Kubatko.
\newblock {HyDe}: A python package for genome-scale hybridization detection.
\newblock {\em Systematic Biology}, 67(5):821--829, 03 2018.

\bibitem{bees}
Silas Bossert, Elizabeth~A. Murray, Alain Pauly, Kyrylo Chernyshov, Se{\'a}n~G.
  Brady, and Bryan~N. Danforth.
\newblock Gene tree estimation error with ultraconserved elements: An empirical
  study on pseudapis bees.
\newblock {\em Systematic Biology}, 70(4):803--821, 12 2020.

\bibitem{beast2}
Remco Bouckaert, Timothy~G. Vaughan, Jo{\"e}lle Barido-Sottani, Sebasti{\'a}n
  Duch{\^e}ne, Mathieu Fourment, Alexandra Gavryushkina, Joseph Heled, Graham
  Jones, Denise K{\"u}hnert, Nicola De~Maio, Michael Matschiner, F{\'a}bio~K.
  Mendes, Nicola~F. M{\"u}ller, Huw~A. Ogilvie, Louis du~Plessis, Alex Popinga,
  Andrew Rambaut, David Rasmussen, Igor Siveroni, Marc~A. Suchard, Chieh-Hsi
  Wu, Dong Xie, Chi Zhang, Tanja Stadler, and Alexei~J. Drummond.
\newblock {BEAST} 2.5: An advanced software platform for {B}ayesian
  evolutionary analysis.
\newblock {\em PLOS Computational Biology}, 15(4):1--28, 04 2019.

\bibitem{TICR-GOF-Julia}
Ruoyi Cai and C{\'e}cile An{\'e}.
\newblock {Assessing the fit of the multi-species network coalescent to
  multi-locus data}.
\newblock {\em Bioinformatics}, 37(5):634--641, 12 2020.

\bibitem{degnan2013anomalous}
James~H. Degnan.
\newblock {Anomalous Unrooted Gene Trees}.
\newblock {\em Systematic Biology}, 62(4):574--590, 05 2013.

\bibitem{Frankel2023}
Lauren~E. Frankel and C{\'e}cile An{\'e}.
\newblock Summary tests of introgression are highly sensitive to rate variation
  across lineages.
\newblock {\em bioRxiv}, 2023.

\bibitem{D3}
Matthew~W. Hahn and Mark~S. Hibbins.
\newblock A three-sample test for introgression.
\newblock {\em Mol Biol Evol}, 36(12):2878--2882, Dec 2019.

\bibitem{Dp}
Jennafer A.~P. Hamlin, Mark~S. Hibbins, and Leonie~C. Moyle.
\newblock Assessing biological factors affecting postspeciation introgression.
\newblock {\em Evol Lett}, 4(2):137--154, Apr 2020.

\bibitem{introgression-wild-tomatoes}
Mark~S. Hibbins and Matthew~W. Hahn.
\newblock The effects of introgression across thousands of quantitative traits
  revealed by gene expression in wild tomatoes.
\newblock {\em PLOS Genetics}, 17(11):1--20, 11 2021.

\bibitem{phylog-approaches-review}
Mark~S. Hibbins and Matthew~W. Hahn.
\newblock {Phylogenomic approaches to detecting and characterizing
  introgression}.
\newblock {\em Genetics}, 220(2), 11 2021.
\newblock iyab173.

\bibitem{ms}
Richard~R. Hudson.
\newblock {Generating samples under a Wright--Fisher neutral model of genetic
  variation }.
\newblock {\em Bioinformatics}, 18(2):337--338, 02 2002.

\bibitem{KongKubatko}
Sungsik Kong and Laura~S. Kubatko.
\newblock Comparative performance of popular methods for hybrid detection using
  genomic data.
\newblock {\em Systematic Biology}, 70(5):891--907, 01 2021.

\bibitem{hyde-paper}
Laura~S. Kubatko and Julia Chifman.
\newblock An invariants-based method for efficient identification of hybrid
  species from large-scale genomic data.
\newblock {\em BMC Evolutionary Biology}, 19(1):112, 2019.

\bibitem{deep-coalescence}
Wayne~P. Maddison.
\newblock Gene trees in species trees.
\newblock {\em Systematic Biology}, 46(3):523--536, 09 1997.

\bibitem{RF-Net2}
Alexey Markin, Sanket Wagle, Tavis~K. Anderson, and Oliver Eulenstein.
\newblock {RF-Net 2: fast inference of virus reassortment and hybridization
  networks}.
\newblock {\em Bioinformatics}, 38(8):2144--2152, 02 2022.

\bibitem{mscquart}
Jonathan~D. Mitchell, Elizabeth~S. Allman, and John~A. Rhodes.
\newblock {Hypothesis testing near singularities and boundaries}.
\newblock {\em Electronic Journal of Statistics}, 13(1):2150 -- 2193, 2019.

\bibitem{phylo-networks-rice}
B.M.E. Moret, L.~Nakhleh, T.~Warnow, C.R. Linder, A.~Tholse, A.~Padolina,
  J.~Sun, and R.~Timme.
\newblock Phylogenetic networks: modeling, reconstructibility, and accuracy.
\newblock {\em IEEE/ACM Transactions on Computational Biology and
  Bioinformatics}, 1(1):13--23, 2004.

\bibitem{naser2019prevalence}
Suha Naser-Khdour, Bui~Quang Minh, Wenqi Zhang, Eric~A Stone, and Robert
  Lanfear.
\newblock {The prevalence and impact of model violations in phylogenetic
  analysis}.
\newblock {\em Genome Biology and Evolution}, 11(12):3341--3352, 09 2019.

\bibitem{iq-tree}
Lam-Tung Nguyen, Heiko~A. Schmidt, Arndt von Haeseler, and Bui~Quang Minh.
\newblock {IQ-TREE}: a fast and effective stochastic algorithm for estimating
  maximum-likelihood phylogenies.
\newblock {\em Mol Biol Evol}, 32(1):268--274, Jan 2015.

\bibitem{Pang2022}
Xiao-Xu Pang and Da-Yong Zhang.
\newblock Impact of ghost introgression on coalescent-based species tree
  inference and estimation of divergence time.
\newblock {\em Systematic Biology}, 07 2022.
\newblock syac047.

\bibitem{pattersonD}
Nick Patterson, Priya Moorjani, Yontao Luo, Swapan Mallick, Nadin Rohland,
  Yiping Zhan, Teri Genschoreck, Teresa Webster, and David Reich.
\newblock Ancient admixture in human history.
\newblock {\em Genetics}, 192(3):1065--1093, 11 2012.

\bibitem{introgression-humans}
Fernando Racimo, Sriram Sankararaman, Rasmus Nielsen, and Emilia
  Huerta-S{\'a}nchez.
\newblock Evidence for archaic adaptive introgression in humans.
\newblock {\em Nat Rev Genet}, 16(6):359--371, Jun 2015.

\bibitem{rambaut-seqgen}
Andrew Rambaut and Nicholas~C. Grass.
\newblock {Seq-Gen: an application for the Monte Carlo simulation of DNA
  sequence evolution along phylogenetic trees}.
\newblock {\em Bioinformatics}, 13(3):235--238, 06 1997.

\bibitem{rannalayang2003}
Bruce Rannala and Ziheng Yang.
\newblock Bayes estimation of species divergence times and ancestral population
  sizes using {DNA} sequences from multiple loci.
\newblock {\em Genetics}, 164(4):1645--1656, Aug 2003.

\bibitem{MSCquartets-RPackage}
John~A. Rhodes, Hector Ba{\~n}os, Jonathan~D. Mitchell, and Elizabeth~S.
  Allman.
\newblock Mscquartets 1.0: quartet methods for species trees and networks under
  the multispecies coalescent model in {R}.
\newblock {\em Bioinformatics}, 37(12):1766--1768, Jul 2021.

\bibitem{quartet-max-cut}
Sagi Snir and Satish Rao.
\newblock {Quartet MaxCut}: A fast algorithm for amalgamating quartet trees.
\newblock {\em Molecular Phylogenetics and Evolution}, 62(1):1--8, 2012.

\bibitem{solisane}
Claudia Sol{\'\i}s-Lemus and C{\'e}cile An{\'e}.
\newblock Inferring phylogenetic networks with maximum pseudolikelihood under
  incomplete lineage sorting.
\newblock {\em PLOS Genetics}, 12(3):e1005896--, 03 2016.

\bibitem{PhyloNetworks}
Claudia Sol{\'\i}s-Lemus, Paul Bastide, and C{\'e}cile An{\'e}.
\newblock {PhyloNetworks}: A package for phylogenetic networks.
\newblock {\em Molecular Biology and Evolution}, 34(12):3292--3298, 09 2017.

\bibitem{raxml}
Alexandros Stamatakis.
\newblock {RAxML} version 8: a tool for phylogenetic analysis and post-analysis
  of large phylogenies.
\newblock {\em Bioinformatics}, 30(9):1312--1313, May 2014.

\bibitem{TICR-original}
Noah W.~M. Stenz, Bret Larget, David~A. Baum, and C{\'e}cile An{\'e}.
\newblock Exploring tree-like and non-tree-like patterns using genome
  sequences: An example using the inbreeding plant species {A}rabidopsis
  thaliana {(L.) Heynh}.
\newblock {\em Systematic Biology}, 64(5):809--823, 06 2015.

\bibitem{introgression-drosophila}
Anton Suvorov, Bernard~Y. Kim, Jeremy Wang, Ellie~E. Armstrong, David Peede,
  Emmanuel~R.R. D'Agostino, Donald~K. Price, Peter~J. Waddell, Michael Lang,
  Virginie Courtier-Orgogozo, Jean~R. David, Dmitri Petrov, Daniel~R. Matute,
  Daniel~R. Schrider, and Aaron~A. Comeault.
\newblock Widespread introgression across a phylogeny of 155 {D}rosophila
  genomes.
\newblock {\em Current Biology}, 32(1):111--123.e5, 2022.

\bibitem{Tricou2022}
Th{\'e}o Tricou, Eric Tannier, and Damien~M. de~Vienne.
\newblock Ghost lineages highly influence the interpretation of introgression
  tests.
\newblock {\em Systematic Biology}, 02 2022.
\newblock syac011.

\bibitem{WenNakleh2018}
Dingqiao Wen and Luay Nakhleh.
\newblock Coestimating reticulate phylogenies and gene trees from multilocus
  sequence data.
\newblock {\em Systematic Biology}, 67(3):439--457, 10 2017.

\bibitem{wen2018inferring}
Dingqiao Wen, Yun Yu, Jiafan Zhu, and Luay Nakhleh.
\newblock Inferring phylogenetic networks using {PhyloNet}.
\newblock {\em Systematic biology}, 67(4):735--740, 2018.

\bibitem{ggtree}
Guangchuang Yu, David~K. Smith, Huachen Zhu, Yi~Guan, and Tommy Tsan-Yuk Lam.
\newblock ggtree: an {R} package for visualization and annotation of
  phylogenetic trees with their covariates and other associated data.
\newblock {\em Methods in Ecology and Evolution}, 8(1):28--36, 2017.

\bibitem{gene-tree-topology-probability-networks}
Yun Yu, James~H. Degnan, and Luay Nakhleh.
\newblock The probability of a gene tree topology within a phylogenetic network
  with applications to hybridization detection.
\newblock {\em PLOS Genetics}, 8(4):1--10, 04 2012.

\bibitem{ASTRAL}
Chao Zhang and Siavash Mirarab.
\newblock Weighting by gene tree uncertainty improves accuracy of quartet-based
  species trees.
\newblock {\em bioRxiv}, 2022.

\bibitem{taper}
Chao Zhang, Yiming Zhao, Edward~L. Braun, and Siavash Mirarab.
\newblock Taper: Pinpointing errors in multiple sequence alignments despite
  varying rates of evolution.
\newblock {\em Methods in Ecology and Evolution}, 12(11):2145--2158, 2021.

\bibitem{Zhang2017}
Chi Zhang, Huw~A. Ogilvie, Alexei~J. Drummond, and Tanja Stadler.
\newblock {B}ayesian inference of species networks from multilocus sequence
  data.
\newblock {\em Molecular Biology and Evolution}, 35(2):504--517, 12 2017.

\bibitem{zhang2018bayesian}
Chi Zhang, Huw~A. Ogilvie, Alexei~J. Drummond, and Tanja Stadler.
\newblock Bayesian inference of species networks from multilocus sequence data.
\newblock {\em Molecular biology and evolution}, 35(2):504--517, 2018.

\end{thebibliography}

\newpage

\appendix
    
\section*{Supplementary figures}

    
\begin{sidewaystable*}[ht]
    \centering
    \begin{tabular}{c|p{1.5cm}|p{2.25cm}|p{5cm}|p{4cm}|p{4cm}}
        Method & Complexity & \revision{Assumptions} & Input & Test & Output \\
        \hline
        TICR & 
        \revision{$O(gn^4)$} & \revision{Equidistant network: No        Number of gene flow events: Any} &
        Population tree and observed qcCFs from gene trees & If observed qcCFs align with expected CFs under ILS-only MSC & p-values from individual quartet tests and
        overall p-value reporting goodness-of-fit to population tree \\
        \hline
        MSCquartets & $O(gn^4)$ & \revision{Equidistant network: No        Number of gene flow events: Any}  & Gene trees & If minor qcCFs align with expected CFs under ILS-only MSC & p-values from individual quartet tests \\
        \hline
        HyDe &
        $O(Ln^3)$ & \revision{Equidistant network: Yes Number of gene flow events: One} &
        Sequences from taxa and outgroup & Hils statistic
        evaluates triples for site pattern frequencies against phylogenetic invariants & 
        List of all triples with 1) proposed hybrid speciation (parental and hybrid lineages), 2) a estimated $\gamma$, and 3) p-value \\
        \hline
        Patterson's D & 
        $O(Ln^3)$ & \revision{Equidistant network: Yes   Number of gene flow events: One} &
        Sequences from taxa and outgroup &
        Excess in site pattern frequencies of $ABBA$ and $BABA$ when expected to be equal & 
        A D-statistic per triple where significant deviation from 0 indicates gene flow \\
        \hline
        D3 & $O(Ln^3)$ & \revision{Equidistant network: Yes    Number of gene flow events: One} & Sequences & If triples reject ILS-only assumption (measured as deviation of D3 from 0) & Distribution of D3 values of triples mapped to z-distribution \\
        \hline
        D$_p$ &  $O(Ln^3)$ & \revision{Equidistant network: Yes 
  Number of gene flow events: One} & Sequences from taxa and outgroup & Excess in site pattern frequencies of $ABBA$ and $BABA$ when expected to be equal & A D$_p$ value per triple where significant deviation from 0 indicates presence and extent of gene flow \\
    \end{tabular}
    \caption{Summary Table of Hybrid Detection Methods}
    \label{tab:summary}
\end{sidewaystable*}
    

\subsection*{Reproducing the power calculations in \cite{KongKubatko}}

\begin{figure*}[!ht]
    \centering
    \includegraphics[width=.5\linewidth]{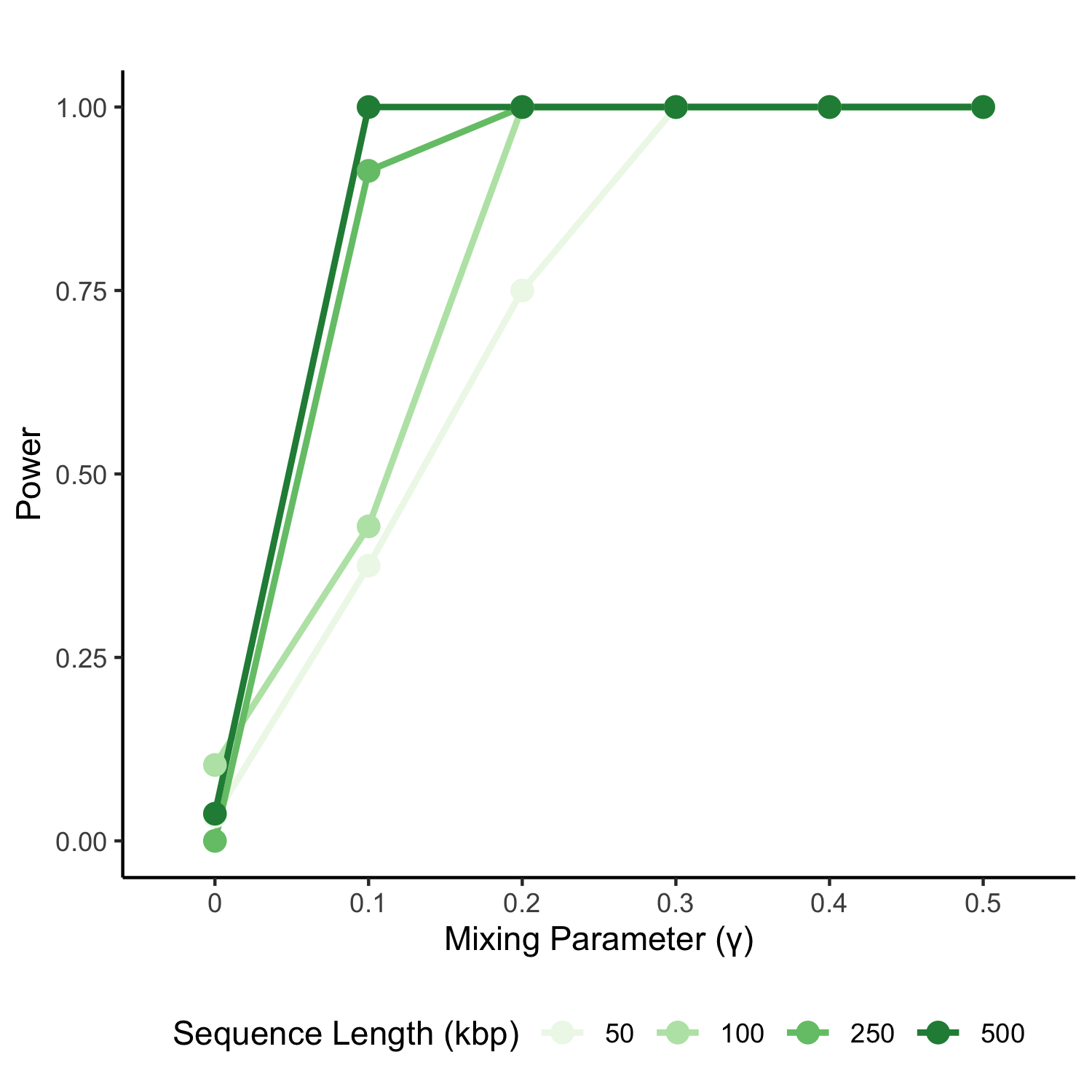}
    \caption{Power of HyDe test on detecting a single hybridization event by mixing parameter. This test is performed on a gene trees generated from a species topology $(O:2.0, ((P1:0.50,\#H1:0::\gamma):0.50,(P2:0.5,(HYB:0.5)\#H1:0::1-\gamma):0.5):1.0)$. This topology has relatively shorter branch lengths than those tested in other networks in this paper. It has four taxa and one hybridization event. }
    \label{fig:hyde_hybridlambda}
\end{figure*}

Our original simulations involved only the networks n10h2, n15h3 and n25h5. We noticed HyDe's low precision and high false negative rates which contradicted the positive results found in \cite{KongKubatko}. We then decided to reproduce the power calculations in \cite{KongKubatko} as shown in Figure \ref{fig:hyde_hybridlambda} and decided to include the networks n4h1, n4h1$_{\textrm{introgression}}$, n5h2 and n8h3 in the simulation study. We discovered that HyDe has good performance when the parent-hybrid relationship is clearly specified, and it is unable to detect hybrid relationships involving ghost lineages. \revision{We note that these simulations were done with hybrid-lambda, rather than ms.}

\subsection*{Simulation figures}

\begin{figure*}[!ht] 
    \centering
    \includegraphics[width=.95\linewidth]{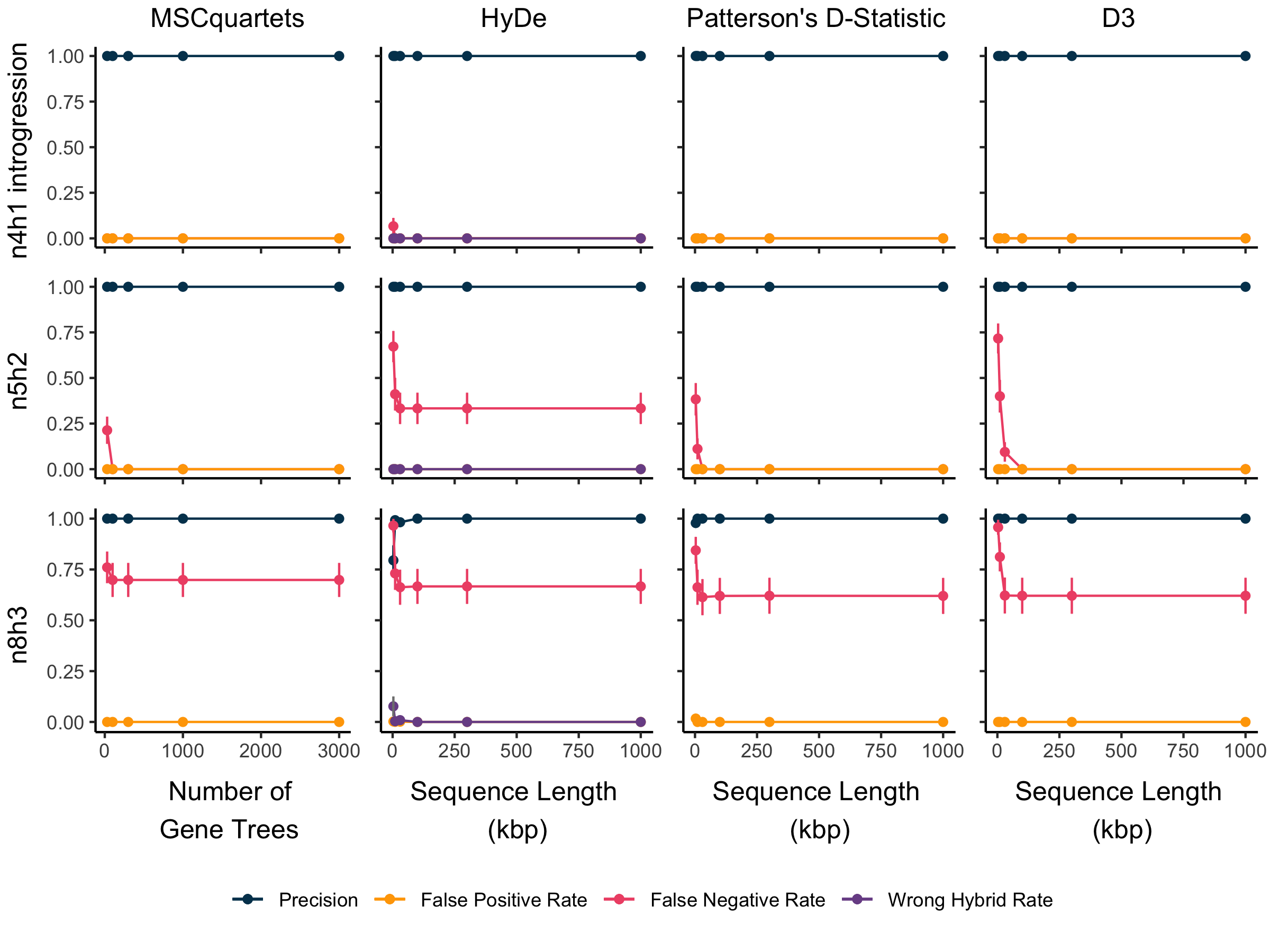}
     \caption{False positive rate (orange), precision (black) and false negative rate (red) for MSCquartets, HyDe, Patterson's D-Statistic, and D3 on the networks: n4h1$_{\textrm{introgression}}$ (network with single shallow hybridization event), n5h2 (network with two overlapping hybridization events) and n8h3 (network with three overlapping hybridization events). All tests are Bonferroni-corrected at a level of significance $\alpha = 0.05 / \textrm{number of tests}$ (see Figure \ref{fig:N458-05} in the \revision{Supplementary materials} for results on $\alpha=0.05$). For HyDe, an additional metric, Wrong Hybrid Rate (purple) describes the rate at which hybridization is detected but is falsely attributed to the incorrect hybrid taxon.
      For the introgression event (n4h1$_{\textrm{introgression}}$), MSCquartets, HyDe, Patterson's D-Statistic, and D3 all behave similarly to the hybrid speciation scenario in n4h1, with near-perfect recovery and identification of hybrid scenarios. Low false negative rates are noted for n5h2 across MSCquartets and D-related methods. For n5h2, HyDe correctly recovers parent-hybrid relationships, but has a slightly elevated false negative rate. Network n8h3 displays elevated false negative rates across all methods. However, precision is high, and false negative rates between all methods are comparable.}
      \label{fig:n4n5n8-supp}
   \end{figure*}

   \begin{figure*}[!ht] 
    \centering
    \includegraphics[width=.95\linewidth]{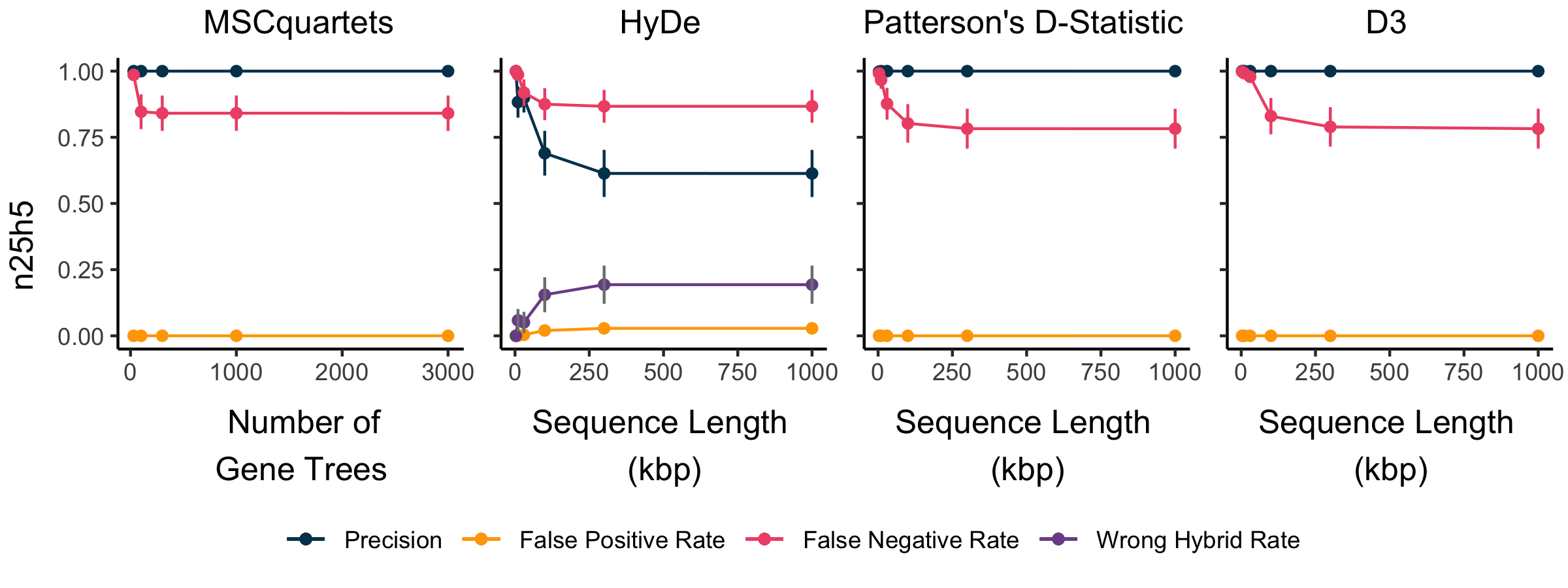}
    \caption{False positive rate (orange), precision (black) and false negative rate (red) for MSCquartets, HyDe, Patterson's D-Statistic, and D3 on the largest network tested with the highest number of hybridization events (n25h5). All tests are Bonferroni-corrected at a level of significance $\alpha = 0.05 / \textrm{number of tests}$ (see Figure \ref{fig:N25-05} in the \revision{Supplementary materials} for results on $\alpha=0.05$). For HyDe, an additional metric, Wrong Hybrid Rate (purple) describes the rate at which hybridization is detected but is falsely attributed to the incorrect hybrid taxon. High precision is noted for all tests that evaluate for hybridization within a triple or quartet. All methods also show a high false negative rate, and low false positive rate. Given that HyDe tests specific parent-hybrid relationships, it has lower precision scores than other methods.}
    \label{fig:n25-supp}
\end{figure*}

\begin{figure*}[!ht]
\begin{subfigure}{.485\linewidth}
  \centering
  \includegraphics[width=.9\linewidth]{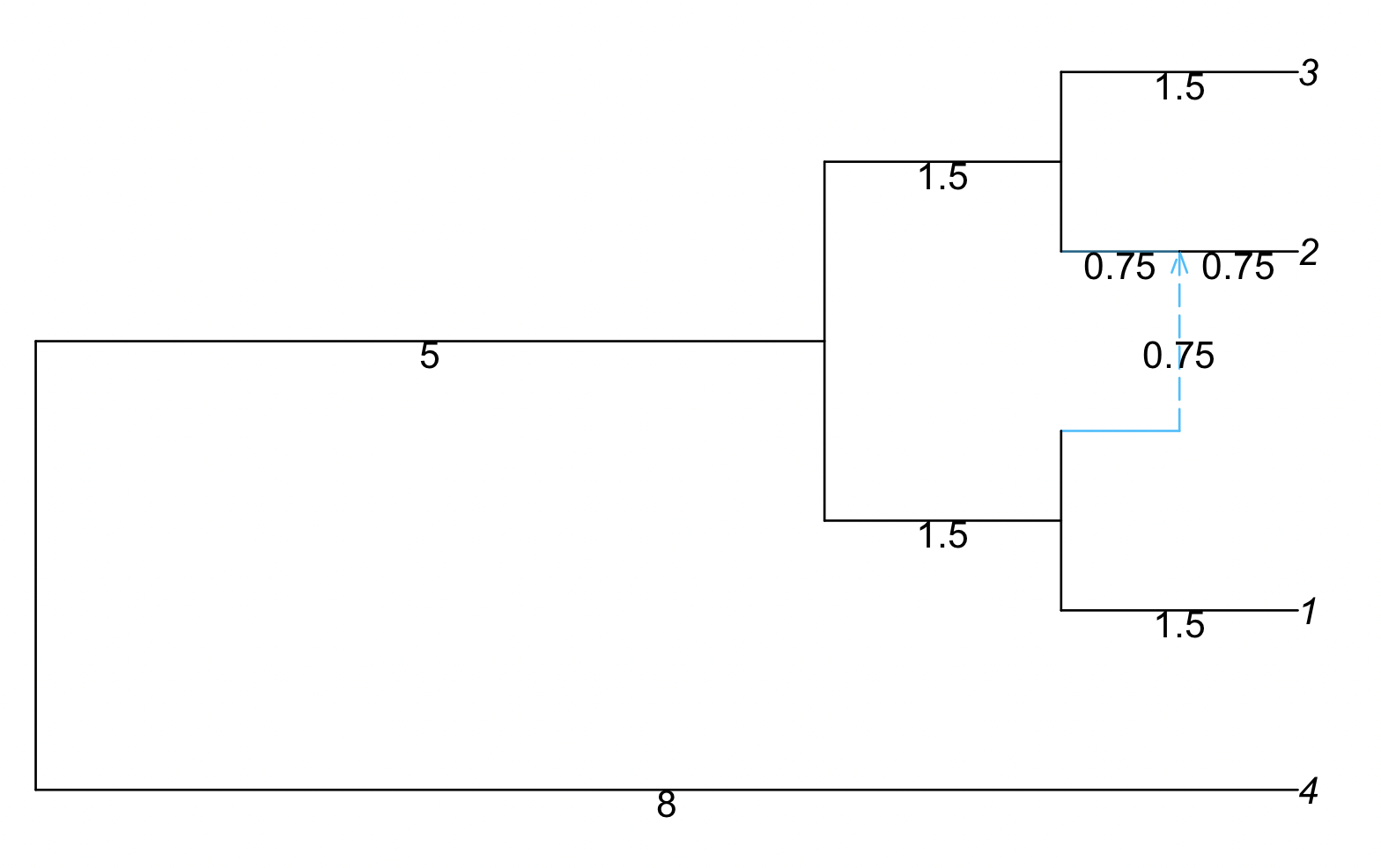}
  \caption{n4h1}
  \label{fig:1a}
\end{subfigure}%
\hfill
\begin{subfigure}{.485\linewidth}
  \centering
  \includegraphics[width=.9\linewidth]{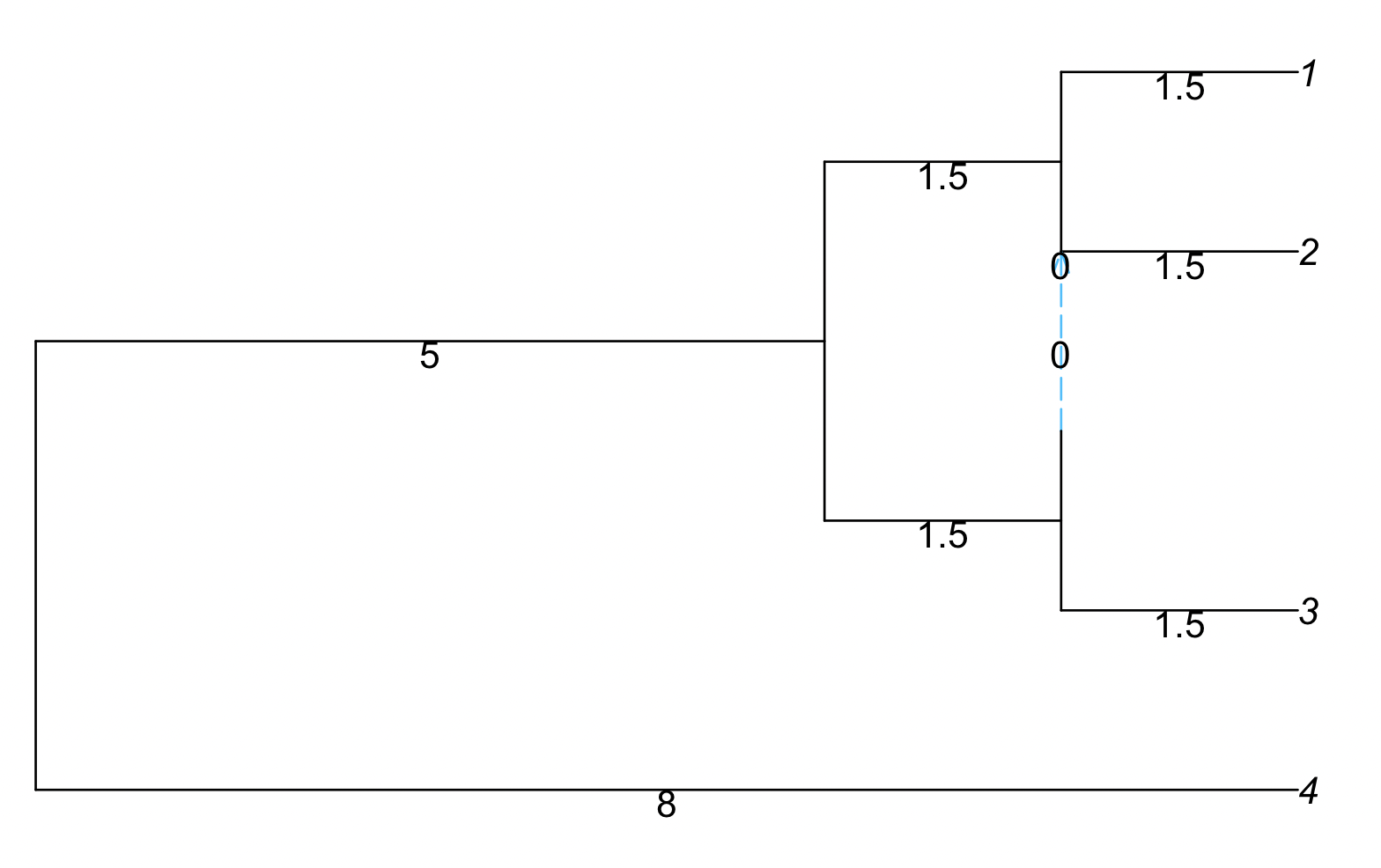}
  \caption{n4h1$_{\textrm{introgression}}$}
  \label{fig:1b}
\end{subfigure}

\begin{subfigure}{.485\linewidth}
  \centering
  \includegraphics[width=.9\linewidth]{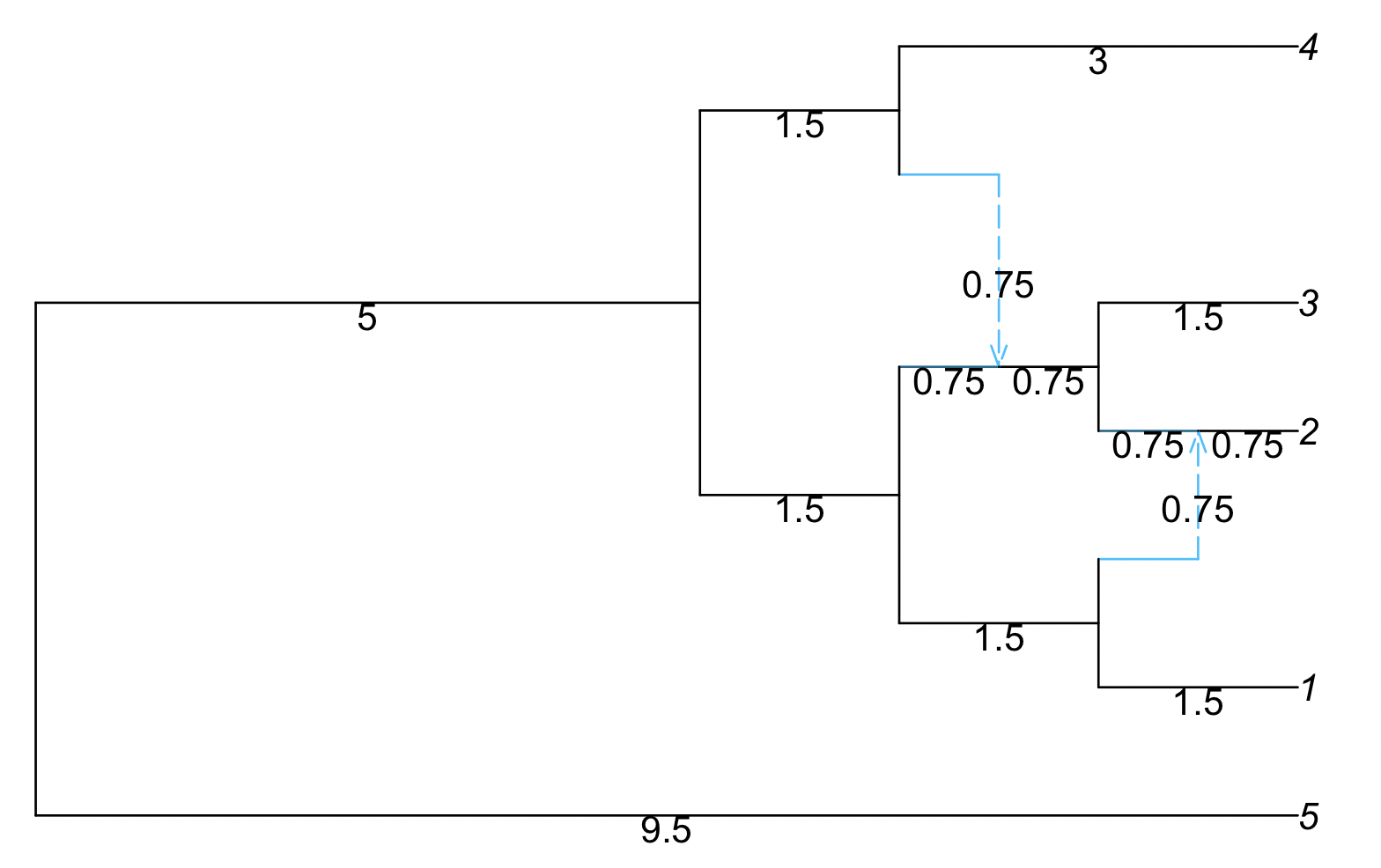}
  \caption{n5h2}
  \label{fig:1f}
\end{subfigure}
\hfill
\begin{subfigure}{.485\linewidth}
  \centering
  \includegraphics[width=.9\linewidth]{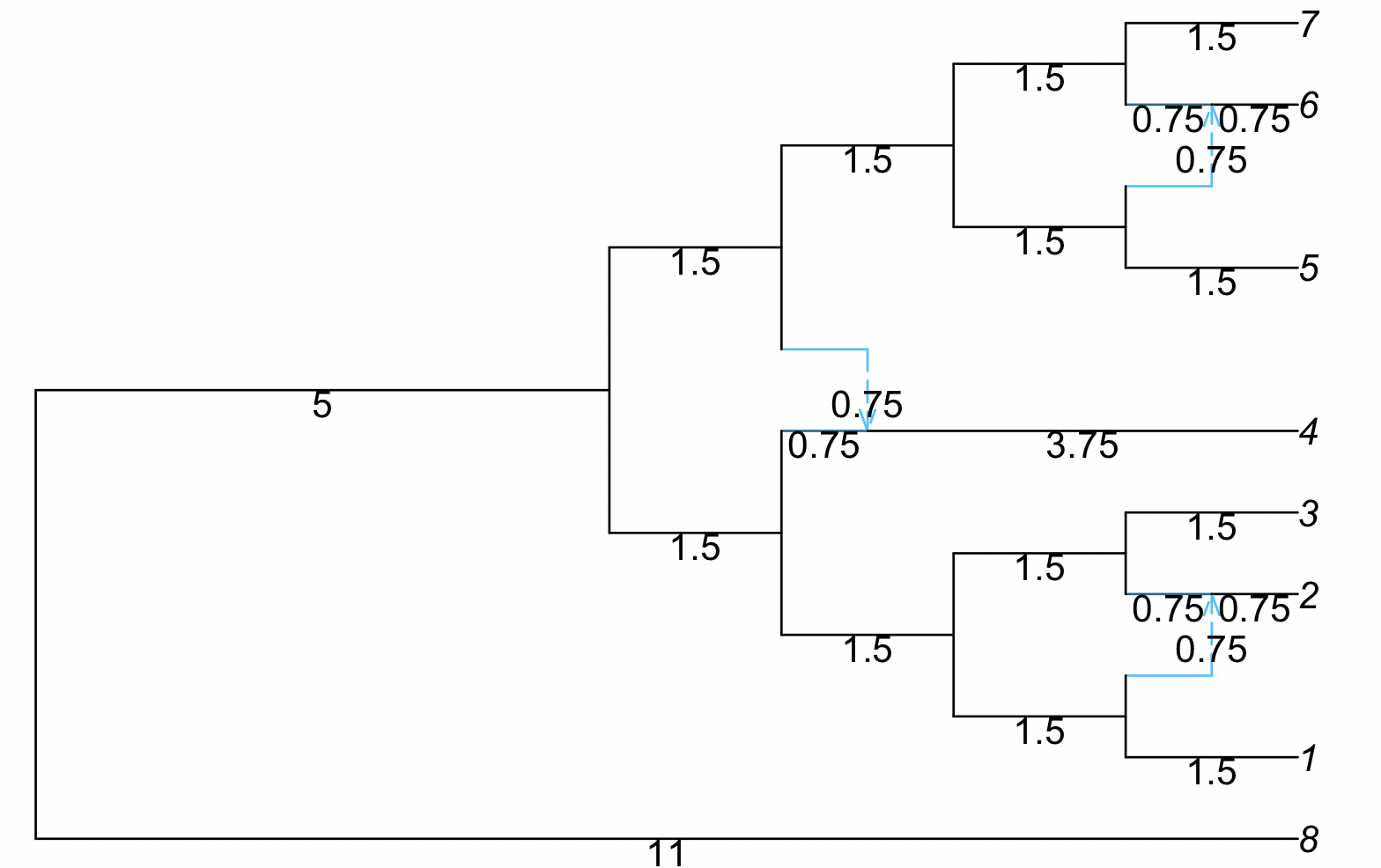}
  \caption{n8h3}
  \label{fig:1e}

\end{subfigure}
\caption{Networks obtained from \cite{KongKubatko}. All branches are labeled in coalescent units.
 (a) n4h1, a four taxon network where taxon 2 arises from ancestral lineages to taxon 1 and taxon 3 (denoted 'parents' in HyDe). This network depicts a hybrid speciation scenario and was simulated with various values for $\gamma \in \{0.0, 0.1, 0.2, 0.3, 0.4, 0.5\}$ (inheritance probability on the minor hybrid edge in blue). (b) n4h1$_{\textrm{introgression}}$ depicts an introgression event where $\gamma = 0.5$ with a single hybrid labeled 2 that has introgressed with taxon 1. 
 (c) n5h2 depicts overlapping hybridizations where one reticulation (ancestral to the (2,3) clade) forms the parental lineage of another reticulation event (taxon 2). (d) n8h3 depicts three singular hybridization events, each affecting single taxa, two of identical depths
 and one deeper event.}
\label{fig:fig-kongkubatkonets}
\end{figure*}

\begin{figure*}[!ht]
\begin{subfigure}{.485\linewidth}
  \centering
  \includegraphics[width=.9\linewidth]{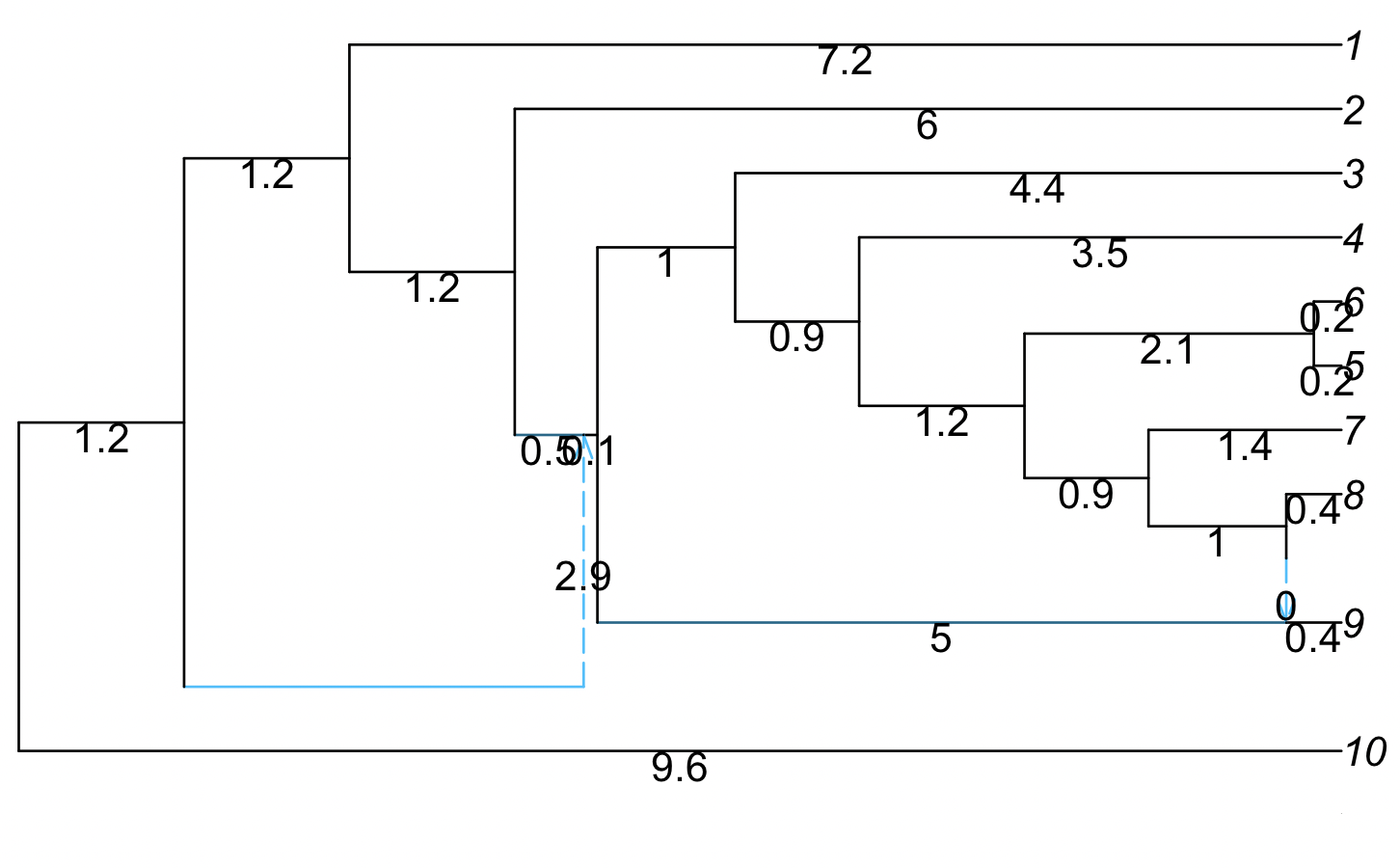}
  \caption{n10h2}
  \label{fig:n10h2}
\end{subfigure}
\hfill
\begin{subfigure}{.485\linewidth}
  \centering
  \includegraphics[width=.9\linewidth]{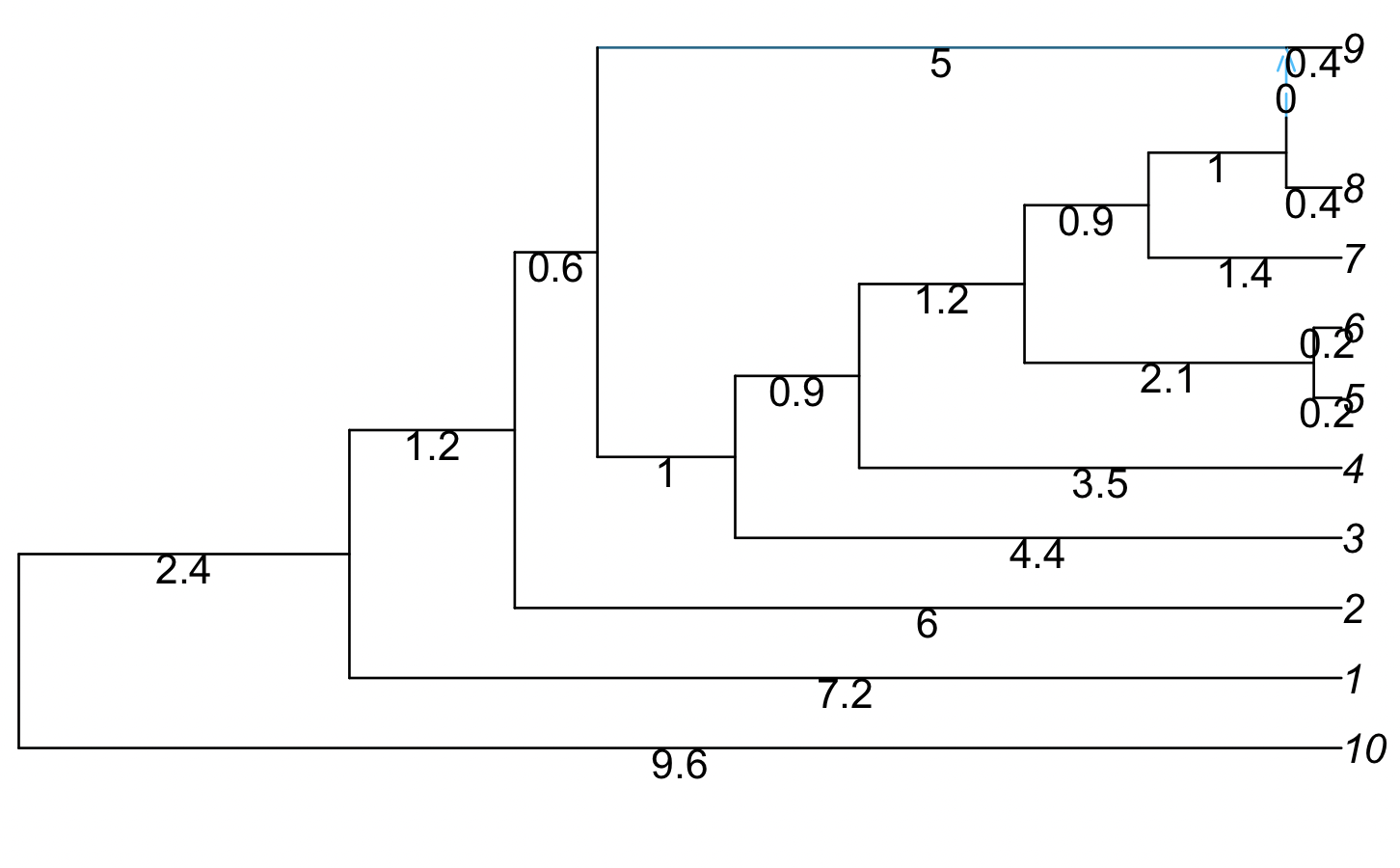}
  \caption{n10h1$_{\textrm{shallow}}$}
  \label{fig:n10h1shallow}
\end{subfigure}
\begin{subfigure}{.485\linewidth}
  \centering
  \includegraphics[width=.9\linewidth]{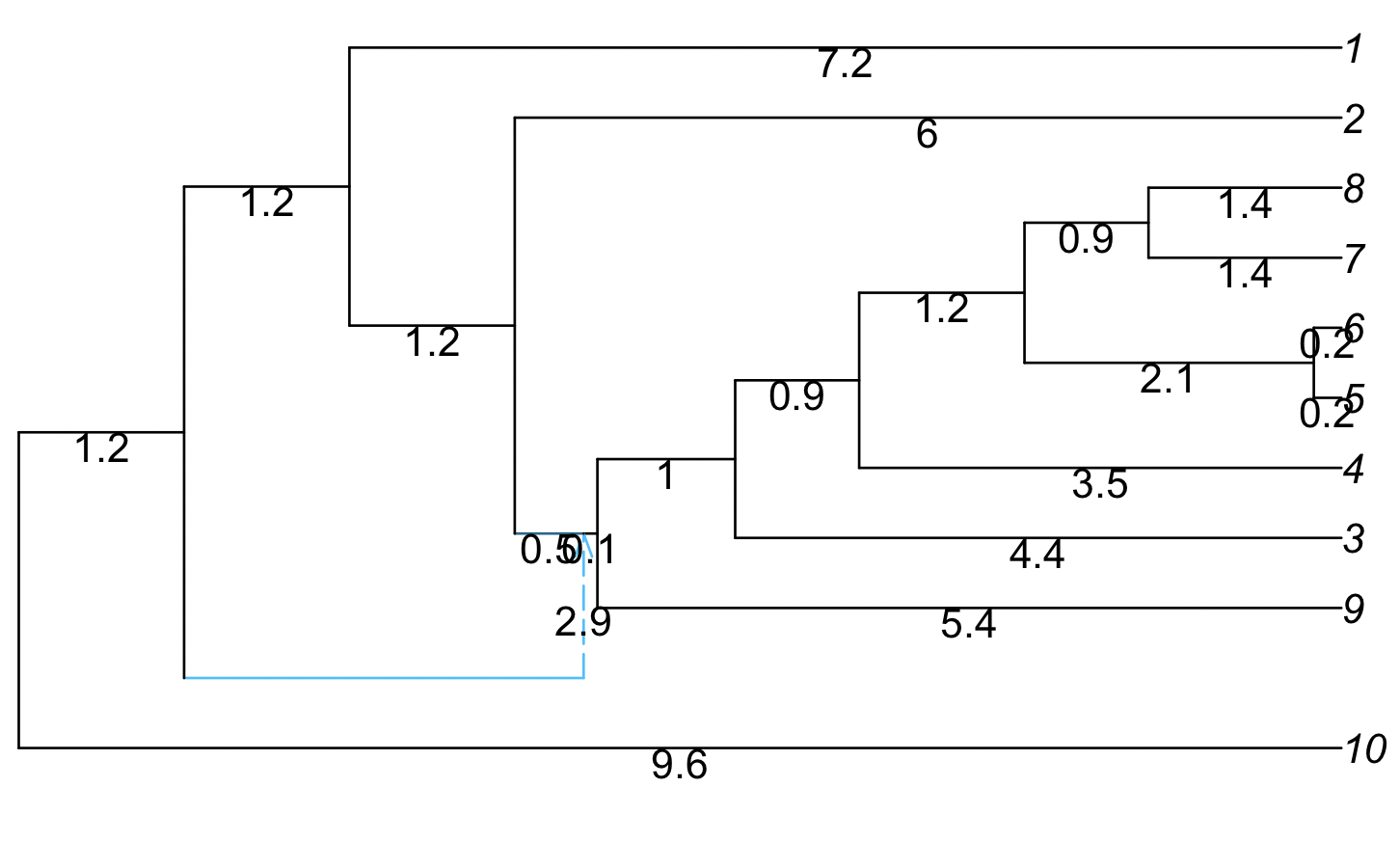}
  \caption{n10h1$_{\textrm{deep}}$}
  \label{fig:n10h1deep}
\end{subfigure}
\hfill
\begin{subfigure}{.485\linewidth}
  \centering
  \includegraphics[width=.9\linewidth]{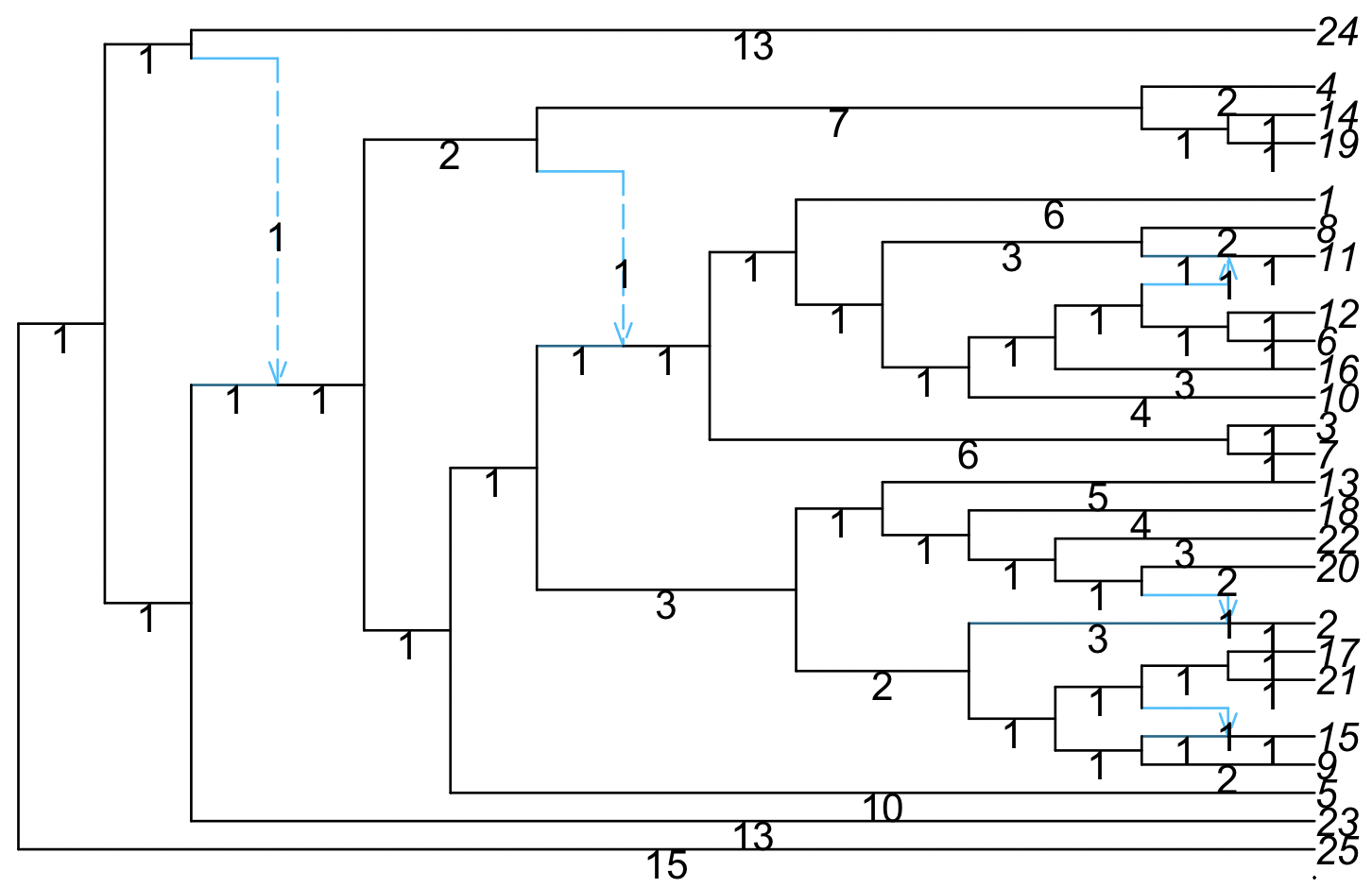}
  \caption{n25h5}
  \label{fig:n25h5}
\end{subfigure}
\caption{(a) Network n10h2 replicated from \cite{solisane} with ten taxa and two hybridization events, its singular hybridizations, (b) n10h1$_{\textrm{shallow}}$ and (c) n10h1$_{\textrm{deep}}$, and the largest network tested, (d) n25h5. All branches are labeled in coalescent units.
 Network n10h1$_{\textrm{shallow}}$ (b) is the n10h2 network containing only the shallowest hybridization ($\gamma = 0.2$), which impacts one taxon, labeled 9. Network n10h1$_{\textrm{deep}}$ (c) is a derivation the n10h2 network containing only the deepest hybridization event ($\gamma = 0.3$), which impacts taxa 3-9. Network n25h5 (d) contains 25 taxa and five hybridization events, three of which impact only a singular downstream taxon ($\gamma = 0.449, 0.395, 0.369$), with two that are deeper in the tree ($\gamma = 0.024, 0.334$), creating overlapping hybridization events.}
\label{fig:fig-n10n25nets}
\end{figure*}

\begin{figure*}[!ht]
\begin{subfigure}{.485\linewidth}
  \centering
  \includegraphics[width=.9\linewidth]{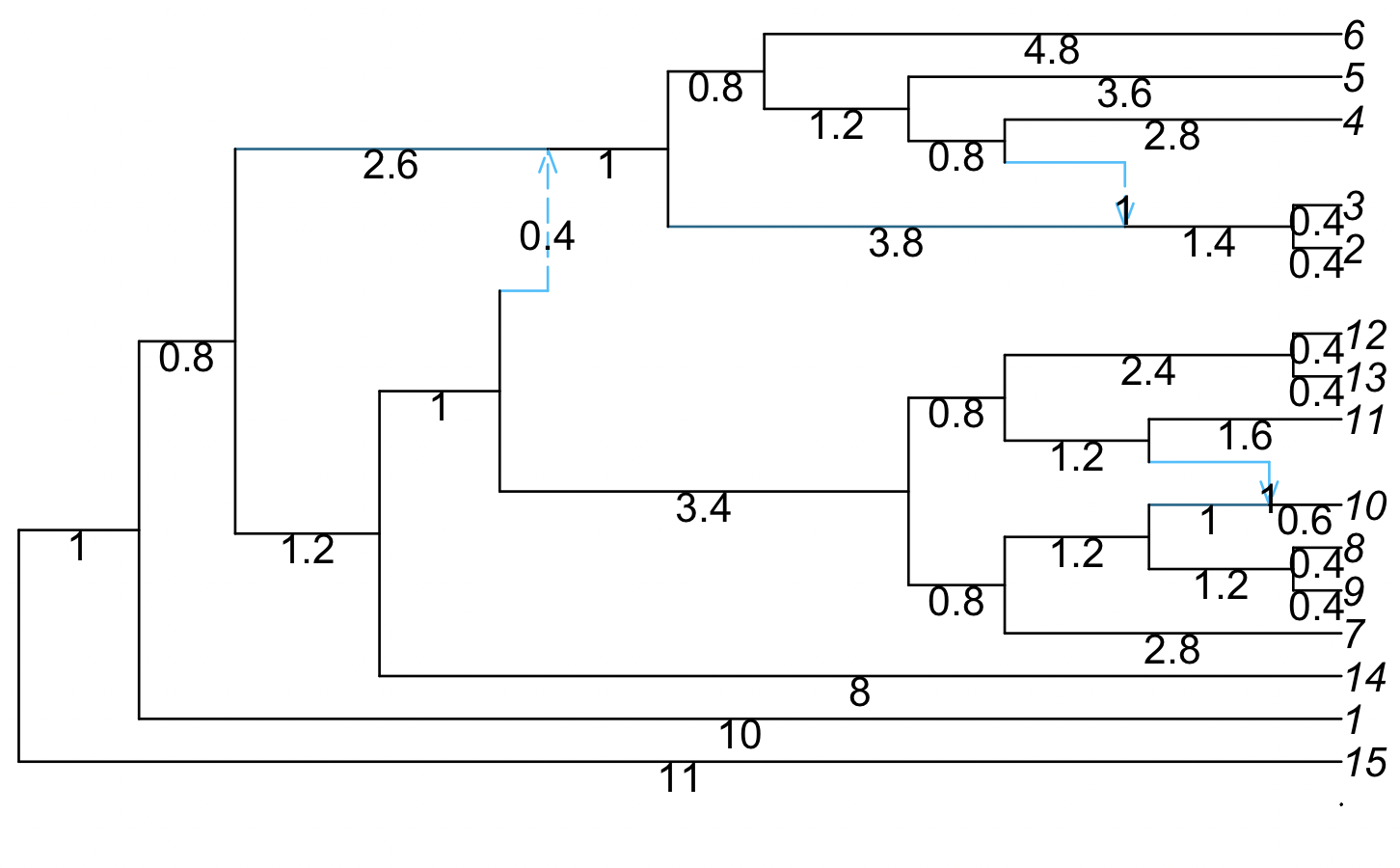}
  \caption{n15h3}
  \label{fig:n15h3}
\end{subfigure}
\hfill
\begin{subfigure}{.485\linewidth}
  \centering
  \includegraphics[width=.9\linewidth]{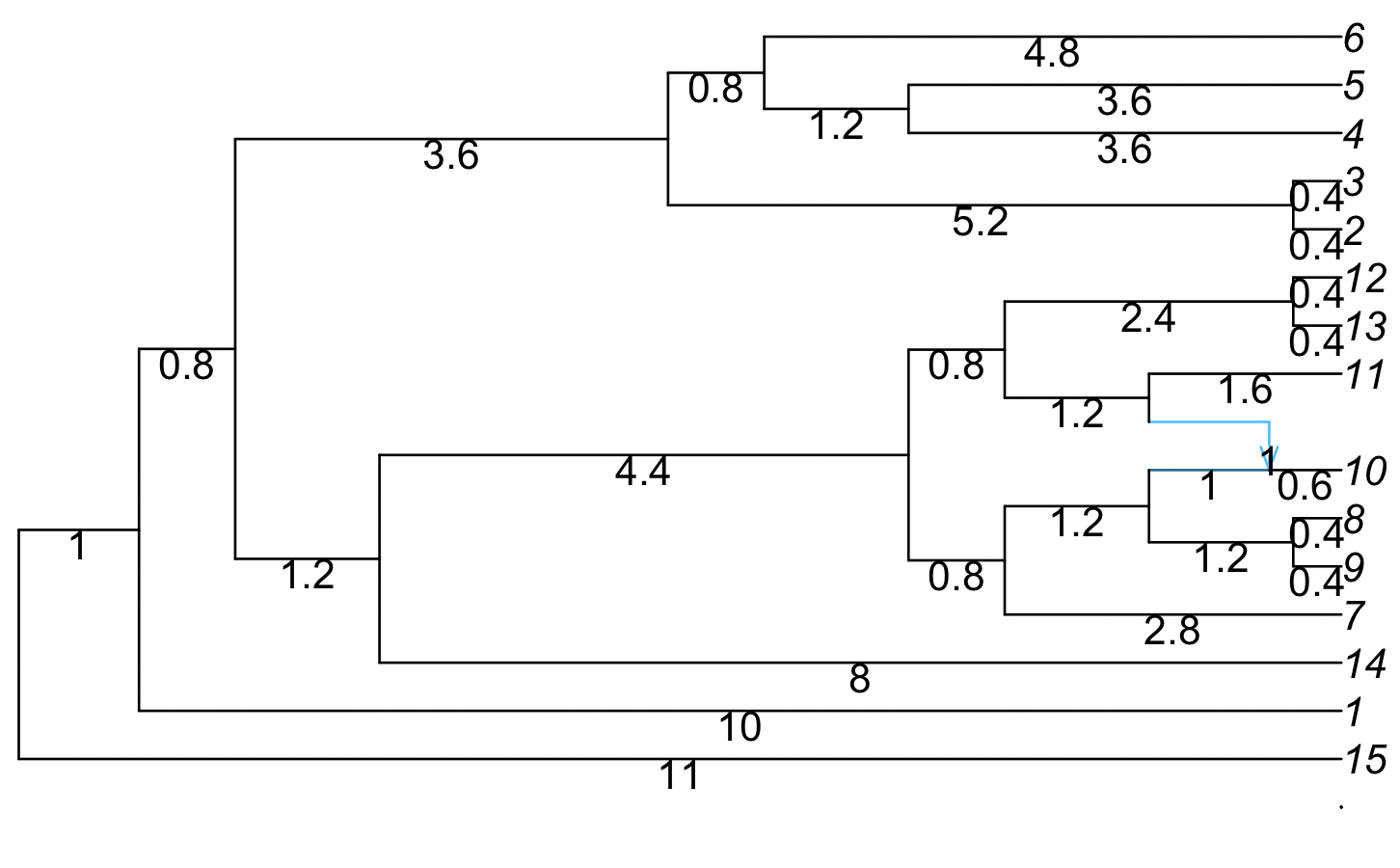}
  \caption{n15h1$_{\textrm{shallow}}$}
  \label{fig:n15h1shallow}
\end{subfigure}
\begin{subfigure}{.485\linewidth}
  \centering
  \includegraphics[width=.9\linewidth]{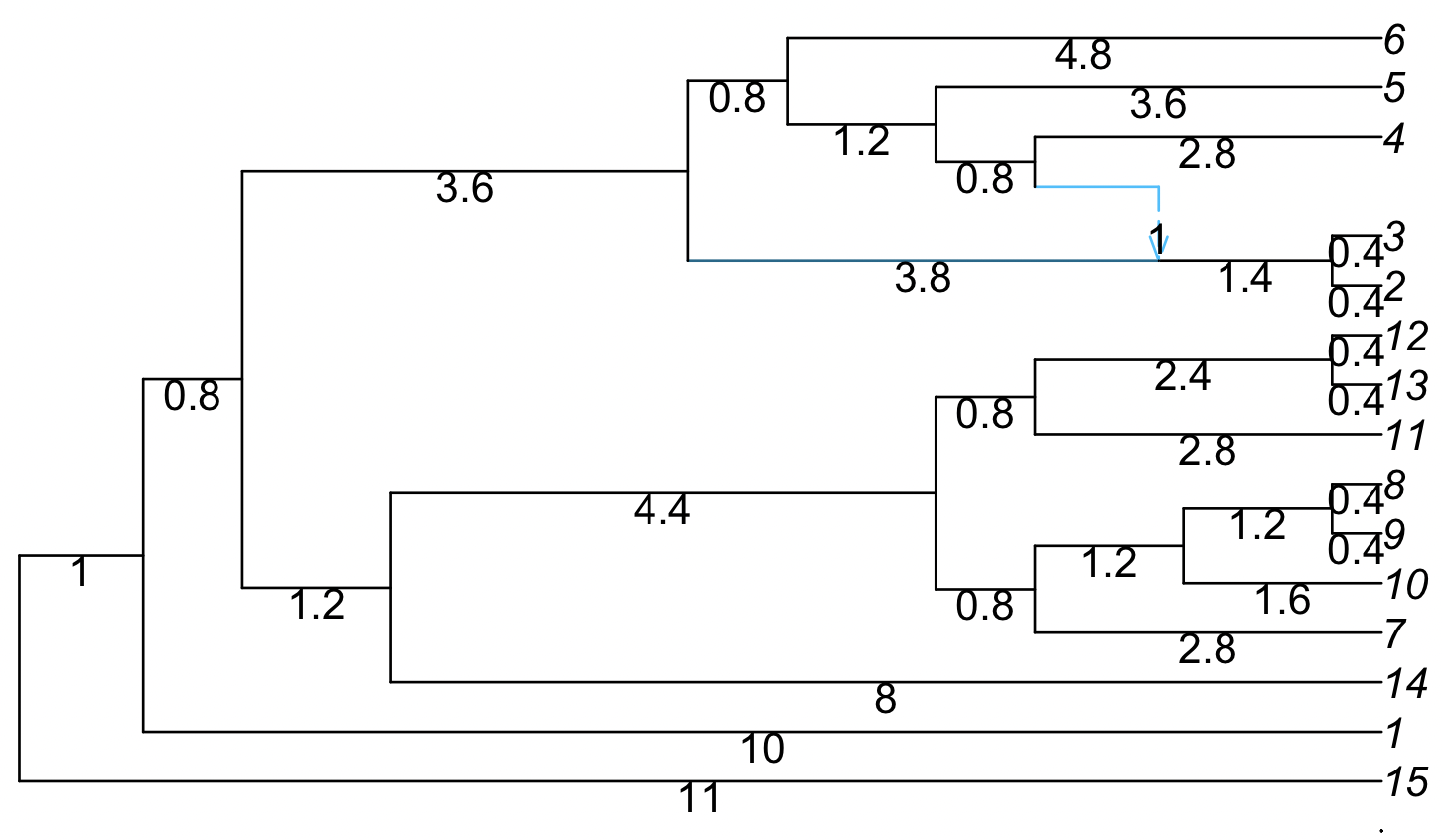}
  \caption{n15h1$_{\textrm{intermediate}}$}
  \label{fig:n15h1intermediate}
\end{subfigure}
\hfill
\begin{subfigure}{.485\linewidth}
  \centering
  \includegraphics[width=.9\linewidth]{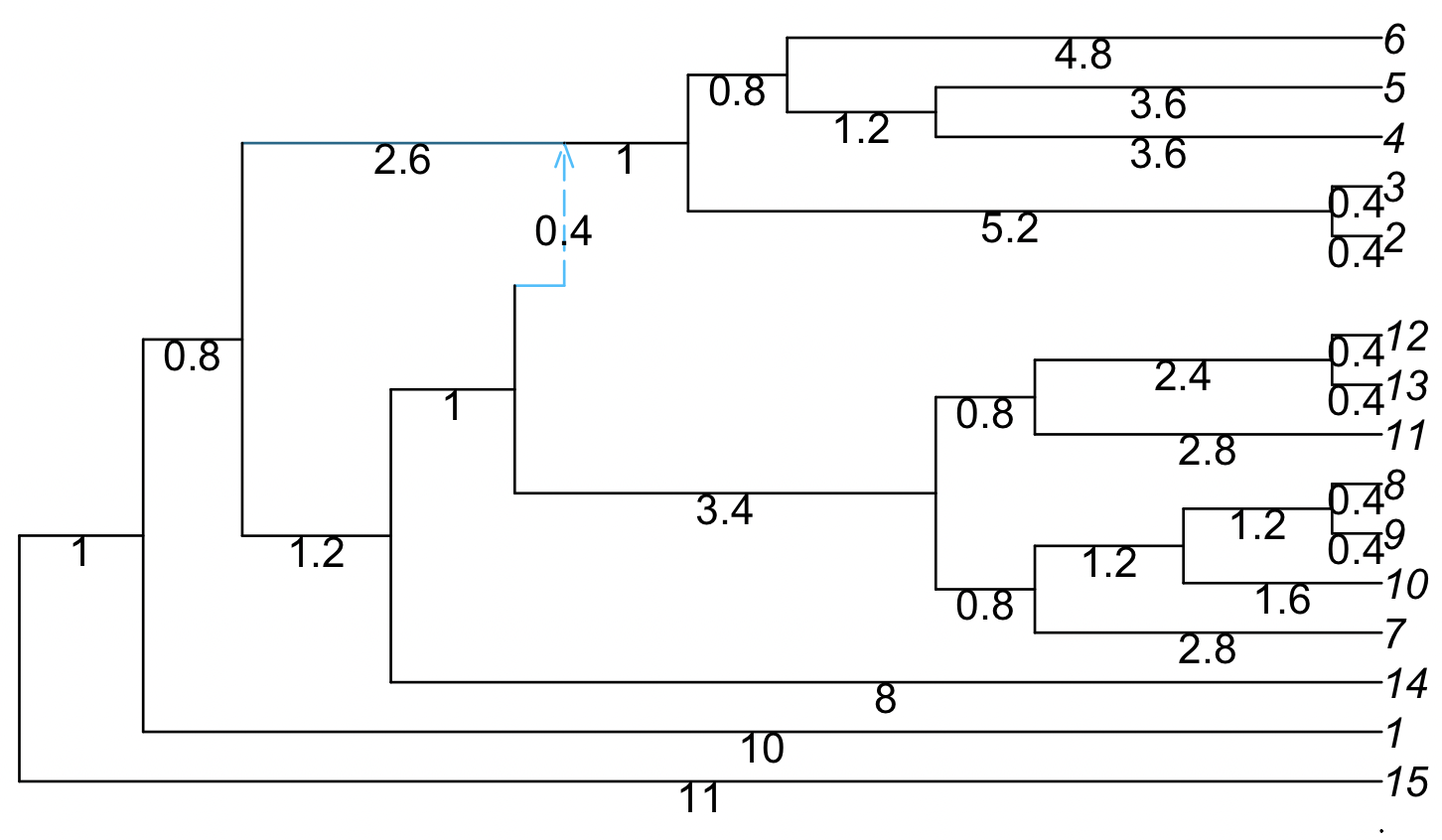}
  \caption{n15h1$_{\textrm{deep}}$}
  \label{fig:n15h1deep}
\end{subfigure}

\caption{(a) Network n15h3 replicated from \cite{solisane} with 15 taxa and 3 hybridizations events, and its singular hybridizations, (b) n15h1$_{\textrm{shallow}}$, (c) n15h1$_{\textrm{intermediate}}$, and (d) n15h1$_{\textrm{deep}}$. All branches are labeled in coalescent units. Network n15h1$_{\textrm{shallow}}$ (b) is the n15h3 network containing only the shallowest hybridization ($\gamma = 0.2$), which impacts only taxon 10. Network n15h1$_{\textrm{intermediate}}$ (c) is the n15h3 network with only the intermediate-depth hybridization ($\gamma = 0.2$) impacting the clade (2,3). Network n15h1$_{\textrm{deep}}$ (d) is the n15h3 network with only the deepest hybridization event ($\gamma = 0.3$). impacting the clade (2,3,4,5,6).}
\label{fig:fig-n15nets}
\end{figure*}

\begin{figure*}[ht]
    \centering
    \includegraphics[width=.95\linewidth]{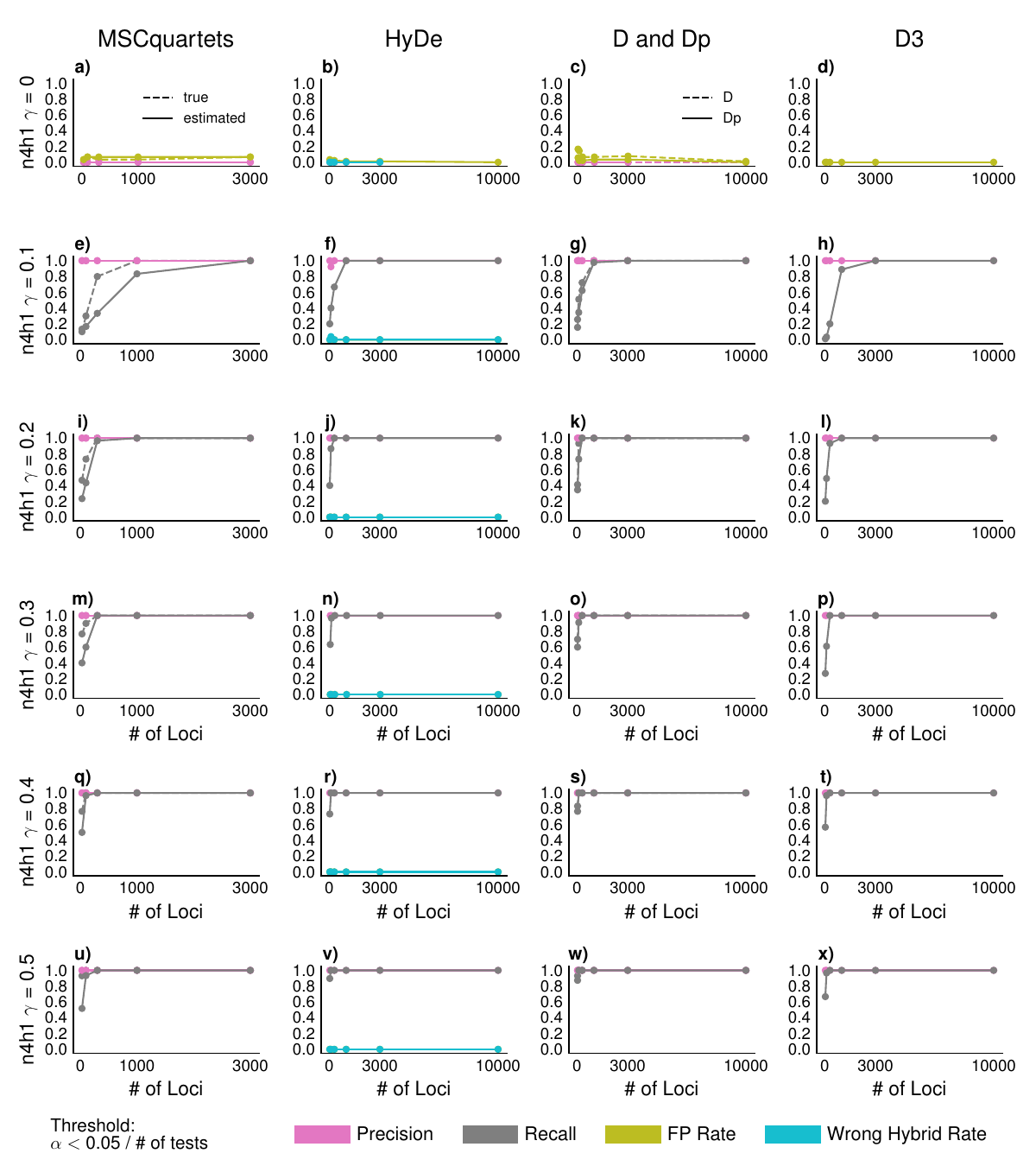}
    \caption{\revision{Precision, recall, false positive rate for MSCquartets (from true and estimated gene trees), HyDe, Patterson's D-Statistic, Dp, and D3 on the n4h1 network with mixing parameters $\gamma \in 0, 0.1, 0.2, 0.3, 0.4, 0.5$, where 0 depicts no gene flow, and 0.5 depicts equal contribution from both parental lineages. All tests are Bonferroni-corrected at a level of significance $\alpha = 0.05 / \textrm{number of tests}$ (see Figure \ref{fig:N4H1-05} in the \revision{Supplementary materials} for results on $\alpha=0.05$). For HyDe, an additional metric, Wrong Hybrid Rate describes the rate at which hybridization is detected but is falsely attributed to the incorrect hybrid taxon.
    In the absence of gene flow, HyDe and D3 have the lowest proportion of false positives, but the difference is not considerable when compared to that of Patterson's D-statistic and MSCquartets. With all other levels of admixture, as both admixture and sample size increases, all tests display an increased recall and decreased false positive rates. In cases where gene flow is present, D3 requires slightly higher sequence lengths to eliminate false negatives when compared to Patterson's D-Statistic. However, given a sufficient sequence length, all tests are able to reliably detect the presence of hybridization.
    }}
\label{fig:n4h1mix}
\end{figure*}

\begin{figure*}[h]
    \includegraphics[width=.95\linewidth]{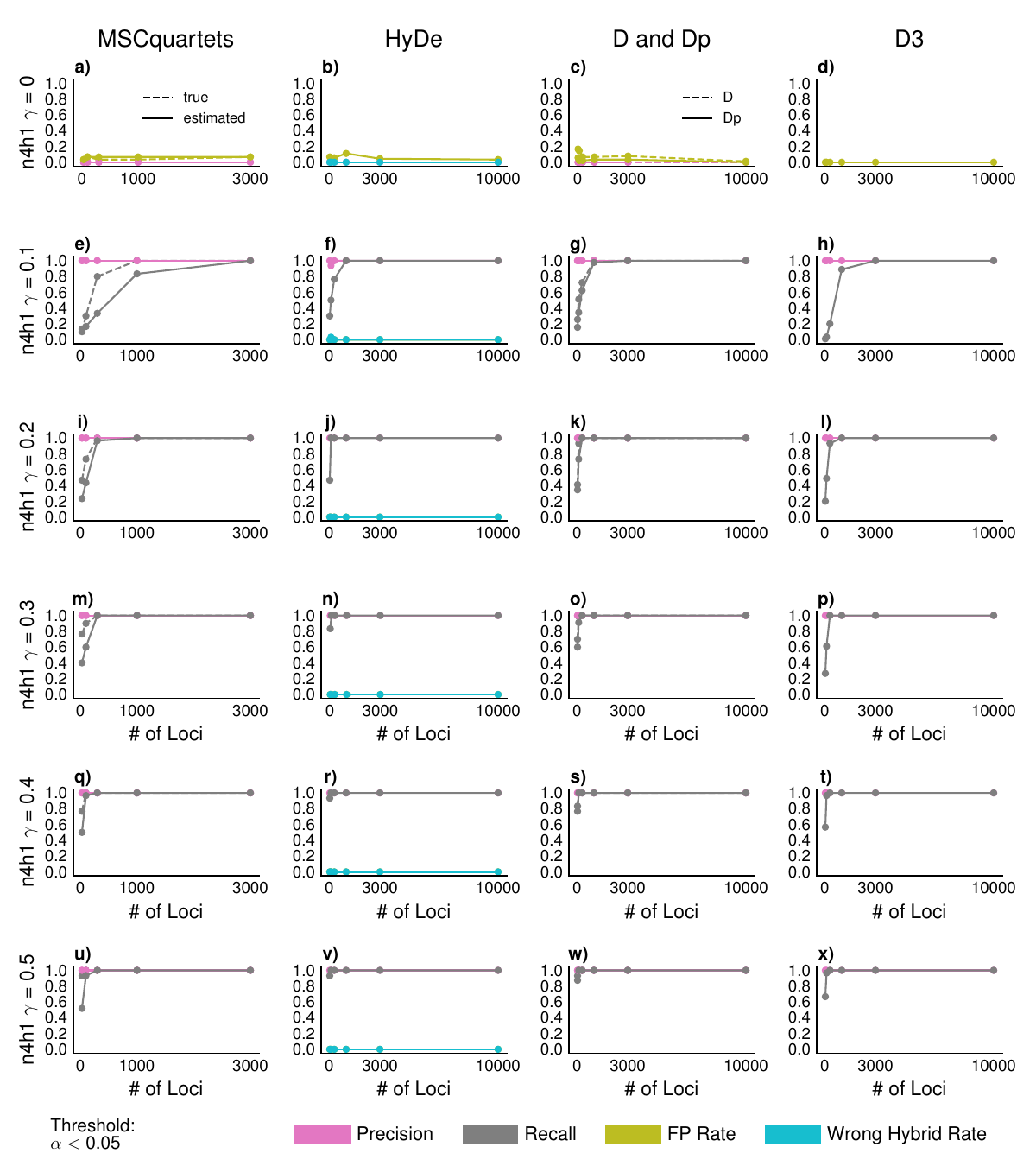}
    \caption{\revision{Precision, recall, false positive rate for MSCquartets (from true and estimated gene trees), HyDe, 's D-Statistic, and D3 on the n4h1 network with mixing parameters $\gamma \in 0, 0.1, 0.2, 0.3, 0.4, 0.5$, where 0 depicts no gene flow, and 0.5 depicts equal contribution from both parental lineages. For HyDe, an additional metric, Wrong Hybrid Rate describes the rate at which hybridization is detected but is falsely attributed to the incorrect hybrid taxon. All tests are at a level of significance $\alpha = 0.05$. Non-zero false positive rates can be seen in the case of zero gene flow for all but D3. As gene flow and amount of genetic information increases, each method displays a decrease in false negative rate.}}
    \label{fig:N4H1-05}
\end{figure*}
    
\begin{figure*}
    \includegraphics[width=.99\linewidth]{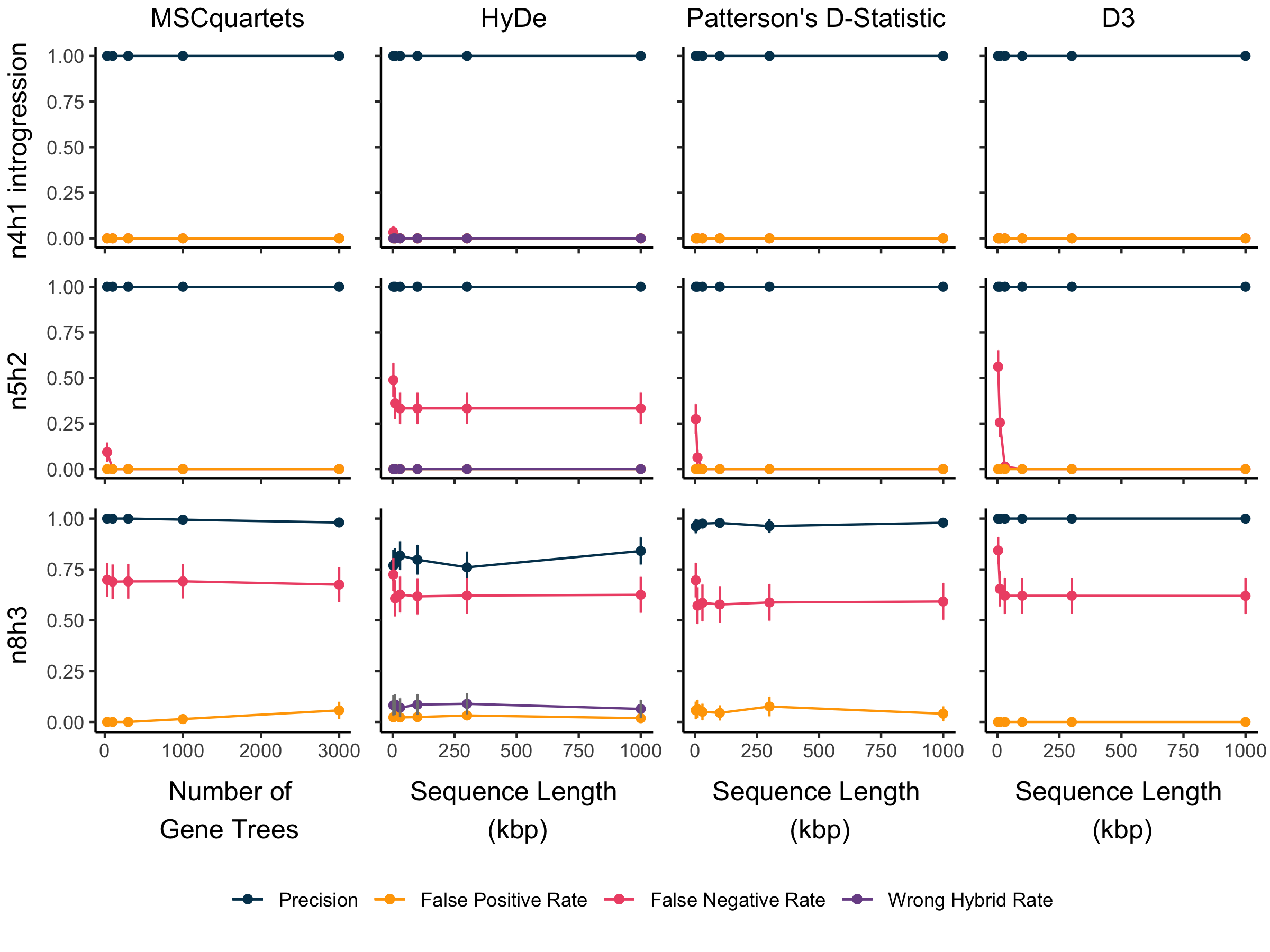}
    \caption{False positive rate (orange), precision (black) and false negative rate (red) for MSCquartets, HyDe, Patterson's D-Statistic, and D3 on the networks: n4h1$_{\textrm{introgression}}$ (network with single shallow hybridization event), n5h2 (network with two overlapping hybridization events) and n8h3 (network with three overlapping hybridization events). All tests are at a level of significance $\alpha = 0.05$. Each method performs well on the introgression event of n4h1, and n5h2 with some false negatives seen with HyDe. Each method has a similar false negative rate for h8h3, with precision near 100\% for all but HyDe, which still had 75\% precision at minimum for all sequence lengths. }
    \label{fig:N458-05}
\end{figure*}
    
\begin{figure*}
    \includegraphics[width=.99\linewidth]{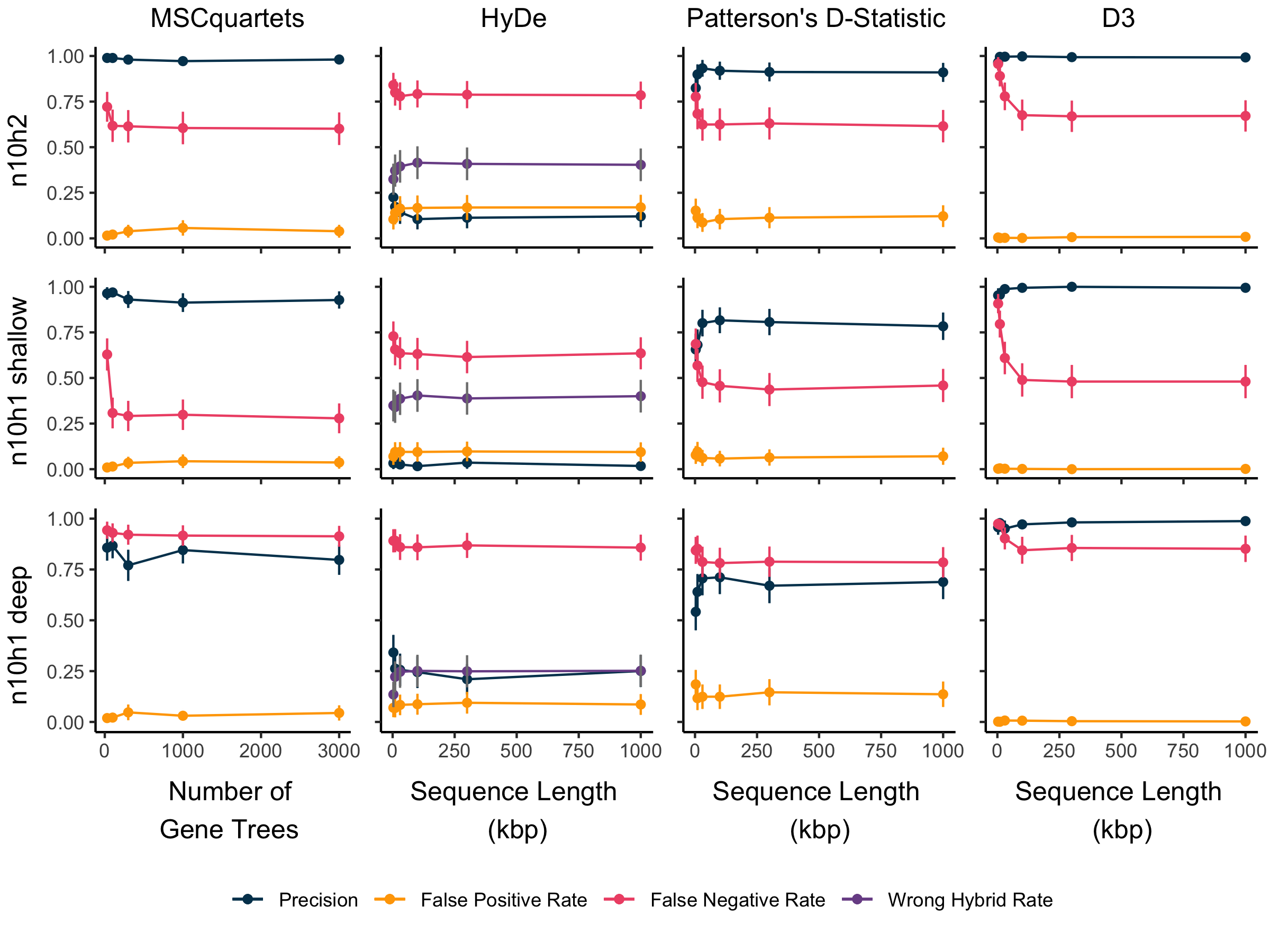}
    \caption{False positive rate (orange), precision (black) and false negative rate (red) for MSCquartets, HyDe, Patterson's D-Statistic, and D3 on the networks:
      n10h2, n10h1$_{\textrm{shallow}}$ (n10h2 with only the shallow hybridization event), n10h1$_{\textrm{deep}}$ (n10h2 with only the deeper hybridization event). All tests are at a level of significance $\alpha = 0.05$. D3 performs well on n10h2 and its singular hybridizations n10h1$_{\textrm{shallow}}$ and n10h1$_{\textrm{deep}}$ relative to other methods given its high precision. In contrast, Patterson's D-Statistic has slightly higher false positives, at the cost of precision. MSCquartets also has increased false positive and false negative rates. HyDe has higher false negative rates for n10h2 and n10h1$_{\textrm{shallow}}$. There are also high false positives (~20\%) in the case of n10h2, with a pattern for hybrid clade misidentification more frequently than precise identification of hybrid clades.}
    \label{fig:N10-05}
\end{figure*}
    
\begin{figure*}
    \includegraphics[width=.99\linewidth]{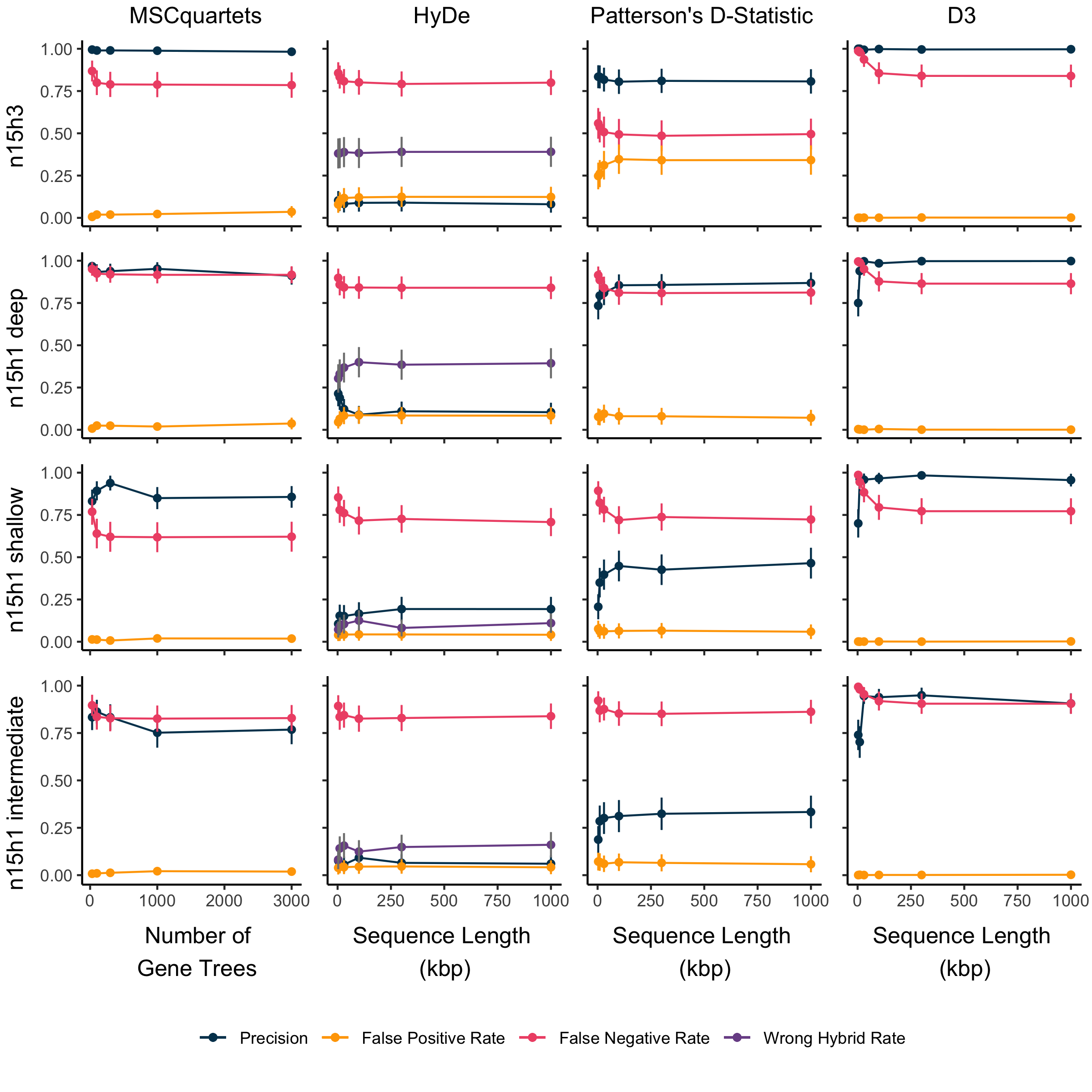}
    \caption{False positive rate (orange), precision (black) and false negative rate (red) for MSCquartets, HyDe, Patterson's D-Statistic, and D3 on the networks:
    n15h3, n15h1$_{\textrm{deep}}$ (n15h3 with only the deepest hybridization event), n15h1$_{\textrm{shallow}}$ (n15h3 with only the shallowest hybridization event), and n15h1$_{\textrm{intermediate}}$ (n15h3 with a hybridization event of intermediate depth). All tests are at a level of significance $\alpha = 0.05$. D3 has the highest and most stable precision across all n15h3 and single hybrid derivatives. False negative rates remain similar across all tests, with the exception of Patterson's D-Statistic on n15h3, which is slightly lower, paired with a higher false positive rate. Patterson's D-Statistic also shows decreased precision scores. MSCquartets' precision remains high ($\>80\%$) in all cases except n15h1$_{\textrm{intermediate}}$. HyDe also has a lower precision, around $10\%-25\%$ across all network formats.}
    \label{fig:N15-05}
\end{figure*}
    
\begin{figure*}
    \includegraphics[width=.99\linewidth]{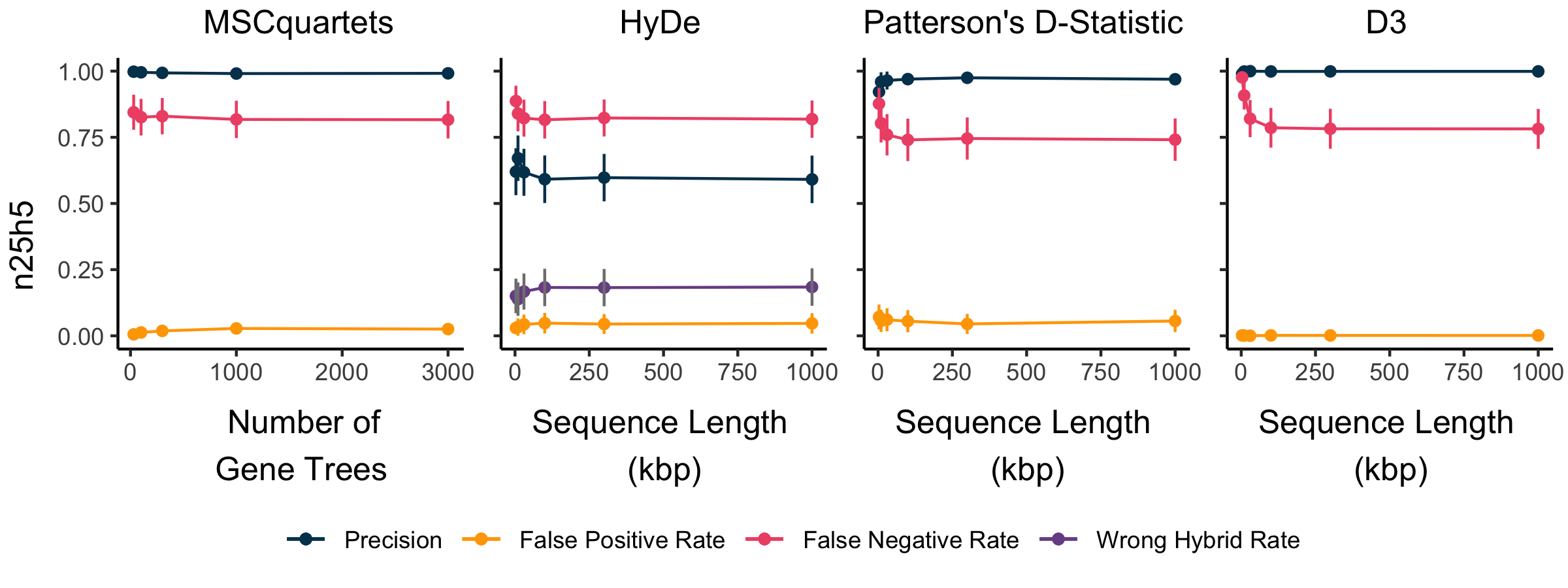}
    \caption{False positive rate (orange), precision (black) and false negative rate (red) for MSCquartets, HyDe, Patterson's D-Statistic, and D3 on the largest network tested with the highest number of hybridization events (n25h5). All tests are at a level of significance $\alpha = 0.05$. MSCquartets, Patterson's D-Statistic, and D3 all behave comparably on n25h5, the largest tested network with three shallow hybridization events. HyDe also shows higher relative precision and lower rates of hybrid clade misattribution.}
    \label{fig:N25-05}
\end{figure*}

\begin{figure*}
    \includegraphics[width=.99\linewidth]{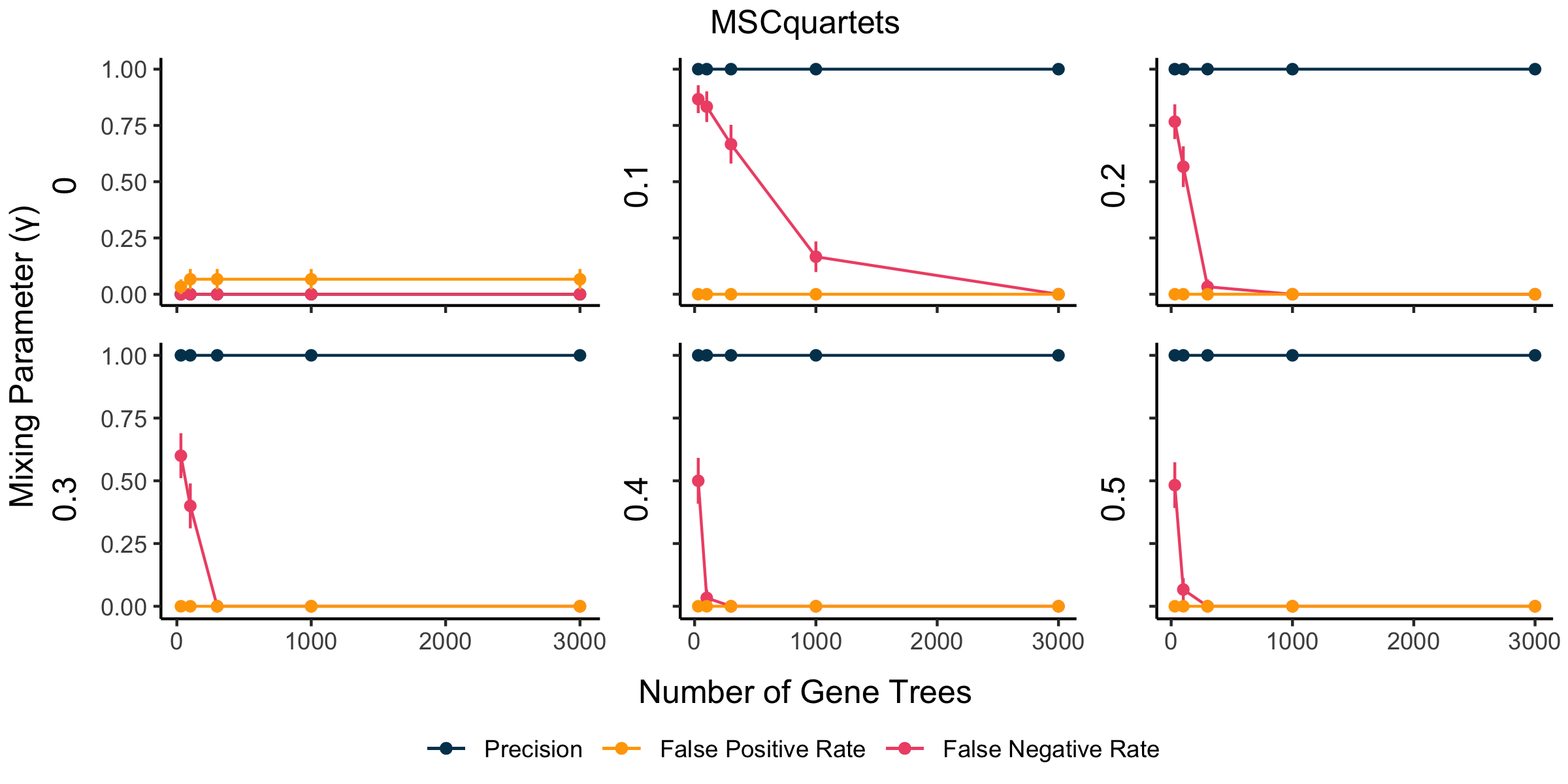}
    \caption{False positive rate (orange), precision (black) and false negative rate (red) for MSCquartets using inputs derived from IQTREE on the n4h1 network with mixing parameters $\gamma \in 0, 0.1, 0.2, 0.3, 0.4, 0.5$, where 0 depicts no gene flow, and 0.5 depicts equal contribution from both parental lineages. In this figure, all tests are at a level of significance $\alpha = 0.05$. }
    \label{fig:IQTREE-msc-n4h1-alpha05}
\end{figure*}

\begin{figure*}
    \includegraphics[width=.99\linewidth]{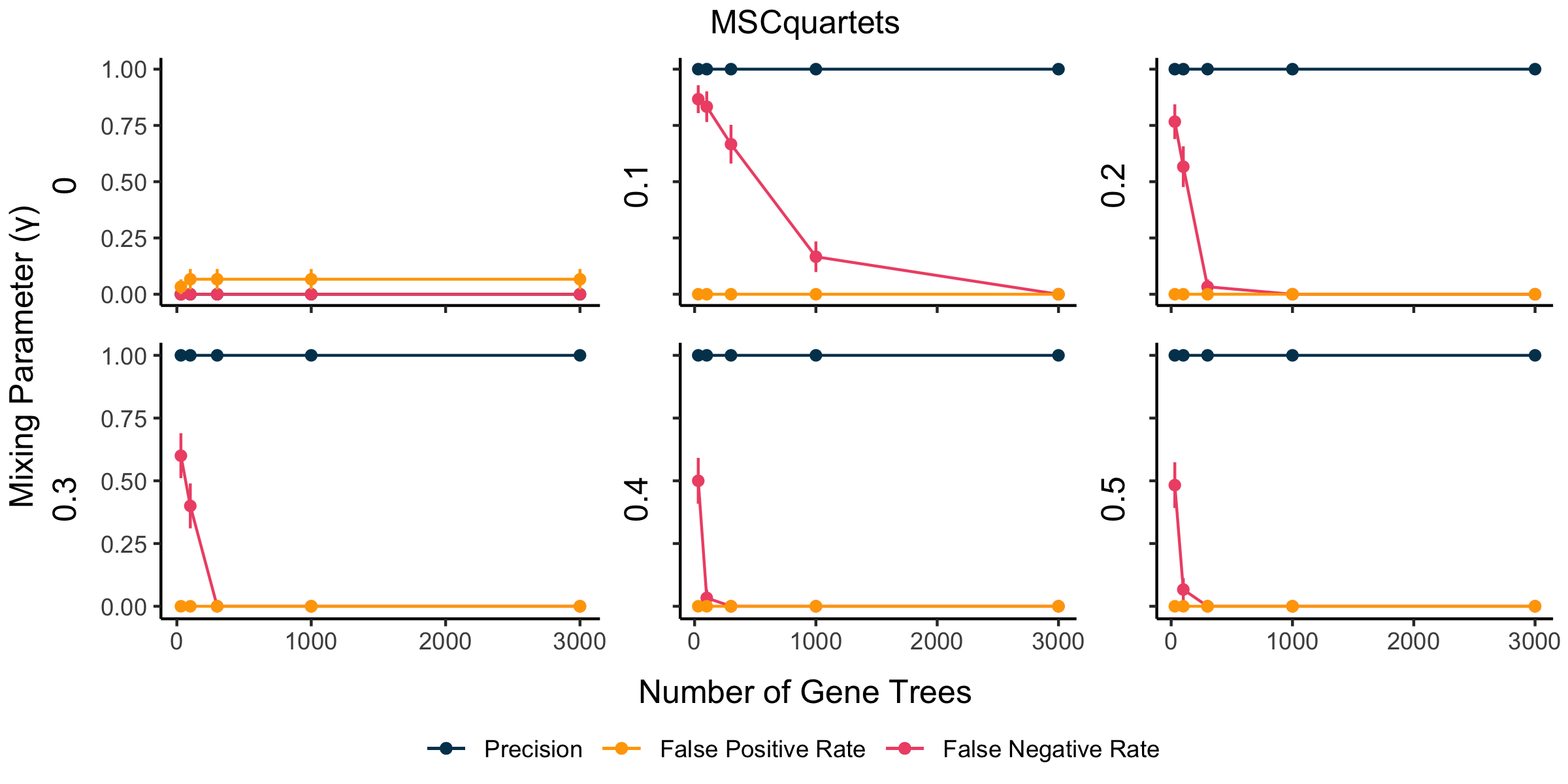}
    \caption{False positive rate (orange), precision (black) and false negative rate (red) for MSCquartets using inputs derived from IQTREE on the n4h1 network with mixing parameters $\gamma \in 0, 0.1, 0.2, 0.3, 0.4, 0.5$, where 0 depicts no gene flow, and 0.5 depicts equal contribution from both parental lineages. In this figure, all tests are Bonferroni-corrected at a level of significance $\alpha = 0.05 / \textrm{number of tests}$. }
    \label{fig:IQTREE-msc-n4h1-bonferroni}
\end{figure*}

\begin{figure*}
    \includegraphics[width=.99\linewidth]{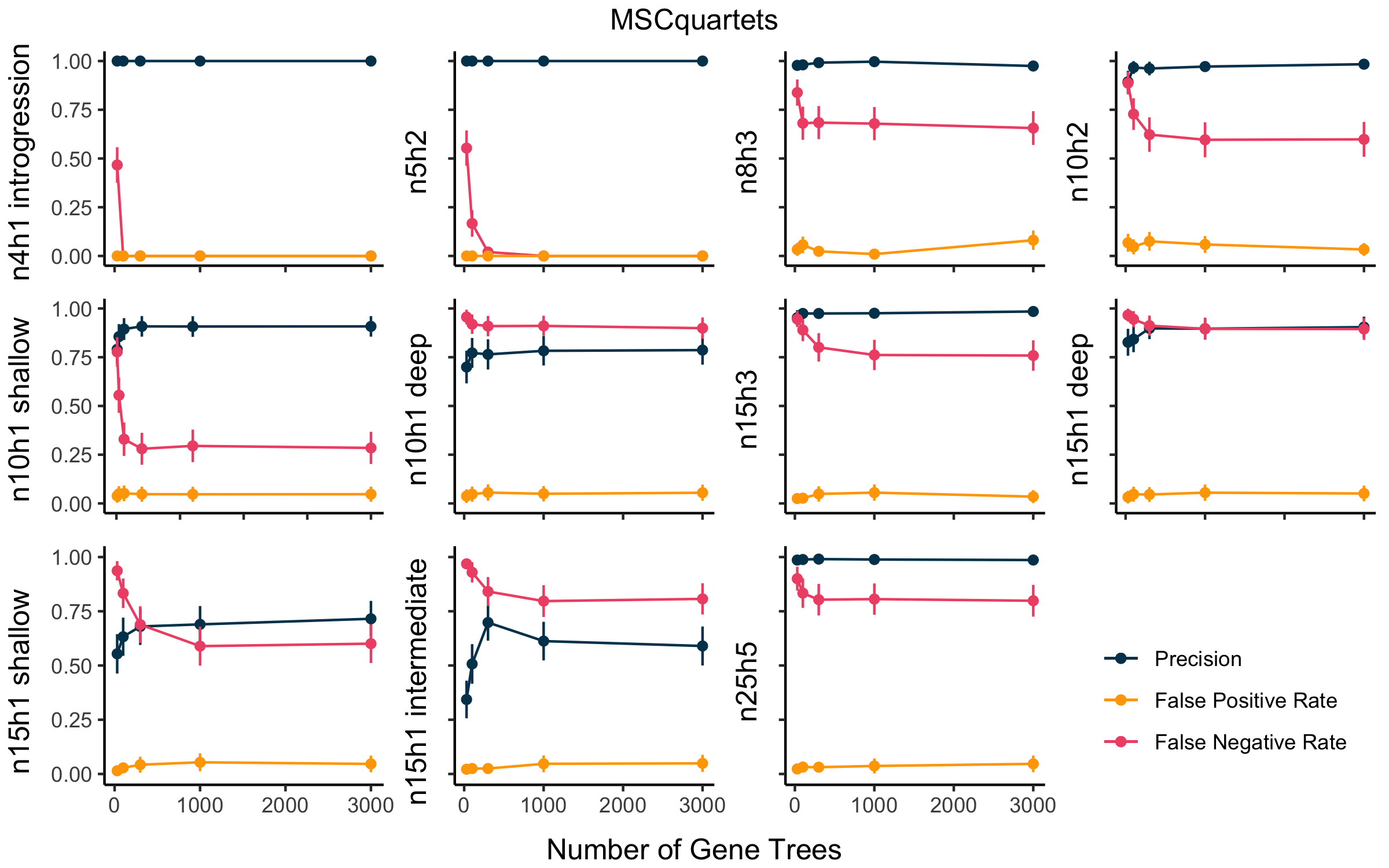}
    \caption{False positive rate (orange), precision (black) and false negative rate (red) for MSCquartets using inputs derived from IQTREE on n4 introgression, n5h2, n8h3, n10h2, n10 shallow, n10 deep, n15h3, n15 shallow, n15 intermediate, n15 deep, and n25h5. In this figure, all tests are at a level of significance $\alpha = 0.05$. }
    \label{fig:IQTREE-msc-n4-n25-alpha05}
\end{figure*}

\begin{figure*}
    \includegraphics[width=.99\linewidth]{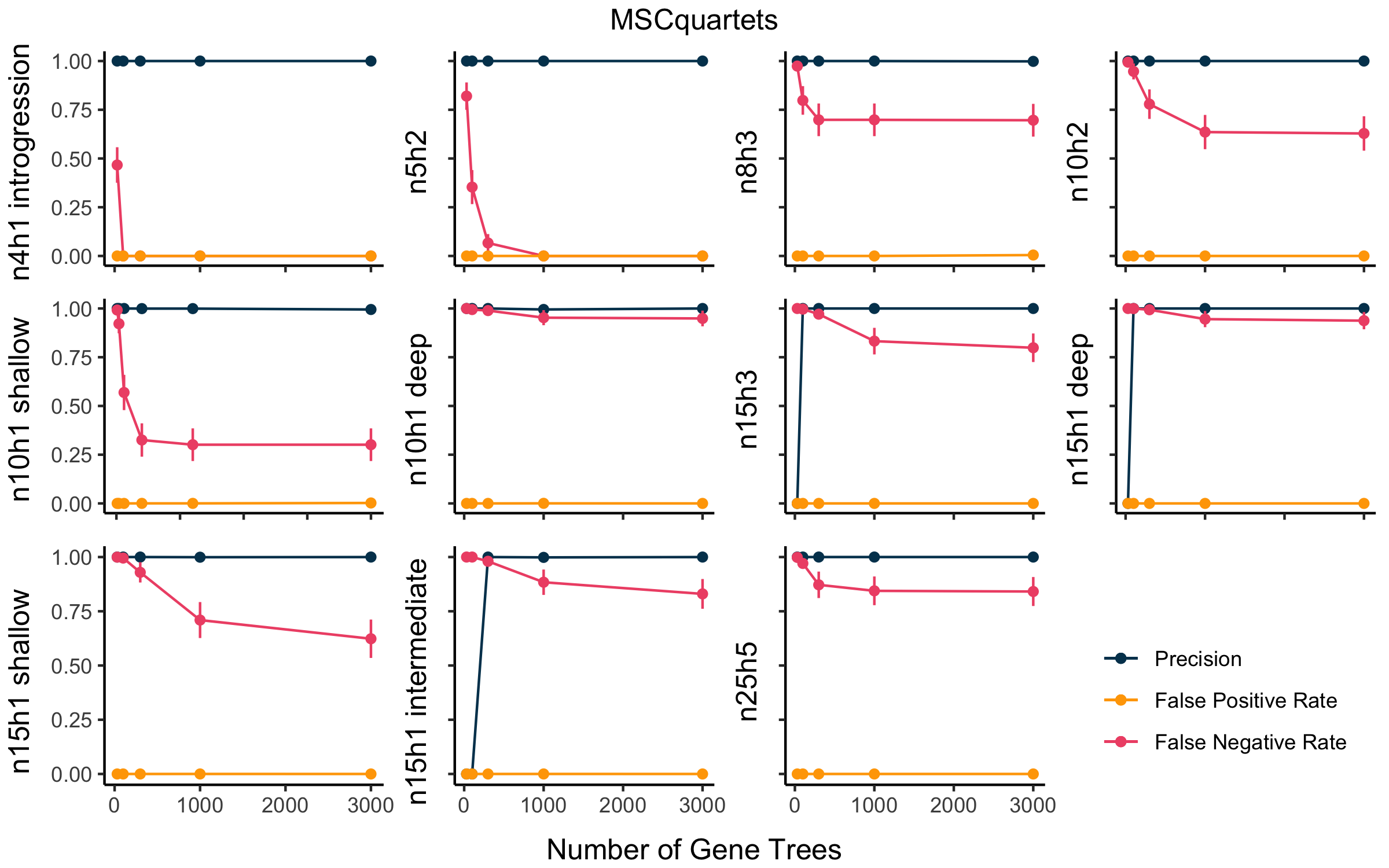}
    \caption{False positive rate (orange), precision (black) and false negative rate (red) for MSCquartets using inputs derived from IQTREE on n4 introgression, n5h2, n8h3, n10h2, n10 shallow, n10 deep, n15h3, n15 shallow, n15 intermediate, n15 deep, and n25h5. In this figure, all tests are Bonferroni-corrected at a level of significance $\alpha = 0.05 / \textrm{number of tests}$.}
    \label{fig:IQTREE-msc-n4-n25-bonferroni}
\end{figure*}

\begin{figure*}
    \centering
    \includegraphics[width=.9\linewidth]{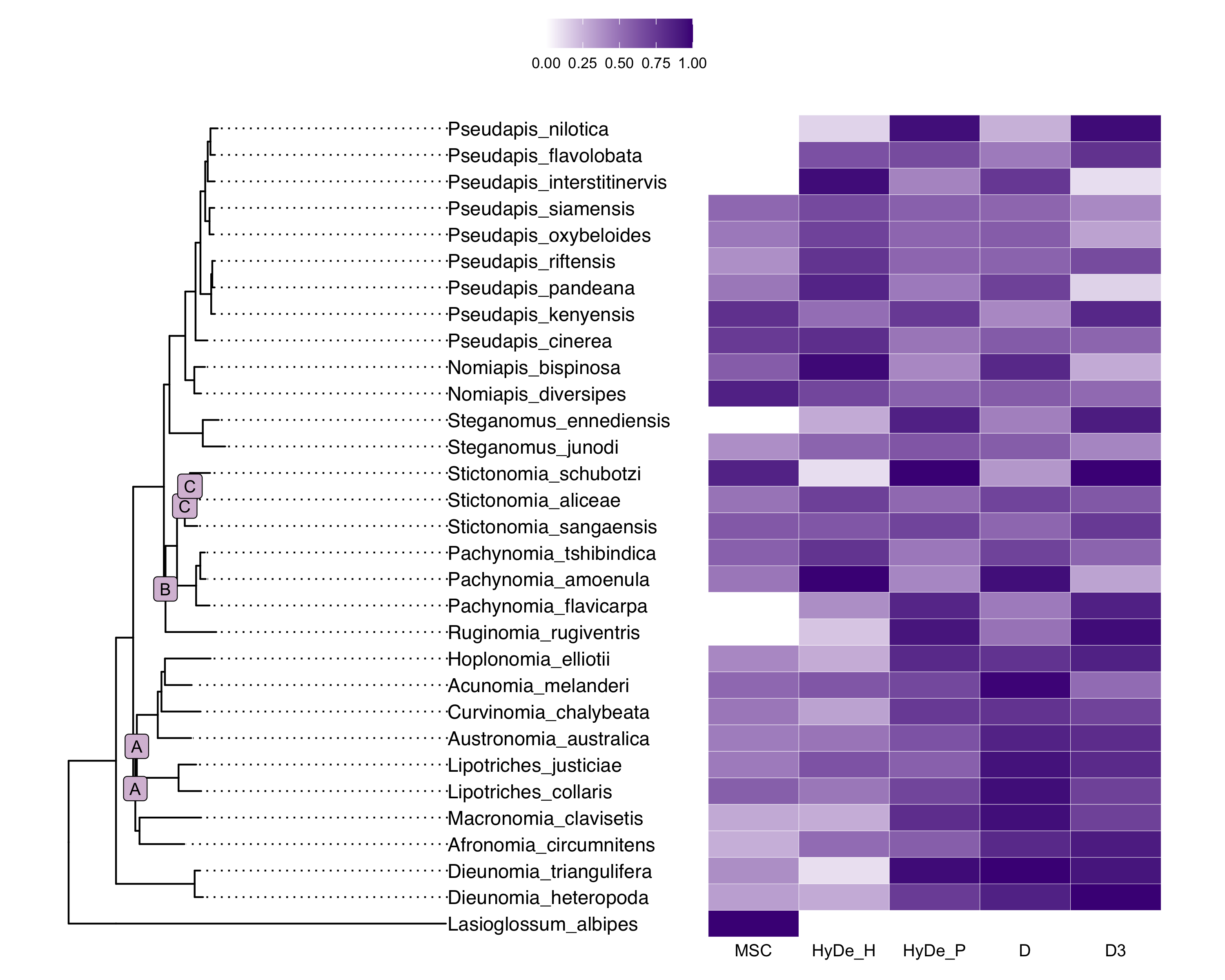}
    \caption{Left: Phylogenetic tree of selected species from the bee subfamily Nomiinae, as provided by \cite{bees}. Letters A-C identify nodes that were identified as conflicting relationships by the analyses conducted in \cite{bees}. Right: Heatmap of frequency of taxa identified as part of hybridization events (scaled between 0 and 1 for comparison across methods). MSCquartets was conducted with gene trees estimated using the IQTree2-GTRG model. Tests are performed at significance level $\alpha=0.05$.}
    \label{fig:bee-fig-alpha}
\end{figure*}

\end{document}